\documentclass[twocolumn,prb,citeautoscript,superscriptaddress,8pt]{revtex4-2}

\usepackage{graphicx}
\usepackage{gensymb}
\usepackage{color}
\usepackage[dvipsnames]{xcolor}
\usepackage{float}
\usepackage{upgreek}
\usepackage{bm}
\usepackage{amsmath,amssymb}
\usepackage{tabularx}
\usepackage{xfrac}
\usepackage{ulem}
\usepackage{comment}
\usepackage{hyperref}

\newcommand{\VB}{\ensuremath{V_{\rm B}^{-}}}
\newcommand{\C}{\ensuremath{{\rm C}_?}}

\newcommand{\bra}[1]{\left\langle#1\right\rvert}
\newcommand{\ket}[1]{\left\lvert#1\right\rangle}

\newcommand{\vect}[1]{\bm{#1}}

\newcommand\add[1]{{\color{black}#1}} 
\newcommand\sub[1]{} 

\begin{document}

\title{Multi-species optically addressable spin defects in a van der Waals material}

\author{Sam~C.~Scholten} 
\thanks{These authors contributed equally to this work.}
\affiliation{School of Science, RMIT University, Melbourne, VIC 3001, Australia}
\affiliation{School of Physics, University of Melbourne, VIC 3010, Australia}
\affiliation{Centre for Quantum Computation and Communication Technology, School of Physics, University of Melbourne, VIC 3010, Australia}

\author{Priya~Singh} 
\thanks{These authors contributed equally to this work.}
\affiliation{School of Science, RMIT University, Melbourne, VIC 3001, Australia}

\author{Alexander~J.~Healey} 
\affiliation{School of Science, RMIT University, Melbourne, VIC 3001, Australia}

\author{Islay~O.~Robertson} 
\affiliation{School of Science, RMIT University, Melbourne, VIC 3001, Australia}

\author{Galya~Haim} 
\affiliation{School of Science, RMIT University, Melbourne, VIC 3001, Australia}
\affiliation{School of Physics, University of Melbourne, VIC 3010, Australia}
\affiliation{Department of Applied Physics, The Hebrew University of Jerusalem, Safra Campus, Givat Ram, Jerusalem 91904, Israel}

\author{Cheng~Tan} 
\affiliation{School of Science, RMIT University, Melbourne, VIC 3001, Australia}

\author{David~A.~Broadway} 
\affiliation{School of Science, RMIT University, Melbourne, VIC 3001, Australia}

\author{Lan~Wang} 
\affiliation{School of Science, RMIT University, Melbourne, VIC 3001, Australia}
\affiliation{Low Dimensional Magnetism and Spintronic Devices Lab,
School of Physics, Hefei University of Technology, Hefei, Anhui 230009, China}

\author{Hiroshi~Abe} 
\affiliation{National Institutes for Quantum Science and Technology (QST), 1233 Watanuki, Takasaki, Gunma 370-1292, Japan}

\author{Takeshi~Ohshima} 
\affiliation{National Institutes for Quantum Science and Technology (QST), 1233 Watanuki, Takasaki, Gunma 370-1292, Japan}
\affiliation{Department of Materials Science, Tohoku University, 6-6-02 Aramaki-Aza, Aoba-ku, Sendai 980-8579, Japan}

\author{Mehran~Kianinia} 
\affiliation{School of Mathematical and Physical Sciences, University of Technology Sydney, Ultimo, NSW 2007, Australia}
\affiliation{ARC Centre of Excellence for Transformative Meta-Optical Systems, University of Technology Sydney, Ultimo, NSW 2007, Australia}

\author{Philipp~Reineck} 
\affiliation{School of Science, RMIT University, Melbourne, VIC 3001, Australia}

\author{Igor~Aharonovich}
\email{igor.aharonovich@uts.edu.au}
\affiliation{School of Mathematical and Physical Sciences, University of Technology Sydney, Ultimo, NSW 2007, Australia}
\affiliation{ARC Centre of Excellence for Transformative Meta-Optical Systems, University of Technology Sydney, Ultimo, NSW 2007, Australia}

\author{Jean-Philippe~Tetienne}
\email{jean-philippe.tetienne@rmit.edu.au}
\affiliation{School of Science, RMIT University, Melbourne, VIC 3001, Australia}

\begin{abstract} 

Optically addressable spin defects hosted in two-dimensional van der Waals materials represent a new frontier for quantum technologies, promising to lead to a new class of ultrathin quantum sensors and simulators. 
Recently, hexagonal boron nitride (hBN) has been shown to host several types of optically addressable spin defects, thus offering a unique opportunity to simultaneously address and utilise various spin species in a single material. 
Here we demonstrate an interplay between two separate spin species within a single hBN crystal, namely $S=1$ boron vacancy defects and \add{carbon-related electron} spins. 
\add{We reveal the $S=\sfrac{1}{2}$ character of the carbon-related defect} and further demonstrate room temperature coherent control and optical readout of both \add{$S=1$ and $S=\sfrac{1}{2}$} spin species.
By tuning the two spin \add{ensembles} into resonance with each other, we observe cross-relaxation indicating strong inter-species dipolar coupling. 
We then demonstrate magnetic imaging using the $S=\sfrac{1}{2}$ defects and leverage their lack of intrinsic quantization axis to \add{probe the magnetic anisotropy} of a test sample. 
Our results establish hBN as a versatile platform for quantum technologies in a van der Waals host at room temperature. 

\end{abstract}

\maketitle 


Hexagonal boron nitride (hBN) has come to prominence as a host material for optically addressable spin defects for quantum technology applications~\cite{GottschollNM2020,Liu2022review,Aharonovich2022,Vaidya2023}. 
The layered van der Waals structure of hBN makes it particularly appealing for nanoscale quantum sensing and imaging, as the spin defects can in principle be confined within hBN flakes just a few atoms thick~\cite{Durand2023}. 
The prospect of two-dimensional (2D) confinement of a dipolar spin system is also attractive for quantum simulations, as it would open the door to realising exotic ground-state phases such as spin liquids~\cite{Yao2018,Davis2023} as well as exploring many-body localization and thermalization in 2D~\cite{Choi2016,Choi2017,Abanin2019,Zu2021,Gong2023}. 
To date, only the negatively charged boron vacancy ($\VB$) defect has been used for such quantum applications~\cite{GottschollNM2020}. 
The $\VB$ defect is a ground-state spin triplet ($S=1$) with a quantization axis along the $c$-axis of the hBN crystal and a zero-field splitting of $D\approx3.45$~GHz between the $m_S=0$ and $m_S=\pm1$ spin sublevels. 
Owing to a spin-dependent intersystem crossing, the electronic spin transitions of the $\VB$ defect, i.e.\ $\ket{0}\leftrightarrow\ket{\pm1}$, can be probed via optically detected magnetic resonance (ODMR).
The sensitivity of these transitions to the defect's environment in turn enables accurate measurements of magnetic fields, temperature and strain, as well as spatial imaging of these fields~\cite{Gottscholl2021NC,Liu2021,HealeyNP2023,HuangNC2022,Lyu2022,Kumar2022,Robertson2023,Gao2023}.

\begin{table*}[t!]
{\begin{tabular}{l|l|l|l|l}
{\bf Sample description} & {\bf Irradiation} & {\bf Presence of $\C$ defects} & {\bf Max. ODMR} & {\bf Sample morphology} \\ 
 &  &  & {\bf contrast} &  {\bf for measurements} \\ 
\hline
Nanopowder & 2 MeV & High density as purchased & $-1.0\%$ (batch     1) & powder film, $\sim$1-10\,$\upmu$m thick (Fig. 1) \\ 
(Graphene Supermarket) & electrons &  & $+1.0\%$ (batch 2) & powder film, $\sim$1-10\,$\upmu$m thick (Fig.~S19) \\ 
\hline
Bulk crystal & 2 MeV & Low density as purchased, & $+0.5\%$ & whole crystal, $\sim$100\,$\upmu$m thick (Fig.~3) \\ 
(HQ Graphene) & electrons & high density after irradiation &  & exfoliated flake, $\sim$1\,$\upmu$m thick (Fig.~2) \\ 
\hline
MOVPE film \cite{Mendelson2021} & none & High density as grown  & $+0.5\%$  & film, $\approx40\,$nm thick (Fig.~4) \\ 
\end{tabular}}
\caption{\add{{\bf Summary of the hBN samples used in this work.} 
Column 3 indicates the qualitative density of $\C$ defects (low/high density means single emitters can/cannot be resolved). Column 4: Maximum CW-ODMR contrast observed for the $\C$ defects. 
}}	
\label{tab:samples}
\end{table*}

\begin{figure*}[t!]
\centering
\includegraphics[width=0.98\textwidth]{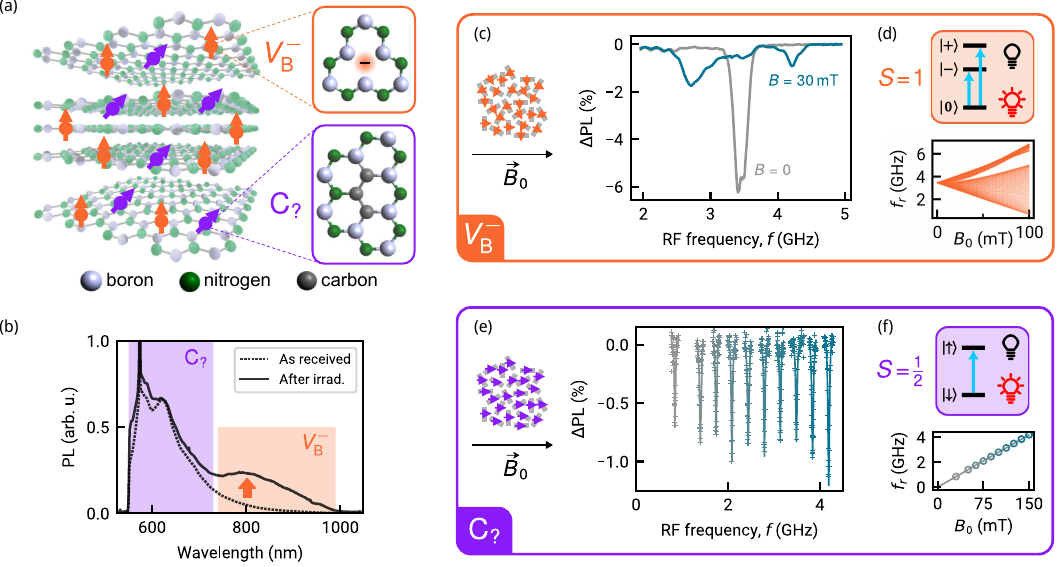}
\caption{{\bf Multiple optically addressable spin defects in hBN.} 
(a)~Schematic representation of two co-existing species of spin defects in hBN: the boron vacancy defect ($\VB$, orange arrows) and a carbon-related defect (referred to as $\C$, purple arrows). 
An example candidate for $\C$ is shown, namely the C$_2$C$_{\rm N}$ defect.
(b)~Photoluminescence (PL) spectra of hBN powder as received (dotted line) and after electron irradiation (solid line), under $\lambda=532$\,nm laser excitation. 
The shaded areas indicate the main emission band of $\C$ (present in the as-received powder) and $\VB$ (created by irradiation).
(c)~Optically detected magnetic resonance (ODMR) spectra of the $\VB$ defects (PL band 750-1000\,nm) in hBN powder at $B_0=0$ (grey line) and $B_0=30$\,mT (blue). 
(d)~Energy level diagram of the $\VB$ spin triplet. 
The spin eigenstates are denoted as $\ket{0,\pm}$ in the general case of an arbitrarily oriented magnetic field. 
The graph shows the calculated spin resonance frequencies $f_r$ as a function of $B_0$ for a range of field orientations. 
(e)~ODMR spectra of the $\C$ defects (PL band 550-700\,nm) in the same hBN powder as in (c), at different field strengths $B_0$ from 30\,mT to 150\,mT (left to right). 
(e)~Energy level diagram of the $\C$ effective spin doublet. 
The graph shows the spin resonance frequency $f_r$ inferred from (e) as a function of $B_0$ (dots). 
The solid line is a linear fit, indicating a $g$-factor of 2.0(1). 
}
\label{fig1}
\end{figure*} 

Meanwhile, several groups have reported the observation of a family of hBN defects emitting primarily at visible wavelengths and exhibiting ODMR with no apparent (or very weak) zero-field splitting~\cite{Mendelson2021,Chejanovsky2021,Stern2022,Guo2023,Yang2023} akin to an effective spin doublet ($S=\sfrac{1}{2}$). 
The exact structure of these defects as well as their spin multiplicity remain unknown, although they are generally believed to be related to carbon impurities~\cite{Mendelson2021,Auburger2021,Jara2021,Golami2022,Li2022}. 
The deterministic creation of these spin defects is also an ongoing challenge~\cite{Mendelson2021,Aharonovich2022,Yang2023} and consequently, these $S=\sfrac{1}{2}$-like defects have not been exploited in sensing applications despite the unique possibilities afforded by the lack of intrinsic quantization axis. 
More generally, having multiple optically addressable spin systems within a single layered solid would open new opportunities for quantum technologies. 

In this Article, we demonstrate the co-existence of two distinct optically addressable spin species in hBN. 
The two spin defects at the heart of this work are represented in Fig.~\ref{fig1}(a). 
The first is the $\VB$ defect, which emits in the near-infrared~\cite{GottschollNM2020}. 
The second spin species is the carbon-related defect emitting in the visible, which we will refer to as $\C$ -- a proposed candidate is the carbon trimer C$_2$C$_{\rm N}$~\cite{Golami2022,Li2022}.
\add{We found that these $\C$ spin defects are present in a variety of hBN samples both commercially sourced (powders and bulk crystals) and lab-grown by metal-organic vapour-phase epitaxy (MOVPE), see Table~\ref{tab:samples}. 
In the following, we leverage the unique characteristics of each sample to demonstrate the isotropic character of the $\C$ defects (using hBN powder), explore the interplay between $\VB$ and $\C$ spins (using a bulk crystal), and demonstrate quantum sensing using the $\C$ defect (using a thin MOVPE film).}

We first consider a commercially sourced hBN powder \add{(see Table~\ref{tab:samples} and Fig.~S2-S5)}. The photoluminescence (PL) spectrum of the as-received powder under laser excitation ($\lambda=532$\,nm wavelength) shows visible PL emission characteristic of the $\C$ defect~\cite{Mendelson2021}, mainly centred around $\lambda=550$-650\,nm with a tail extending in the near-infrared up to 900\,nm [Fig.~\ref{fig1}(b)]. 
To create $\VB$ defects, the as-received powder was irradiated with high-energy electrons, causing the appearance of a broad near-infrared emission peak centered around 820\,nm characteristic~\cite{GottschollNM2020} of the $\VB$ defect [Fig.~\ref{fig1}(b)]. 

Next, we used radiofrequency (RF) fields to probe the spin transitions of the defects via spin-dependent PL.
A continuous-wave \add{(CW)} ODMR spectrum of the $S=1$ $\VB$ defect ensemble is obtained by collecting the $\VB$ emission (750-1000\,nm) as a function of RF frequency [Fig.~\ref{fig1}(c)].
Under zero magnetic field ($B_0=0$), a single resonance at $f_r=D\approx3.45$~GHz is observed corresponding to the nearly degenerate electronic spin transitions $\ket{0}\leftrightarrow\ket{\pm1}$ [see energy level diagram in Fig.~\ref{fig1}(d)]. 
Upon applying a magnetic field, this central resonance splits into two broad resonances due to the Zeeman effect, where the significant broadening is the result of the random orientation of the defects in the powder, as shown in the resonance frequencies calculated for a range of orientations [Fig.~\ref{fig1}(d)].

On the other hand, the ODMR spectrum of the $\C$ defects (550-700\,nm) shows a single resonance at a frequency $f_r$ that scales linearly with the applied field $B_0$ [Fig.~\ref{fig1}(e)], thus resembling a $S=\sfrac{1}{2}$ electronic system. 
Fitting the data with $hf_r=g_C\mu_BB_0$ where $h$ is Planck's constant and $\mu_B$ the Bohr magneton, yields a $g$-factor of $g_C=2.0(1)$ [Fig.~\ref{fig1}(f)]. 
\add{The powdered nature of the sample and the relatively large orientation-averaged ODMR contrast (of about $-1\%$) provide a strong indication that the mechanism underpinning ODMR is intrinsically isotropic, i.e.\ independent of the direction of the applied field relative to the crystal orientation, rather than relying on an intersystem crossing in a $S\geq1$ spin system as was suggested previously~\cite{Stern2022}.}

\begin{figure}[tb!]
\centering
\includegraphics[width=1\linewidth]{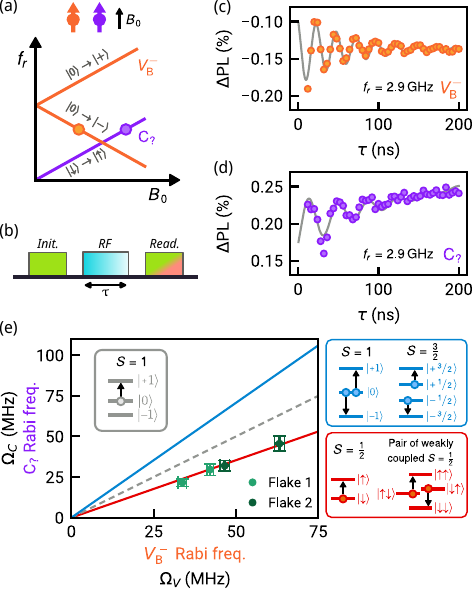}
\caption{{\bf Coherent spin control at room temperature and $S=\sfrac{1}{2}$ \add{character} of the $\C$ defect.} 
(a)~Diagram depicting the spin resonance frequencies $f_r$ of the $\VB$ and $\C$ defects as a function of the magnetic field strength $B_0$ applied parallel to the $c$-axis of the hBN crystal. 
(b)~Pulse sequence used for the Rabi measurement.
(c,d)~Rabi oscillations of (c) the $\VB$ defects at $B_0=20$\,mT and of (d) the $\C$ defects at $B_0=100$\,mT in an exfoliated single-crystal hBN flake. 
The resonance frequency being driven is $f_r=2.9$\,GHz in both cases \add{[see circles in (a)]}.
(e)~Rabi frequency of the $\C$ defects ($\Omega_C$) as a function of the Rabi frequency of the $\VB$ defects ($\Omega_V$). 
Measurements taken from two different flakes are shown, see raw data in Fig.~S9. 
The error bars correspond to one standard error in the Rabi fit, see details in the SI, Sec.~V.B and Fig.~S10-S11. 
The solid lines correspond to the expectation under different assumptions for the spin multiplicity of the $\C$ defects: \add{a single $S=\sfrac{1}{2}$ or a pair of weakly coupled $S=\sfrac{1}{2}$ particles (red line); a pure $S=1$ or $S=\sfrac{3}{2}$ system with negligible zero-field splitting (blue line).} 
The dashed line is the $\Omega_C=\Omega_V$ reference corresponding to driving a single transition of a $S=1$ system.}
\label{fig2}
\end{figure}

\add{To gain further insights into the nature of the $\C$ spin defect and the origin of the ODMR contrast, we now use a flake exfoliated from a bulk electron-irradiated hBN crystal (see Table~\ref{tab:samples} and Fig. S6-S7, here the ODMR contrast is about $+0.5\%$). 
The single crystal orientation allows us to align the magnetic field with the $\VB$ spin's quantization axis and drive a single spin transition, e.g.\ $\ket{0}\rightarrow\ket{-1}$ [Fig.~\ref{fig2}(a)], whose coherent dynamics can then be precisely compared to that of the $\C$ spins. 
First,} we perform a Rabi experiment whereby each spin ensemble is driven by a resonant RF pulse of variable duration and the corresponding PL monitored [Fig.~\ref{fig2}(b)]. 
Rabi oscillations are observed both for the $\VB$ ensemble [Fig.~\ref{fig2}(c)] and the $\C$ ensemble [Fig.~\ref{fig2}(d)], demonstrating coherent control and optical readout of two distinct spin species within the same host material, at room temperature. 

\add{We can now address the question of the spin multiplicity of the $\C$ defect, making use of the fact that the Rabi flopping frequency ($\Omega_C$) directly depends on the spin quantum number ($S$) of the driven species according to $h\Omega_C=f(S)g_C\mu_BB_1$ where $B_1$ is the magnitude of the driving field and $f(S)$ is a dimensionless function of $S$. 
To calibrate $B_1$, we measure the Rabi frequency of the $S=1$ $\VB$ ensemble, which is given by $h\Omega_V=g_V\mu_BB_1/\sqrt{2}$ where $g_V=2.00\approx g_C$ is the $\VB$ $g$-factor~\cite{GottschollNM2020}, and compare it to $\Omega_C$, measured at the same resonance frequency $f_r$ [Fig.~\ref{fig2}(a)] to ensure $B_1$ is identical in both Rabi measurements. 
We find that $\Omega_C$ is consistently smaller than $\Omega_V$, with a ratio $\Omega_C/\Omega_V\approx1/\sqrt{2}$ that reveals the $S=\sfrac{1}{2}$ character of the $\C$ defect's magnetic resonance [Fig.~\ref{fig2}(e)]. 
Consequently, we can rule out a pure $S=1$ or $S=3/2$ system, which would give $\Omega_C/\Omega_V=\sqrt{2}$ (blue box, see SI Sec.~V.A for a detailed discussion).

Rather than a single $S=\sfrac{1}{2}$ system, which is not normally expected to give rise to ODMR, we propose a more specific interpretation whereby the $\C$ defect would instead involve a pair of weakly coupled $S=\sfrac{1}{2}$ particles (e.g.\ two unpaired electrons). 
Such a system also satisfies $\Omega_C/\Omega_V=1/\sqrt{2}$~\cite{McCamey2010}, while providing a more natural explanation for the ODMR contrast by invoking different photon absorption/emission probabilities for the singlet and triplet configurations of the spin pair~\cite{Davies1988,Boehme2003,McCamey2008}. 
In this scenario, the optical readout would distinguish the parallel spin states (pure triplet states $\ket{\uparrow\uparrow}$ and $\ket{\downarrow\downarrow}$) from the anti-parallel states (singlet-triplet mixtures of $\ket{\uparrow\downarrow}$ and $\ket{\downarrow\uparrow}$). 
In the SI, we provide evidence supporting the spin pair theory (Sec.~V.E and Fig.~S12).
}

\begin{figure*}[tb!]
\centering
\includegraphics[width=1.0\linewidth]{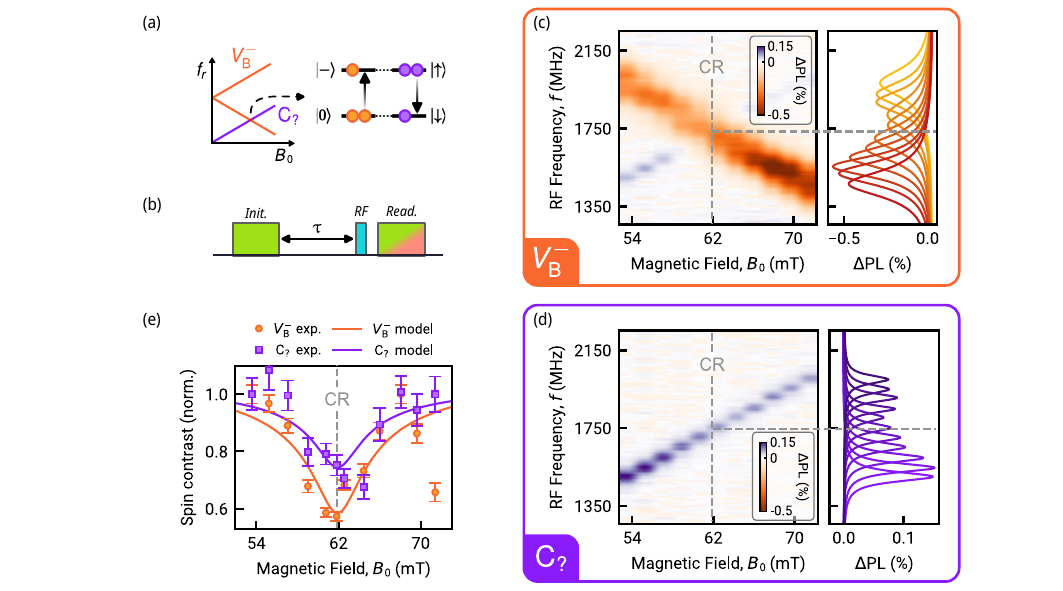}
\caption{{\bf Cross-relaxation between two optically addressable spin species.} 
(a)~Diagram illustrating the cross-relaxation (CR) resonance condition between the $\VB$ and $\C$ spin ensembles, at which the ensembles exchange energy. \add{The populations (circles) are for illustrative purpose only, as the sign and level of spin polarisation of the $\C$ spins is unknown.}   
(b)~Schematic of the pulsed ODMR sequence, which includes an interaction time $\tau=350$\,ns before applying the RF pulse. 
(c,d)~Pulsed ODMR spectra as a function of $B_0$ for the $\VB$ (c) and $\C$ (d) defects near the CR resonance. 
Lorentzian fits to the individual ODMR spectra are plot together on the right with different colours for the different values of $B_0$. The corresponding raw data is shown in Fig.~S13. 
(e)~Spin contrast of each spin ensemble as a function of $B_0$, corresponding to the ODMR contrast extracted from (c,d) after a linear slope was subtracted to remove RF power variations across the frequency range, normalised to 1 away from the CR resonance. 
The solid lines are the result of a model of the CR effect, see main text.}
\label{fig3}
\end{figure*}

\add{So far we have used the $\VB$ and $\C$ spin ensembles independently.}
We now study the dynamics and interplay between the two spin species.
To this end, we use again a single-crystal hBN sample containing both $\VB$ and $\C$ spins, and apply a magnetic field to tune the $\ket{0}\leftrightarrow\ket{-1}$ transition of the $\VB$ spins in resonance with the $\ket{\uparrow}\leftrightarrow\ket{\downarrow}$ transition of the $\C$ spins [Fig.~\ref{fig3}(a)\add{, here $\ket{\uparrow}$ and $\ket{\downarrow}$ refer to the states of either of the spin pair partners}]. 
At the resonant field of $B_0=hD/[(g_C+g_V)\mu_B]\approx62$\,mT, the two spin ensembles can exchange energy causing an increased relaxation of their respective spin populations [Fig.~\ref{fig3}(a)], a process known as cross-relaxation (CR)~\cite{Hall2016,Broadway2018}.

To probe the CR resonance, \add{we use a fixed} free interaction time $\tau$ between the initial laser pulse (which \add{partially} polarises both spin species) and the probe RF pulse [Fig.~\ref{fig3}(b)], \add{while varying the field strength $B_0$. 
For each value of $B_0$, the RF frequency of the probe pulse is scanned to construct a pulsed ODMR spectrum [Fig.~\ref{fig3}(c,d)] and extract a normalised spin contrast [plot against $B_0$ in Fig.~\ref{fig3}(e)]. 
This spin contrast is a measure of the amount of spin polarisation averaged over the sequence, which depends on the CR coupling present during the free interaction time $\tau$ as well as during the laser pulse where CR competes with optically induced spin polarisation.}
\add{Examining the $\VB$ spin contrast first, we see that} the $\VB$ ensemble experiences a reduced \add{spin polarisation} at the CR resonance [Fig.~\ref{fig3}(c)], indicating coupling with a bath of $S=\sfrac{1}{2}$ electron spins, which includes the $\C$ defects~\cite{Haykal2022DecoherenceNitride,Baber2022}. 
Crucially, the $\C$ defects also exhibit a reduced \add{spin polarisation} at the CR resonance [Fig.~\ref{fig3}(d)], providing unambiguous evidence of $\C$-$\VB$ coupling.

We model the CR effect by considering the coupling of each spin to its nearest neighbour from the opposite spin species (see details in the SI, Sec.~VI.B and Fig.~S14). \add{For simplicity we assume the system to be initialised into the interacting state. 
We note that changing the initialisation using a combination of fast magnetic field control and RF pulses~\cite{Fuchs2011} could in principle reveal the direction of spin polarisation of the $\C$ defect.} 
For the $\C$ spins, the mean distance to the nearest $\VB$ is approximately 5\,nm based on the estimated density of $[\VB]=10^{18}$\,cm$^{-3}$. 
The resulting model with no free parameter [purple line in Fig.~\ref{fig3}(e)] is in good agreement with the data, confirming that we do indeed detect CR between the $\C$ and $\VB$ spin ensembles. 
For the $\VB$ case, we leave the total density of $S=\sfrac{1}{2}$ spins free; best fit to the data is obtained for $[S=\sfrac{1}{2}]=5\times10^{18}$\,cm$^{-3}$ \add{corresponding to a mean distance between nearest neighbours of approximately 3\,nm. 
This cross-relaxation experiment thus reveals the high density of $S=\sfrac{1}{2}$ spins in this hBN sample, and could be applied to determine the spin content of hBN flakes subject to various treatments. 
More generally,} the demonstrated coupling between two optically addressable spin species of different multiplicities \add{could be a useful new resource for future quantum technologies}.

\begin{figure*}[tb!]
\centering
\includegraphics[width=1\linewidth]{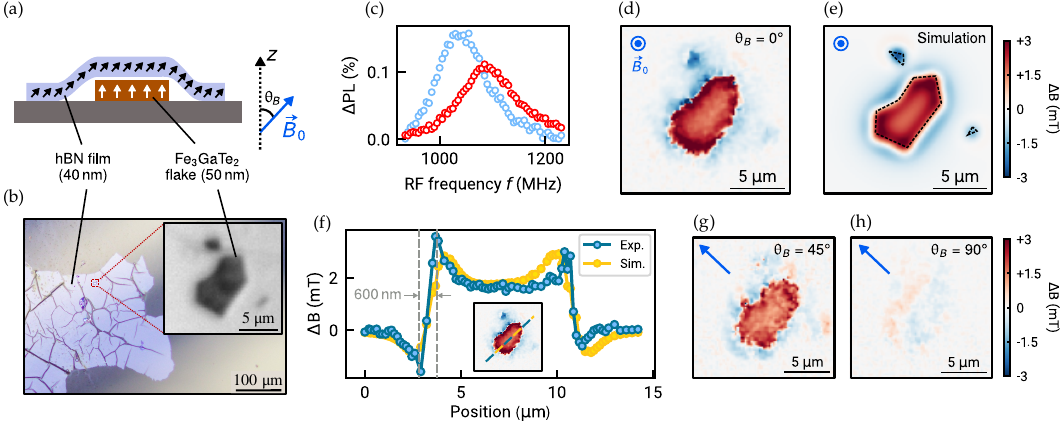}
\caption{\add{{\bf Magnetic imaging with $S=\sfrac{1}{2}$ defects in hBN.} 
(a)~Schematic of the experiment. A 40-nm-thick hBN film containing $\C$ spins (black arrows) is placed on top of a Fe$_3$GaTe$_2$ flake exhibiting perpendicular magnetic anisotropy (white arrows). 
The external field $\vec{B}_0$ sets the quantization axis of the $\C$ spins. 
(b)~Optical micrograph of the hBN film covering multiple Fe$_3$GaTe$_2$ flakes. 
Inset: magnified view of the Fe$_3$GaTe$_2$ flake studied in (c-h). 
(c)~ODMR spectra taken at the centre of the flake (red data) and outside the flake (black), with $B_0\approx35$\,mT pointing along the $z$ direction ($\theta_B=0^\circ$). 
The dashed lines are Lorentzian fits. 
(d)~Stray magnetic field image of a Fe$_3$GaTe$_2$ obtained with the $\C$ defects at room temperature.
(e)~Simulated stray field for a uniformly magnetized flake (see details in SI Sec.~VII.C).
(f)~Line profile of the measured and simulated stray field across the flake, taken along the dashed line displayed in (e). 
(g,h)~Stray field images obtained with the same applied field magnitude $B_0\approx35$\,mT as in (d) but with an angle (g) $\theta_B=45^\circ$ and (h) $\theta_B=90^\circ$. 
The blue arrow in (g,h) indicates the direction of the in-plane component of $\vec{B}_0$.
}}
\label{fig4}
\end{figure*}

Having established the $S=\sfrac{1}{2}$ \add{character} of the $\C$ defect, we now explore its potential for quantum sensing and imaging. 
The key advantage of the $\C$ spin is its lack of intrinsic quantization axis, allowing magnetometry to be performed under any direction of the external magnetic field.
This feature is in contrast to spin defects with uniaxial anisotropy such as the $\VB$ defect or the nitrogen-vacancy centre in diamond which are restricted to a narrow range of field directions and magnitudes due to spin mixing caused by transverse fields~\cite{Tetienne2012}. 

\add{To demonstrate magnetic imaging, we used a 40-nm-thick hBN film grown by MOVPE with precursors tuned to create a dense and uniform ensemble of $\C$ spin defects~\cite{Chugh2018,Mendelson2021} (see Table~\ref{tab:samples} and Fig.~S8). 
Importantly, large flakes (hundreds of micrometres laterally) of the continuous wafer-scale film can be transferred from the growth substrate to the target sample [Fig.~\ref{fig4}(a) and~S15] while preserving a uniform, well-defined thickness, an advantage over exfoliation from a bulk hBN crystal.

As a test magnetic sample, we used micron-sized flakes of Fe$_3$GaTe$_2$, a ferromagnetic van der Waals material with a Curie temperature just above room temperature and a perpendicular magnetic anisotropy~\cite{Zhang2022}. 
A single piece of the hBN film enables us to cover a multitude of Fe$_3$GaTe$_2$ flakes [Fig.~\ref{fig4}(b)] which can then be magnetically imaged by performing widefield ODMR measurements of the \C~defects. 
ODMR spectra on and off the magnetic flakes reveal local Zeeman shifts of up to $\Delta f_r\approx100$\,MHz [Fig.~\ref{fig4}(c)], corresponding to stray fields $\Delta B=h\Delta f_r/g_C\mu_B\approx3$\,mT. 
Since the magnitude of the applied field is  $B_0\approx35\,{\rm mT}\gg \Delta B$, the measured sample's stray field $\Delta B$ corresponds to the projection along the direction of $\vec{B}_0$.

By fitting the ODMR spectrum at each pixel of the image, we can construct a 
$\Delta B$ map of a Fe$_3$GaTe$_2$ flake, for example in Fig.~\ref{fig4}(d) where $\vec{B}_0$ is along the easy axis of magnetisation ($\theta_B=0^\circ$). 
The flake appears uniformly magnetised, as confirmed by the good agreement with the simulation [Fig.~\ref{fig4}(e)] which assumed a uniform out-of-plane magnetisation of $M_s=160$\,kA/m, in line with the expected value at room temperature~\cite{Zhang2022}. 
A more detailed magnetisation map of the flake can be reconstructed from the data through a reverse propagation method [Fig.~S18]. 
A line cut across the flake [Fig.~\ref{fig4}(f)] indicates a spatial resolution of about 600\,nm in good agreement with the diffraction limit of our setup, which could be improved to 300 nm using a higher numerical-aperture objective.

As $\vec{B}_0$ is rotated away from the easy axis, the stray field becomes weaker overall [Fig.~\ref{fig4}(g,h)] and nearly vanishes in the purely in-plane case ($\theta_B=90^\circ$). 
This indicates the formation of multiple domains of opposite signs within the flake in the absence of a stabilising out-of-plane field, consistent with the findings in Ref.~\cite{Zhang2022}. 
Note that the pattern of the residual stray field at $\theta_B=90^\circ$ suggests the remanent magnetisation still points along the easy axis (see Fig.~S17), confirming the strong out-of-plane anisotropy of the material. Angle-dependent images of additional flakes are shown in Fig.~S16.

These measurements illustrate the utility of the isotropic $\C$ defect to study micron-sized magnetic samples. 
The omnidirectional magnetometry allows the application of magnetic fields along arbitrary easy or hard axes and spanning a broad range of magnitudes from about 10\,mT (below which the ODMR contrast drops) up to 300\,mT or more (only limited by the RF electronics), allowing the precise determination of unknown anisotropies. 
As an example, it could be applied to test the potential existence of easy axes within in-plane ferromagnets such as monolayer CrCl$_3$, which is believed to exhibit 2D-XY ferromagnetism~\cite{Bedoya2021} but has not been systematically investigated as a function of the azimuthal angle of $\vec{B}_0$. 
Such measurements are generally inaccessible or impractical with current stray field techniques.

The shot-noise-limited magnetic sensitivity of our 40-nm-thick MOVPE film is about 100\,$\upmu$T/$\sqrt{\rm Hz}$ per (500\,nm)$^2$ pixel (see SI, Sec.~VII.E), which is an order of magnitude better than that obtained with $\VB$ defects in flakes of similar thickness~\cite{HealeyNP2023,HuangNC2022} thanks to the higher brightness of the $\C$ defects. 
Crucially, this sensitivity is already sufficient to image atomically thin ferromagnets and investigate the emerging field of Moir{\'e} magnetism~\cite{Huang2023}, for example, and could be boosted by using advanced sensing protocols~\cite{Rizzato2023,Ramsay2023} and potentially by optimising sample fabrication.

Besides the unique omnidirectional magnetometry capability and the relative ease of fabrication compared to existing quantum sensing platforms like diamond, another advantage of the hBN platform is the possibility of leveraging the multiple spin species available. 
In particular, the $\C$ defect is an ideal complement to $\VB$-based temperature and strain sensing~\cite{Gottscholl2021NC} as it is insensitive to variations in those quantities (an expected consequence of its $S=\sfrac{1}{2}$ character and negligible orbital magnetic moment), and remains a reliable magnetometer regardless of sensor geometry (for instance in the presence of flake rippling or powder averaging). 
In this sense, a co-present $\C$ ensemble can be viewed as augmenting a $\VB$ ensemble tasked with temperature and strain sensing, with orientation-independent magnetic sensitivity. 
We envisage that this dual-spin multi-modal imaging capability will present new opportunities for studying phase transitions where the magnetic properties of a sample can be spatially correlated with the temperature profile or mechanical properties of a sample, and illustrate this capability with a proof-of-principle experiment, see Fig.~S19 and S20.}

More generally, the co-existence of two distinct spin ensembles within a single host material, which can both be optically initialised, read out, and coherently driven, at room temperature, distinguishes hBN from established material platforms such as diamond and silicon carbide in which only one type of optically addressable spin defect can generally be stabilised in a given sample~\cite{Wolfowicz2021QuantumDefects}. 
\add{Combined with the prospect of 2D confinement afforded by the layered structure of hBN, which appears realistic given the recent observation of $\VB$~defects in few-layer hBN~\cite{Durand2023}, 
hBN emerges as a rich and promising platform for future quantum technologies.}

\section*{Data availability}

The data supporting the findings of this study are available within the paper and its supplementary information files.

\begin{acknowledgments}
This work was supported by the Australian Research Council (ARC) through grants CE170100012, CE170100039, CE200100010, FT200100073, FT220100053, DE200100279, DE230100192, and DP220100178, and by the Office of Naval Research Global (N62909-22-1-2028). 
The work was performed in part at the RMIT Micro Nano Research Facility (MNRF) in the Victorian Node of the Australian National Fabrication Facility (ANFF). 
The authors acknowledge the facilities, and the scientific and technical assistance of the RMIT Microscopy \& Microanalysis Facility (RMMF), a linked laboratory of Microscopy Australia, enabled by NCRIS. 
We thank Hoe Tan from the Australian National University for providing the MOVPE films.
S.C.S gratefully acknowledges the support of an Ernst and Grace Matthaei scholarship.
I.O.R. is supported by an Australian Government Research Training Program Scholarship.
G.H. is supported by the University of Melbourne through a Melbourne Research Scholarship.
P.R. acknowledges support through an RMIT University Vice-Chancellor’s Research Fellowship. 
Part of this study was supported by QST President's Strategic Grant ``QST International Research Initiative''.
\end{acknowledgments}
\bibliography{references}

\begin{thebibliography}{64}%
\makeatletter
\providecommand \@ifxundefined [1]{%
 \@ifx{#1\undefined}
}%
\providecommand \@ifnum [1]{%
 \ifnum #1\expandafter \@firstoftwo
 \else \expandafter \@secondoftwo
 \fi
}%
\providecommand \@ifx [1]{%
 \ifx #1\expandafter \@firstoftwo
 \else \expandafter \@secondoftwo
 \fi
}%
\providecommand \natexlab [1]{#1}%
\providecommand \enquote  [1]{``#1''}%
\providecommand \bibnamefont  [1]{#1}%
\providecommand \bibfnamefont [1]{#1}%
\providecommand \citenamefont [1]{#1}%
\providecommand \href@noop [0]{\@secondoftwo}%
\providecommand \href [0]{\begingroup \@sanitize@url \@href}%
\providecommand \@href[1]{\@@startlink{#1}\@@href}%
\providecommand \@@href[1]{\endgroup#1\@@endlink}%
\providecommand \@sanitize@url [0]{\catcode `\\12\catcode `\$12\catcode
  `\&12\catcode `\#12\catcode `\^12\catcode `\_12\catcode `\%12\relax}%
\providecommand \@@startlink[1]{}%
\providecommand \@@endlink[0]{}%
\providecommand \url  [0]{\begingroup\@sanitize@url \@url }%
\providecommand \@url [1]{\endgroup\@href {#1}{\urlprefix }}%
\providecommand \urlprefix  [0]{URL }%
\providecommand \Eprint [0]{\href }%
\providecommand \doibase [0]{https://doi.org/}%
\providecommand \selectlanguage [0]{\@gobble}%
\providecommand \bibinfo  [0]{\@secondoftwo}%
\providecommand \bibfield  [0]{\@secondoftwo}%
\providecommand \translation [1]{[#1]}%
\providecommand \BibitemOpen [0]{}%
\providecommand \bibitemStop [0]{}%
\providecommand \bibitemNoStop [0]{.\EOS\space}%
\providecommand \EOS [0]{\spacefactor3000\relax}%
\providecommand \BibitemShut  [1]{\csname bibitem#1\endcsname}%
\let\auto@bib@innerbib\@empty
\bibitem [{\citenamefont {Gottscholl}\ \emph {et~al.}(2020)\citenamefont
  {Gottscholl}, \citenamefont {Kianinia}, \citenamefont {Soltamov},
  \citenamefont {Orlinskii}, \citenamefont {Mamin}, \citenamefont {Bradac},
  \citenamefont {Kasper}, \citenamefont {Krambrock}, \citenamefont {Sperlich},
  \citenamefont {Toth}, \citenamefont {Aharonovich},\ and\ \citenamefont
  {Dyakonov}}]{GottschollNM2020}%
  \BibitemOpen
  \bibfield  {author} {\bibinfo {author} {\bibfnamefont {A.}~\bibnamefont
  {Gottscholl}}, \bibinfo {author} {\bibfnamefont {M.}~\bibnamefont
  {Kianinia}}, \bibinfo {author} {\bibfnamefont {V.}~\bibnamefont {Soltamov}},
  \bibinfo {author} {\bibfnamefont {S.}~\bibnamefont {Orlinskii}}, \bibinfo
  {author} {\bibfnamefont {G.}~\bibnamefont {Mamin}}, \bibinfo {author}
  {\bibfnamefont {C.}~\bibnamefont {Bradac}}, \bibinfo {author} {\bibfnamefont
  {C.}~\bibnamefont {Kasper}}, \bibinfo {author} {\bibfnamefont
  {K.}~\bibnamefont {Krambrock}}, \bibinfo {author} {\bibfnamefont
  {A.}~\bibnamefont {Sperlich}}, \bibinfo {author} {\bibfnamefont
  {M.}~\bibnamefont {Toth}}, \bibinfo {author} {\bibfnamefont {I.}~\bibnamefont
  {Aharonovich}},\ and\ \bibinfo {author} {\bibfnamefont {V.}~\bibnamefont
  {Dyakonov}},\ }\bibfield  {title} {\bibinfo {title} {Initialization and
  read-out of intrinsic spin defects in a van der waals crystal at room
  temperature},\ }\href {https://doi.org/10.1038/s41563-020-0619-6} {\bibfield
  {journal} {\bibinfo  {journal} {Nature Materials}\ }\textbf {\bibinfo
  {volume} {19}},\ \bibinfo {pages} {540} (\bibinfo {year} {2020})}\BibitemShut
  {NoStop}%
\bibitem [{\citenamefont {Liu}\ \emph {et~al.}(2022{\natexlab{a}})\citenamefont
  {Liu}, \citenamefont {Guo}, \citenamefont {Yu}, \citenamefont {Meng},
  \citenamefont {Li}, \citenamefont {Yang}, \citenamefont {Wang}, \citenamefont
  {Zeng}, \citenamefont {Xie}, \citenamefont {Li}, \citenamefont {Wang},
  \citenamefont {Xu}, \citenamefont {Wang}, \citenamefont {Tang}, \citenamefont
  {Li},\ and\ \citenamefont {Guo}}]{Liu2022review}%
  \BibitemOpen
  \bibfield  {author} {\bibinfo {author} {\bibfnamefont {W.}~\bibnamefont
  {Liu}}, \bibinfo {author} {\bibfnamefont {N.-J.}\ \bibnamefont {Guo}},
  \bibinfo {author} {\bibfnamefont {S.}~\bibnamefont {Yu}}, \bibinfo {author}
  {\bibfnamefont {Y.}~\bibnamefont {Meng}}, \bibinfo {author} {\bibfnamefont
  {Z.-P.}\ \bibnamefont {Li}}, \bibinfo {author} {\bibfnamefont {Y.-Z.}\
  \bibnamefont {Yang}}, \bibinfo {author} {\bibfnamefont {Z.-A.}\ \bibnamefont
  {Wang}}, \bibinfo {author} {\bibfnamefont {X.-D.}\ \bibnamefont {Zeng}},
  \bibinfo {author} {\bibfnamefont {L.-K.}\ \bibnamefont {Xie}}, \bibinfo
  {author} {\bibfnamefont {Q.}~\bibnamefont {Li}}, \bibinfo {author}
  {\bibfnamefont {J.-F.}\ \bibnamefont {Wang}}, \bibinfo {author}
  {\bibfnamefont {J.-S.}\ \bibnamefont {Xu}}, \bibinfo {author} {\bibfnamefont
  {Y.-T.}\ \bibnamefont {Wang}}, \bibinfo {author} {\bibfnamefont {J.-S.}\
  \bibnamefont {Tang}}, \bibinfo {author} {\bibfnamefont {C.-F.}\ \bibnamefont
  {Li}},\ and\ \bibinfo {author} {\bibfnamefont {G.-C.}\ \bibnamefont {Guo}},\
  }\bibfield  {title} {\bibinfo {title} {Spin-active defects in hexagonal boron
  nitride},\ }\href {https://doi.org/10.1088/2633-4356/ac7e9f} {\bibfield
  {journal} {\bibinfo  {journal} {Materials for Quantum Technology}\ }\textbf
  {\bibinfo {volume} {2}},\ \bibinfo {pages} {032002} (\bibinfo {year}
  {2022}{\natexlab{a}})}\BibitemShut {NoStop}%
\bibitem [{\citenamefont {Aharonovich}\ \emph {et~al.}(2022)\citenamefont
  {Aharonovich}, \citenamefont {Tetienne},\ and\ \citenamefont
  {Toth}}]{Aharonovich2022}%
  \BibitemOpen
  \bibfield  {author} {\bibinfo {author} {\bibfnamefont {I.}~\bibnamefont
  {Aharonovich}}, \bibinfo {author} {\bibfnamefont {J.-P.}\ \bibnamefont
  {Tetienne}},\ and\ \bibinfo {author} {\bibfnamefont {M.}~\bibnamefont
  {Toth}},\ }\bibfield  {title} {\bibinfo {title} {Quantum emitters in
  hexagonal boron nitride},\ }\href
  {https://doi.org/10.1021/acs.nanolett.2c03743} {\bibfield  {journal}
  {\bibinfo  {journal} {Nano Letters}\ }\textbf {\bibinfo {volume} {22}},\
  \bibinfo {pages} {9227} (\bibinfo {year} {2022})}\BibitemShut {NoStop}%
\bibitem [{\citenamefont {Vaidya}\ \emph {et~al.}(2023)\citenamefont {Vaidya},
  \citenamefont {Gao}, \citenamefont {Dikshit}, \citenamefont {Aharonovich},\
  and\ \citenamefont {Li}}]{Vaidya2023}%
  \BibitemOpen
  \bibfield  {author} {\bibinfo {author} {\bibfnamefont {S.}~\bibnamefont
  {Vaidya}}, \bibinfo {author} {\bibfnamefont {X.}~\bibnamefont {Gao}},
  \bibinfo {author} {\bibfnamefont {S.}~\bibnamefont {Dikshit}}, \bibinfo
  {author} {\bibfnamefont {I.}~\bibnamefont {Aharonovich}},\ and\ \bibinfo
  {author} {\bibfnamefont {T.}~\bibnamefont {Li}},\ }\bibfield  {title}
  {\bibinfo {title} {Quantum sensing and imaging with spin defects in hexagonal
  boron nitride},\ }\href {https://doi.org/10.1080/23746149.2023.2206049}
  {\bibfield  {journal} {\bibinfo  {journal} {Advances in Physics: X}\ }\textbf
  {\bibinfo {volume} {8}},\ \bibinfo {pages} {2206049} (\bibinfo {year}
  {2023})}\BibitemShut {NoStop}%
\bibitem [{\citenamefont {Durand}\ \emph {et~al.}(2023)\citenamefont {Durand},
  \citenamefont {{Clua-Provost}}, \citenamefont {Fabre}, \citenamefont {Kumar},
  \citenamefont {Li}, \citenamefont {Edgar}, \citenamefont {Udvarhelyi},
  \citenamefont {Gali}, \citenamefont {Marie}, \citenamefont {Robert},
  \citenamefont {G{\'e}rard}, \citenamefont {Gil}, \citenamefont {Cassabois},\
  and\ \citenamefont {Jacques}}]{Durand2023}%
  \BibitemOpen
  \bibfield  {author} {\bibinfo {author} {\bibfnamefont {A.}~\bibnamefont
  {Durand}}, \bibinfo {author} {\bibfnamefont {T.}~\bibnamefont
  {{Clua-Provost}}}, \bibinfo {author} {\bibfnamefont {F.}~\bibnamefont
  {Fabre}}, \bibinfo {author} {\bibfnamefont {P.}~\bibnamefont {Kumar}},
  \bibinfo {author} {\bibfnamefont {J.}~\bibnamefont {Li}}, \bibinfo {author}
  {\bibfnamefont {J.~H.}\ \bibnamefont {Edgar}}, \bibinfo {author}
  {\bibfnamefont {P.}~\bibnamefont {Udvarhelyi}}, \bibinfo {author}
  {\bibfnamefont {A.}~\bibnamefont {Gali}}, \bibinfo {author} {\bibfnamefont
  {X.}~\bibnamefont {Marie}}, \bibinfo {author} {\bibfnamefont
  {C.}~\bibnamefont {Robert}}, \bibinfo {author} {\bibfnamefont {J.~M.}\
  \bibnamefont {G{\'e}rard}}, \bibinfo {author} {\bibfnamefont
  {B.}~\bibnamefont {Gil}}, \bibinfo {author} {\bibfnamefont {G.}~\bibnamefont
  {Cassabois}},\ and\ \bibinfo {author} {\bibfnamefont {V.}~\bibnamefont
  {Jacques}},\ }\bibfield  {title} {\bibinfo {title} {Optically-active spin
  defects in few-layer thick hexagonal boron nitride},\ }\href
  {https://doi.org/10.48550/arXiv.2304.12071} {\bibfield  {journal} {\bibinfo
  {journal} {Preprint}\ ,\ \bibinfo {pages} {arXiv:2304.12071}} (\bibinfo
  {year} {2023})}\BibitemShut {NoStop}%
\bibitem [{\citenamefont {Yao}\ \emph {et~al.}(2018)\citenamefont {Yao},
  \citenamefont {Zaletel}, \citenamefont {{Stamper-Kurn}},\ and\ \citenamefont
  {Vishwanath}}]{Yao2018}%
  \BibitemOpen
  \bibfield  {author} {\bibinfo {author} {\bibfnamefont {N.~Y.}\ \bibnamefont
  {Yao}}, \bibinfo {author} {\bibfnamefont {M.~P.}\ \bibnamefont {Zaletel}},
  \bibinfo {author} {\bibfnamefont {D.~M.}\ \bibnamefont {{Stamper-Kurn}}},\
  and\ \bibinfo {author} {\bibfnamefont {A.}~\bibnamefont {Vishwanath}},\
  }\bibfield  {title} {\bibinfo {title} {A quantum dipolar spin liquid},\
  }\href {https://doi.org/10.1038/s41567-017-0030-7} {\bibfield  {journal}
  {\bibinfo  {journal} {Nature Physics}\ }\textbf {\bibinfo {volume} {14}},\
  \bibinfo {pages} {405} (\bibinfo {year} {2018})}\BibitemShut {NoStop}%
\bibitem [{\citenamefont {Davis}\ \emph {et~al.}(2023)\citenamefont {Davis},
  \citenamefont {Ye}, \citenamefont {Machado}, \citenamefont {Meynell},
  \citenamefont {Wu}, \citenamefont {Mittiga}, \citenamefont {Schenken},
  \citenamefont {Joos}, \citenamefont {Kobrin}, \citenamefont {Lyu},
  \citenamefont {Wang}, \citenamefont {Bluvstein}, \citenamefont {Choi},
  \citenamefont {Zu}, \citenamefont {Jayich},\ and\ \citenamefont
  {Yao}}]{Davis2023}%
  \BibitemOpen
  \bibfield  {author} {\bibinfo {author} {\bibfnamefont {E.~J.}\ \bibnamefont
  {Davis}}, \bibinfo {author} {\bibfnamefont {B.}~\bibnamefont {Ye}}, \bibinfo
  {author} {\bibfnamefont {F.}~\bibnamefont {Machado}}, \bibinfo {author}
  {\bibfnamefont {S.~A.}\ \bibnamefont {Meynell}}, \bibinfo {author}
  {\bibfnamefont {W.}~\bibnamefont {Wu}}, \bibinfo {author} {\bibfnamefont
  {T.}~\bibnamefont {Mittiga}}, \bibinfo {author} {\bibfnamefont
  {W.}~\bibnamefont {Schenken}}, \bibinfo {author} {\bibfnamefont
  {M.}~\bibnamefont {Joos}}, \bibinfo {author} {\bibfnamefont {B.}~\bibnamefont
  {Kobrin}}, \bibinfo {author} {\bibfnamefont {Y.}~\bibnamefont {Lyu}},
  \bibinfo {author} {\bibfnamefont {Z.}~\bibnamefont {Wang}}, \bibinfo {author}
  {\bibfnamefont {D.}~\bibnamefont {Bluvstein}}, \bibinfo {author}
  {\bibfnamefont {S.}~\bibnamefont {Choi}}, \bibinfo {author} {\bibfnamefont
  {C.}~\bibnamefont {Zu}}, \bibinfo {author} {\bibfnamefont {A.~C.~B.}\
  \bibnamefont {Jayich}},\ and\ \bibinfo {author} {\bibfnamefont {N.~Y.}\
  \bibnamefont {Yao}},\ }\bibfield  {title} {\bibinfo {title} {Probing
  many-body dynamics in a two-dimensional dipolar spin ensemble},\ }\href
  {https://doi.org/10.1038/s41567-023-01944-5} {\bibfield  {journal} {\bibinfo
  {journal} {Nature Physics}\ }\textbf {\bibinfo {volume} {19}},\ \bibinfo
  {pages} {836} (\bibinfo {year} {2023})}\BibitemShut {NoStop}%
\bibitem [{\citenamefont {Choi}\ \emph {et~al.}(2016)\citenamefont {Choi},
  \citenamefont {Hild}, \citenamefont {Zeiher}, \citenamefont {Schau{\ss}},
  \citenamefont {{Rubio-Abadal}}, \citenamefont {Yefsah}, \citenamefont
  {Khemani}, \citenamefont {Huse}, \citenamefont {Bloch},\ and\ \citenamefont
  {Gross}}]{Choi2016}%
  \BibitemOpen
  \bibfield  {author} {\bibinfo {author} {\bibfnamefont {J.-y.}\ \bibnamefont
  {Choi}}, \bibinfo {author} {\bibfnamefont {S.}~\bibnamefont {Hild}}, \bibinfo
  {author} {\bibfnamefont {J.}~\bibnamefont {Zeiher}}, \bibinfo {author}
  {\bibfnamefont {P.}~\bibnamefont {Schau{\ss}}}, \bibinfo {author}
  {\bibfnamefont {A.}~\bibnamefont {{Rubio-Abadal}}}, \bibinfo {author}
  {\bibfnamefont {T.}~\bibnamefont {Yefsah}}, \bibinfo {author} {\bibfnamefont
  {V.}~\bibnamefont {Khemani}}, \bibinfo {author} {\bibfnamefont {D.~A.}\
  \bibnamefont {Huse}}, \bibinfo {author} {\bibfnamefont {I.}~\bibnamefont
  {Bloch}},\ and\ \bibinfo {author} {\bibfnamefont {C.}~\bibnamefont {Gross}},\
  }\bibfield  {title} {\bibinfo {title} {Exploring the many-body localization
  transition in two dimensions},\ }\href
  {https://doi.org/10.1126/science.aaf8834} {\bibfield  {journal} {\bibinfo
  {journal} {Science (New York, N.Y.)}\ }\textbf {\bibinfo {volume} {352}},\
  \bibinfo {pages} {1547} (\bibinfo {year} {2016})}\BibitemShut {NoStop}%
\bibitem [{\citenamefont {Choi}\ \emph {et~al.}(2017)\citenamefont {Choi},
  \citenamefont {Choi}, \citenamefont {Landig}, \citenamefont {Kucsko},
  \citenamefont {Zhou}, \citenamefont {Isoya}, \citenamefont {Jelezko},
  \citenamefont {Onoda}, \citenamefont {Sumiya}, \citenamefont {Khemani} \emph
  {et~al.}}]{Choi2017}%
  \BibitemOpen
  \bibfield  {author} {\bibinfo {author} {\bibfnamefont {S.}~\bibnamefont
  {Choi}}, \bibinfo {author} {\bibfnamefont {J.}~\bibnamefont {Choi}}, \bibinfo
  {author} {\bibfnamefont {R.}~\bibnamefont {Landig}}, \bibinfo {author}
  {\bibfnamefont {G.}~\bibnamefont {Kucsko}}, \bibinfo {author} {\bibfnamefont
  {H.}~\bibnamefont {Zhou}}, \bibinfo {author} {\bibfnamefont {J.}~\bibnamefont
  {Isoya}}, \bibinfo {author} {\bibfnamefont {F.}~\bibnamefont {Jelezko}},
  \bibinfo {author} {\bibfnamefont {S.}~\bibnamefont {Onoda}}, \bibinfo
  {author} {\bibfnamefont {H.}~\bibnamefont {Sumiya}}, \bibinfo {author}
  {\bibfnamefont {V.}~\bibnamefont {Khemani}}, \emph {et~al.},\ }\bibfield
  {title} {\bibinfo {title} {Observation of discrete time-crystalline order in
  a disordered dipolar many-body system},\ }\href
  {https://doi.org/10.1038/nature21426} {\bibfield  {journal} {\bibinfo
  {journal} {Nature}\ }\textbf {\bibinfo {volume} {543}},\ \bibinfo {pages}
  {221} (\bibinfo {year} {2017})}\BibitemShut {NoStop}%
\bibitem [{\citenamefont {Abanin}\ \emph {et~al.}(2019)\citenamefont {Abanin},
  \citenamefont {Altman}, \citenamefont {Bloch},\ and\ \citenamefont
  {Serbyn}}]{Abanin2019}%
  \BibitemOpen
  \bibfield  {author} {\bibinfo {author} {\bibfnamefont {D.~A.}\ \bibnamefont
  {Abanin}}, \bibinfo {author} {\bibfnamefont {E.}~\bibnamefont {Altman}},
  \bibinfo {author} {\bibfnamefont {I.}~\bibnamefont {Bloch}},\ and\ \bibinfo
  {author} {\bibfnamefont {M.}~\bibnamefont {Serbyn}},\ }\bibfield  {title}
  {\bibinfo {title} {Colloquium: Many-body localization, thermalization, and
  entanglement},\ }\href {https://doi.org/10.1103/RevModPhys.91.021001}
  {\bibfield  {journal} {\bibinfo  {journal} {Reviews of Modern Physics}\
  }\textbf {\bibinfo {volume} {91}},\ \bibinfo {pages} {021001} (\bibinfo
  {year} {2019})}\BibitemShut {NoStop}%
\bibitem [{\citenamefont {Zu}\ \emph {et~al.}(2021)\citenamefont {Zu},
  \citenamefont {Machado}, \citenamefont {Ye}, \citenamefont {Choi},
  \citenamefont {Kobrin}, \citenamefont {Mittiga}, \citenamefont {Hsieh},
  \citenamefont {Bhattacharyya}, \citenamefont {Markham}, \citenamefont
  {Twitchen} \emph {et~al.}}]{Zu2021}%
  \BibitemOpen
  \bibfield  {author} {\bibinfo {author} {\bibfnamefont {C.}~\bibnamefont
  {Zu}}, \bibinfo {author} {\bibfnamefont {F.}~\bibnamefont {Machado}},
  \bibinfo {author} {\bibfnamefont {B.}~\bibnamefont {Ye}}, \bibinfo {author}
  {\bibfnamefont {S.}~\bibnamefont {Choi}}, \bibinfo {author} {\bibfnamefont
  {B.}~\bibnamefont {Kobrin}}, \bibinfo {author} {\bibfnamefont
  {T.}~\bibnamefont {Mittiga}}, \bibinfo {author} {\bibfnamefont
  {S.}~\bibnamefont {Hsieh}}, \bibinfo {author} {\bibfnamefont
  {P.}~\bibnamefont {Bhattacharyya}}, \bibinfo {author} {\bibfnamefont
  {M.}~\bibnamefont {Markham}}, \bibinfo {author} {\bibfnamefont
  {D.}~\bibnamefont {Twitchen}}, \emph {et~al.},\ }\bibfield  {title} {\bibinfo
  {title} {Emergent hydrodynamics in a strongly interacting dipolar spin
  ensemble},\ }\href {https://doi.org/10.1038/s41586-021-03763-1} {\bibfield
  {journal} {\bibinfo  {journal} {Nature}\ }\textbf {\bibinfo {volume} {597}},\
  \bibinfo {pages} {45} (\bibinfo {year} {2021})}\BibitemShut {NoStop}%
\bibitem [{\citenamefont {Gong}\ \emph {et~al.}(2023)\citenamefont {Gong},
  \citenamefont {He}, \citenamefont {Gao}, \citenamefont {Ju}, \citenamefont
  {Liu}, \citenamefont {Ye}, \citenamefont {Henriksen}, \citenamefont {Li},\
  and\ \citenamefont {Zu}}]{Gong2023}%
  \BibitemOpen
  \bibfield  {author} {\bibinfo {author} {\bibfnamefont {R.}~\bibnamefont
  {Gong}}, \bibinfo {author} {\bibfnamefont {G.}~\bibnamefont {He}}, \bibinfo
  {author} {\bibfnamefont {X.}~\bibnamefont {Gao}}, \bibinfo {author}
  {\bibfnamefont {P.}~\bibnamefont {Ju}}, \bibinfo {author} {\bibfnamefont
  {Z.}~\bibnamefont {Liu}}, \bibinfo {author} {\bibfnamefont {B.}~\bibnamefont
  {Ye}}, \bibinfo {author} {\bibfnamefont {E.~A.}\ \bibnamefont {Henriksen}},
  \bibinfo {author} {\bibfnamefont {T.}~\bibnamefont {Li}},\ and\ \bibinfo
  {author} {\bibfnamefont {C.}~\bibnamefont {Zu}},\ }\bibfield  {title}
  {\bibinfo {title} {Coherent dynamics of strongly interacting electronic spin
  defects in hexagonal boron nitride},\ }\href
  {https://doi.org/10.1038/s41467-023-39115-y} {\bibfield  {journal} {\bibinfo
  {journal} {Nature Communications}\ }\textbf {\bibinfo {volume} {14}},\
  \bibinfo {pages} {3299} (\bibinfo {year} {2023})}\BibitemShut {NoStop}%
\bibitem [{\citenamefont {Gottscholl}\ \emph {et~al.}(2021)\citenamefont
  {Gottscholl}, \citenamefont {Diez}, \citenamefont {Soltamov}, \citenamefont
  {Kasper}, \citenamefont {Krausse}, \citenamefont {Sperlich}, \citenamefont
  {Kianinia}, \citenamefont {Bradac}, \citenamefont {Aharonovich},\ and\
  \citenamefont {Dyakonov}}]{Gottscholl2021NC}%
  \BibitemOpen
  \bibfield  {author} {\bibinfo {author} {\bibfnamefont {A.}~\bibnamefont
  {Gottscholl}}, \bibinfo {author} {\bibfnamefont {M.}~\bibnamefont {Diez}},
  \bibinfo {author} {\bibfnamefont {V.}~\bibnamefont {Soltamov}}, \bibinfo
  {author} {\bibfnamefont {C.}~\bibnamefont {Kasper}}, \bibinfo {author}
  {\bibfnamefont {D.}~\bibnamefont {Krausse}}, \bibinfo {author} {\bibfnamefont
  {A.}~\bibnamefont {Sperlich}}, \bibinfo {author} {\bibfnamefont
  {M.}~\bibnamefont {Kianinia}}, \bibinfo {author} {\bibfnamefont
  {C.}~\bibnamefont {Bradac}}, \bibinfo {author} {\bibfnamefont
  {I.}~\bibnamefont {Aharonovich}},\ and\ \bibinfo {author} {\bibfnamefont
  {V.}~\bibnamefont {Dyakonov}},\ }\bibfield  {title} {\bibinfo {title} {Spin
  defects in {{hBN}} as promising temperature, pressure and magnetic field
  quantum sensors},\ }\href {https://doi.org/10.1038/s41467-021-24725-1}
  {\bibfield  {journal} {\bibinfo  {journal} {Nature Communications}\ }\textbf
  {\bibinfo {volume} {12}},\ \bibinfo {pages} {4480} (\bibinfo {year}
  {2021})}\BibitemShut {NoStop}%
\bibitem [{\citenamefont {Liu}\ \emph {et~al.}(2021)\citenamefont {Liu},
  \citenamefont {Li}, \citenamefont {Yang}, \citenamefont {Yu}, \citenamefont
  {Meng}, \citenamefont {Wang}, \citenamefont {Li}, \citenamefont {Guo},
  \citenamefont {Yan}, \citenamefont {Li}, \citenamefont {Wang}, \citenamefont
  {Xu}, \citenamefont {Wang}, \citenamefont {Tang}, \citenamefont {Li},\ and\
  \citenamefont {Guo}}]{Liu2021}%
  \BibitemOpen
  \bibfield  {author} {\bibinfo {author} {\bibfnamefont {W.}~\bibnamefont
  {Liu}}, \bibinfo {author} {\bibfnamefont {Z.-P.}\ \bibnamefont {Li}},
  \bibinfo {author} {\bibfnamefont {Y.-Z.}\ \bibnamefont {Yang}}, \bibinfo
  {author} {\bibfnamefont {S.}~\bibnamefont {Yu}}, \bibinfo {author}
  {\bibfnamefont {Y.}~\bibnamefont {Meng}}, \bibinfo {author} {\bibfnamefont
  {Z.-A.}\ \bibnamefont {Wang}}, \bibinfo {author} {\bibfnamefont {Z.-C.}\
  \bibnamefont {Li}}, \bibinfo {author} {\bibfnamefont {N.-J.}\ \bibnamefont
  {Guo}}, \bibinfo {author} {\bibfnamefont {F.-F.}\ \bibnamefont {Yan}},
  \bibinfo {author} {\bibfnamefont {Q.}~\bibnamefont {Li}}, \bibinfo {author}
  {\bibfnamefont {J.-F.}\ \bibnamefont {Wang}}, \bibinfo {author}
  {\bibfnamefont {J.-S.}\ \bibnamefont {Xu}}, \bibinfo {author} {\bibfnamefont
  {Y.-T.}\ \bibnamefont {Wang}}, \bibinfo {author} {\bibfnamefont {J.-S.}\
  \bibnamefont {Tang}}, \bibinfo {author} {\bibfnamefont {C.-F.}\ \bibnamefont
  {Li}},\ and\ \bibinfo {author} {\bibfnamefont {G.-C.}\ \bibnamefont {Guo}},\
  }\bibfield  {title} {\bibinfo {title} {Temperature-dependent energy-level
  shifts of spin defects in hexagonal boron nitride},\ }\href
  {https://doi.org/10.1021/acsphotonics.1c00320} {\bibfield  {journal}
  {\bibinfo  {journal} {ACS Photonics}\ }\textbf {\bibinfo {volume} {8}},\
  \bibinfo {pages} {1889} (\bibinfo {year} {2021})}\BibitemShut {NoStop}%
\bibitem [{\citenamefont {Healey}\ \emph {et~al.}(2023)\citenamefont {Healey},
  \citenamefont {Scholten}, \citenamefont {Yang}, \citenamefont {Scott},
  \citenamefont {Abrahams}, \citenamefont {Robertson}, \citenamefont {Hou},
  \citenamefont {Guo}, \citenamefont {Rahman}, \citenamefont {Lu},
  \citenamefont {Kianinia}, \citenamefont {Aharonovich},\ and\ \citenamefont
  {Tetienne}}]{HealeyNP2023}%
  \BibitemOpen
  \bibfield  {author} {\bibinfo {author} {\bibfnamefont {A.~J.}\ \bibnamefont
  {Healey}}, \bibinfo {author} {\bibfnamefont {S.~C.}\ \bibnamefont
  {Scholten}}, \bibinfo {author} {\bibfnamefont {T.}~\bibnamefont {Yang}},
  \bibinfo {author} {\bibfnamefont {J.~A.}\ \bibnamefont {Scott}}, \bibinfo
  {author} {\bibfnamefont {G.~J.}\ \bibnamefont {Abrahams}}, \bibinfo {author}
  {\bibfnamefont {I.~O.}\ \bibnamefont {Robertson}}, \bibinfo {author}
  {\bibfnamefont {X.~F.}\ \bibnamefont {Hou}}, \bibinfo {author} {\bibfnamefont
  {Y.~F.}\ \bibnamefont {Guo}}, \bibinfo {author} {\bibfnamefont
  {S.}~\bibnamefont {Rahman}}, \bibinfo {author} {\bibfnamefont
  {Y.}~\bibnamefont {Lu}}, \bibinfo {author} {\bibfnamefont {M.}~\bibnamefont
  {Kianinia}}, \bibinfo {author} {\bibfnamefont {I.}~\bibnamefont
  {Aharonovich}},\ and\ \bibinfo {author} {\bibfnamefont {J.-P.}\ \bibnamefont
  {Tetienne}},\ }\bibfield  {title} {\bibinfo {title} {Quantum microscopy with
  van der waals heterostructures},\ }\href
  {https://doi.org/10.1038/s41567-022-01815-5} {\bibfield  {journal} {\bibinfo
  {journal} {Nature Physics}\ }\textbf {\bibinfo {volume} {19}},\ \bibinfo
  {pages} {87} (\bibinfo {year} {2023})}\BibitemShut {NoStop}%
\bibitem [{\citenamefont {Huang}\ \emph {et~al.}(2022)\citenamefont {Huang},
  \citenamefont {Zhou}, \citenamefont {Chen}, \citenamefont {Lu}, \citenamefont
  {McLaughlin}, \citenamefont {Li}, \citenamefont {Alghamdi}, \citenamefont
  {Djugba}, \citenamefont {Shi}, \citenamefont {Wang},\ and\ \citenamefont
  {Du}}]{HuangNC2022}%
  \BibitemOpen
  \bibfield  {author} {\bibinfo {author} {\bibfnamefont {M.}~\bibnamefont
  {Huang}}, \bibinfo {author} {\bibfnamefont {J.}~\bibnamefont {Zhou}},
  \bibinfo {author} {\bibfnamefont {D.}~\bibnamefont {Chen}}, \bibinfo {author}
  {\bibfnamefont {H.}~\bibnamefont {Lu}}, \bibinfo {author} {\bibfnamefont
  {N.~J.}\ \bibnamefont {McLaughlin}}, \bibinfo {author} {\bibfnamefont
  {S.}~\bibnamefont {Li}}, \bibinfo {author} {\bibfnamefont {M.}~\bibnamefont
  {Alghamdi}}, \bibinfo {author} {\bibfnamefont {D.}~\bibnamefont {Djugba}},
  \bibinfo {author} {\bibfnamefont {J.}~\bibnamefont {Shi}}, \bibinfo {author}
  {\bibfnamefont {H.}~\bibnamefont {Wang}},\ and\ \bibinfo {author}
  {\bibfnamefont {C.~R.}\ \bibnamefont {Du}},\ }\bibfield  {title} {\bibinfo
  {title} {Wide field imaging of van der waals ferromagnet {{Fe3GeTe2}} by spin
  defects in hexagonal boron nitride},\ }\href
  {https://doi.org/10.1038/s41467-022-33016-2} {\bibfield  {journal} {\bibinfo
  {journal} {Nature Communications}\ }\textbf {\bibinfo {volume} {13}},\
  \bibinfo {pages} {5369} (\bibinfo {year} {2022})}\BibitemShut {NoStop}%
\bibitem [{\citenamefont {Lyu}\ \emph {et~al.}(2022)\citenamefont {Lyu},
  \citenamefont {Tan}, \citenamefont {Wu}, \citenamefont {Zhang}, \citenamefont
  {Zhang}, \citenamefont {Mu}, \citenamefont {{Z{\'u}{\~n}iga-P{\'e}rez}},
  \citenamefont {Cai},\ and\ \citenamefont {Gao}}]{Lyu2022}%
  \BibitemOpen
  \bibfield  {author} {\bibinfo {author} {\bibfnamefont {X.}~\bibnamefont
  {Lyu}}, \bibinfo {author} {\bibfnamefont {Q.}~\bibnamefont {Tan}}, \bibinfo
  {author} {\bibfnamefont {L.}~\bibnamefont {Wu}}, \bibinfo {author}
  {\bibfnamefont {C.}~\bibnamefont {Zhang}}, \bibinfo {author} {\bibfnamefont
  {Z.}~\bibnamefont {Zhang}}, \bibinfo {author} {\bibfnamefont
  {Z.}~\bibnamefont {Mu}}, \bibinfo {author} {\bibfnamefont {J.}~\bibnamefont
  {{Z{\'u}{\~n}iga-P{\'e}rez}}}, \bibinfo {author} {\bibfnamefont
  {H.}~\bibnamefont {Cai}},\ and\ \bibinfo {author} {\bibfnamefont
  {W.}~\bibnamefont {Gao}},\ }\bibfield  {title} {\bibinfo {title} {Strain
  quantum sensing with spin defects in hexagonal boron nitride},\ }\href
  {https://doi.org/10.1021/acs.nanolett.2c01722} {\bibfield  {journal}
  {\bibinfo  {journal} {Nano Letters}\ }\textbf {\bibinfo {volume} {22}},\
  \bibinfo {pages} {6553} (\bibinfo {year} {2022})}\BibitemShut {NoStop}%
\bibitem [{\citenamefont {Kumar}\ \emph {et~al.}(2022)\citenamefont {Kumar},
  \citenamefont {Fabre}, \citenamefont {Durand}, \citenamefont
  {{Clua-Provost}}, \citenamefont {Li}, \citenamefont {Edgar}, \citenamefont
  {Rougemaille}, \citenamefont {Coraux}, \citenamefont {Marie}, \citenamefont
  {Renucci}, \citenamefont {Robert}, \citenamefont {{Robert-Philip}},
  \citenamefont {Gil}, \citenamefont {Cassabois}, \citenamefont {Finco},\ and\
  \citenamefont {Jacques}}]{Kumar2022}%
  \BibitemOpen
  \bibfield  {author} {\bibinfo {author} {\bibfnamefont {P.}~\bibnamefont
  {Kumar}}, \bibinfo {author} {\bibfnamefont {F.}~\bibnamefont {Fabre}},
  \bibinfo {author} {\bibfnamefont {A.}~\bibnamefont {Durand}}, \bibinfo
  {author} {\bibfnamefont {T.}~\bibnamefont {{Clua-Provost}}}, \bibinfo
  {author} {\bibfnamefont {J.}~\bibnamefont {Li}}, \bibinfo {author}
  {\bibfnamefont {J.}~\bibnamefont {Edgar}}, \bibinfo {author} {\bibfnamefont
  {N.}~\bibnamefont {Rougemaille}}, \bibinfo {author} {\bibfnamefont
  {J.}~\bibnamefont {Coraux}}, \bibinfo {author} {\bibfnamefont
  {X.}~\bibnamefont {Marie}}, \bibinfo {author} {\bibfnamefont
  {P.}~\bibnamefont {Renucci}}, \bibinfo {author} {\bibfnamefont
  {C.}~\bibnamefont {Robert}}, \bibinfo {author} {\bibfnamefont
  {I.}~\bibnamefont {{Robert-Philip}}}, \bibinfo {author} {\bibfnamefont
  {B.}~\bibnamefont {Gil}}, \bibinfo {author} {\bibfnamefont {G.}~\bibnamefont
  {Cassabois}}, \bibinfo {author} {\bibfnamefont {A.}~\bibnamefont {Finco}},\
  and\ \bibinfo {author} {\bibfnamefont {V.}~\bibnamefont {Jacques}},\
  }\bibfield  {title} {\bibinfo {title} {Magnetic imaging with spin defects in
  hexagonal boron nitride},\ }\href
  {https://doi.org/10.1103/PhysRevApplied.18.L061002} {\bibfield  {journal}
  {\bibinfo  {journal} {Physical Review Applied}\ }\textbf {\bibinfo {volume}
  {18}},\ \bibinfo {pages} {L061002} (\bibinfo {year} {2022})}\BibitemShut
  {NoStop}%
\bibitem [{\citenamefont {Robertson}\ \emph {et~al.}(2023)\citenamefont
  {Robertson}, \citenamefont {Scholten}, \citenamefont {Singh}, \citenamefont
  {Healey}, \citenamefont {Meneses}, \citenamefont {Reineck}, \citenamefont
  {Abe}, \citenamefont {Ohshima}, \citenamefont {Kianinia}, \citenamefont
  {Aharonovich},\ and\ \citenamefont {Tetienne}}]{Robertson2023}%
  \BibitemOpen
  \bibfield  {author} {\bibinfo {author} {\bibfnamefont {I.~O.}\ \bibnamefont
  {Robertson}}, \bibinfo {author} {\bibfnamefont {S.~C.}\ \bibnamefont
  {Scholten}}, \bibinfo {author} {\bibfnamefont {P.}~\bibnamefont {Singh}},
  \bibinfo {author} {\bibfnamefont {A.~J.}\ \bibnamefont {Healey}}, \bibinfo
  {author} {\bibfnamefont {F.}~\bibnamefont {Meneses}}, \bibinfo {author}
  {\bibfnamefont {P.}~\bibnamefont {Reineck}}, \bibinfo {author} {\bibfnamefont
  {H.}~\bibnamefont {Abe}}, \bibinfo {author} {\bibfnamefont {T.}~\bibnamefont
  {Ohshima}}, \bibinfo {author} {\bibfnamefont {M.}~\bibnamefont {Kianinia}},
  \bibinfo {author} {\bibfnamefont {I.}~\bibnamefont {Aharonovich}},\ and\
  \bibinfo {author} {\bibfnamefont {J.-P.}\ \bibnamefont {Tetienne}},\
  }\bibfield  {title} {\bibinfo {title} {Detection of paramagnetic spins with
  an ultrathin van der waals quantum sensor},\ }\href
  {https://doi.org/10.1021/acsnano.3c01678} {\bibfield  {journal} {\bibinfo
  {journal} {ACS Nano}\ }\textbf {\bibinfo {volume} {17}},\ \bibinfo {pages}
  {13408} (\bibinfo {year} {2023})}\BibitemShut {NoStop}%
\bibitem [{\citenamefont {Gao}\ \emph {et~al.}(2023)\citenamefont {Gao},
  \citenamefont {Vaidya}, \citenamefont {Ju}, \citenamefont {Dikshit},
  \citenamefont {Shen}, \citenamefont {Chen},\ and\ \citenamefont
  {Li}}]{Gao2023}%
  \BibitemOpen
  \bibfield  {author} {\bibinfo {author} {\bibfnamefont {X.}~\bibnamefont
  {Gao}}, \bibinfo {author} {\bibfnamefont {S.}~\bibnamefont {Vaidya}},
  \bibinfo {author} {\bibfnamefont {P.}~\bibnamefont {Ju}}, \bibinfo {author}
  {\bibfnamefont {S.}~\bibnamefont {Dikshit}}, \bibinfo {author} {\bibfnamefont
  {K.}~\bibnamefont {Shen}}, \bibinfo {author} {\bibfnamefont {Y.~P.}\
  \bibnamefont {Chen}},\ and\ \bibinfo {author} {\bibfnamefont
  {T.}~\bibnamefont {Li}},\ }\bibfield  {title} {\bibinfo {title} {Quantum
  sensing of paramagnetic spins in liquids with spin qubits in hexagonal boron
  nitride},\ }\href {https://doi.org/10.1021/acsphotonics.3c00621} {\bibfield
  {journal} {\bibinfo  {journal} {ACS Photonics}\ }\textbf {\bibinfo {volume}
  {10}},\ \bibinfo {pages} {2894} (\bibinfo {year} {2023})}\BibitemShut
  {NoStop}%
\bibitem [{\citenamefont {Mendelson}\ \emph {et~al.}(2021)\citenamefont
  {Mendelson}, \citenamefont {Chugh}, \citenamefont {Reimers}, \citenamefont
  {Cheng}, \citenamefont {Gottscholl}, \citenamefont {Long}, \citenamefont
  {Mellor}, \citenamefont {Zettl}, \citenamefont {Dyakonov}, \citenamefont
  {Beton}, \citenamefont {Novikov}, \citenamefont {Jagadish}, \citenamefont
  {Tan}, \citenamefont {Ford}, \citenamefont {Toth}, \citenamefont {Bradac},\
  and\ \citenamefont {Aharonovich}}]{Mendelson2021}%
  \BibitemOpen
  \bibfield  {author} {\bibinfo {author} {\bibfnamefont {N.}~\bibnamefont
  {Mendelson}}, \bibinfo {author} {\bibfnamefont {D.}~\bibnamefont {Chugh}},
  \bibinfo {author} {\bibfnamefont {J.~R.}\ \bibnamefont {Reimers}}, \bibinfo
  {author} {\bibfnamefont {T.~S.}\ \bibnamefont {Cheng}}, \bibinfo {author}
  {\bibfnamefont {A.}~\bibnamefont {Gottscholl}}, \bibinfo {author}
  {\bibfnamefont {H.}~\bibnamefont {Long}}, \bibinfo {author} {\bibfnamefont
  {C.~J.}\ \bibnamefont {Mellor}}, \bibinfo {author} {\bibfnamefont
  {A.}~\bibnamefont {Zettl}}, \bibinfo {author} {\bibfnamefont
  {V.}~\bibnamefont {Dyakonov}}, \bibinfo {author} {\bibfnamefont {P.~H.}\
  \bibnamefont {Beton}}, \bibinfo {author} {\bibfnamefont {S.~V.}\ \bibnamefont
  {Novikov}}, \bibinfo {author} {\bibfnamefont {C.}~\bibnamefont {Jagadish}},
  \bibinfo {author} {\bibfnamefont {H.~H.}\ \bibnamefont {Tan}}, \bibinfo
  {author} {\bibfnamefont {M.~J.}\ \bibnamefont {Ford}}, \bibinfo {author}
  {\bibfnamefont {M.}~\bibnamefont {Toth}}, \bibinfo {author} {\bibfnamefont
  {C.}~\bibnamefont {Bradac}},\ and\ \bibinfo {author} {\bibfnamefont
  {I.}~\bibnamefont {Aharonovich}},\ }\bibfield  {title} {\bibinfo {title}
  {Identifying carbon as the source of visible single-photon emission from
  hexagonal boron nitride},\ }\href
  {https://doi.org/10.1038/s41563-020-00850-y} {\bibfield  {journal} {\bibinfo
  {journal} {Nature Materials}\ }\textbf {\bibinfo {volume} {20}},\ \bibinfo
  {pages} {321} (\bibinfo {year} {2021})}\BibitemShut {NoStop}%
\bibitem [{\citenamefont {Chejanovsky}\ \emph {et~al.}(2021)\citenamefont
  {Chejanovsky}, \citenamefont {Mukherjee}, \citenamefont {Geng}, \citenamefont
  {Chen}, \citenamefont {Kim}, \citenamefont {Denisenko}, \citenamefont
  {Finkler}, \citenamefont {Taniguchi}, \citenamefont {Watanabe}, \citenamefont
  {Dasari} \emph {et~al.}}]{Chejanovsky2021}%
  \BibitemOpen
  \bibfield  {author} {\bibinfo {author} {\bibfnamefont {N.}~\bibnamefont
  {Chejanovsky}}, \bibinfo {author} {\bibfnamefont {A.}~\bibnamefont
  {Mukherjee}}, \bibinfo {author} {\bibfnamefont {J.}~\bibnamefont {Geng}},
  \bibinfo {author} {\bibfnamefont {Y.-C.}\ \bibnamefont {Chen}}, \bibinfo
  {author} {\bibfnamefont {Y.}~\bibnamefont {Kim}}, \bibinfo {author}
  {\bibfnamefont {A.}~\bibnamefont {Denisenko}}, \bibinfo {author}
  {\bibfnamefont {A.}~\bibnamefont {Finkler}}, \bibinfo {author} {\bibfnamefont
  {T.}~\bibnamefont {Taniguchi}}, \bibinfo {author} {\bibfnamefont
  {K.}~\bibnamefont {Watanabe}}, \bibinfo {author} {\bibfnamefont {D.~B.~R.}\
  \bibnamefont {Dasari}}, \emph {et~al.},\ }\bibfield  {title} {\bibinfo
  {title} {Single-spin resonance in a van der waals embedded paramagnetic
  defect},\ }\href {https://doi.org/10.1038/s41563-021-00979-4} {\bibfield
  {journal} {\bibinfo  {journal} {Nature Materials}\ }\textbf {\bibinfo
  {volume} {20}},\ \bibinfo {pages} {1079} (\bibinfo {year}
  {2021})}\BibitemShut {NoStop}%
\bibitem [{\citenamefont {Stern}\ \emph {et~al.}(2022)\citenamefont {Stern},
  \citenamefont {Gu}, \citenamefont {Jarman}, \citenamefont {Eizagirre~Barker},
  \citenamefont {Mendelson}, \citenamefont {Chugh}, \citenamefont {Schott},
  \citenamefont {Tan}, \citenamefont {Sirringhaus}, \citenamefont
  {Aharonovich},\ and\ \citenamefont {Atature}}]{Stern2022}%
  \BibitemOpen
  \bibfield  {author} {\bibinfo {author} {\bibfnamefont {H.~L.}\ \bibnamefont
  {Stern}}, \bibinfo {author} {\bibfnamefont {Q.}~\bibnamefont {Gu}}, \bibinfo
  {author} {\bibfnamefont {J.}~\bibnamefont {Jarman}}, \bibinfo {author}
  {\bibfnamefont {S.}~\bibnamefont {Eizagirre~Barker}}, \bibinfo {author}
  {\bibfnamefont {N.}~\bibnamefont {Mendelson}}, \bibinfo {author}
  {\bibfnamefont {D.}~\bibnamefont {Chugh}}, \bibinfo {author} {\bibfnamefont
  {S.}~\bibnamefont {Schott}}, \bibinfo {author} {\bibfnamefont {H.~H.}\
  \bibnamefont {Tan}}, \bibinfo {author} {\bibfnamefont {H.}~\bibnamefont
  {Sirringhaus}}, \bibinfo {author} {\bibfnamefont {I.}~\bibnamefont
  {Aharonovich}},\ and\ \bibinfo {author} {\bibfnamefont {M.}~\bibnamefont
  {Atature}},\ }\bibfield  {title} {\bibinfo {title} {Room-temperature
  optically detected magnetic resonance of single defects in hexagonal boron
  nitride},\ }\href {https://doi.org/10.1038/s41467-022-28169-z} {\bibfield
  {journal} {\bibinfo  {journal} {Nature Communications}\ }\textbf {\bibinfo
  {volume} {13}},\ \bibinfo {pages} {618} (\bibinfo {year} {2022})}\BibitemShut
  {NoStop}%
\bibitem [{\citenamefont {Guo}\ \emph {et~al.}(2023)\citenamefont {Guo},
  \citenamefont {Li}, \citenamefont {Liu}, \citenamefont {Yang}, \citenamefont
  {Zeng}, \citenamefont {Yu}, \citenamefont {Meng}, \citenamefont {Li},
  \citenamefont {Wang}, \citenamefont {Xie}, \citenamefont {Ge}, \citenamefont
  {Wang}, \citenamefont {Li}, \citenamefont {Xu}, \citenamefont {Wang},
  \citenamefont {Tang}, \citenamefont {Gali}, \citenamefont {Li},\ and\
  \citenamefont {Guo}}]{Guo2023}%
  \BibitemOpen
  \bibfield  {author} {\bibinfo {author} {\bibfnamefont {N.-J.}\ \bibnamefont
  {Guo}}, \bibinfo {author} {\bibfnamefont {S.}~\bibnamefont {Li}}, \bibinfo
  {author} {\bibfnamefont {W.}~\bibnamefont {Liu}}, \bibinfo {author}
  {\bibfnamefont {Y.-Z.}\ \bibnamefont {Yang}}, \bibinfo {author}
  {\bibfnamefont {X.-D.}\ \bibnamefont {Zeng}}, \bibinfo {author}
  {\bibfnamefont {S.}~\bibnamefont {Yu}}, \bibinfo {author} {\bibfnamefont
  {Y.}~\bibnamefont {Meng}}, \bibinfo {author} {\bibfnamefont {Z.-P.}\
  \bibnamefont {Li}}, \bibinfo {author} {\bibfnamefont {Z.-A.}\ \bibnamefont
  {Wang}}, \bibinfo {author} {\bibfnamefont {L.-K.}\ \bibnamefont {Xie}},
  \bibinfo {author} {\bibfnamefont {R.-C.}\ \bibnamefont {Ge}}, \bibinfo
  {author} {\bibfnamefont {J.-F.}\ \bibnamefont {Wang}}, \bibinfo {author}
  {\bibfnamefont {Q.}~\bibnamefont {Li}}, \bibinfo {author} {\bibfnamefont
  {J.-S.}\ \bibnamefont {Xu}}, \bibinfo {author} {\bibfnamefont {Y.-T.}\
  \bibnamefont {Wang}}, \bibinfo {author} {\bibfnamefont {J.-S.}\ \bibnamefont
  {Tang}}, \bibinfo {author} {\bibfnamefont {A.}~\bibnamefont {Gali}}, \bibinfo
  {author} {\bibfnamefont {C.-F.}\ \bibnamefont {Li}},\ and\ \bibinfo {author}
  {\bibfnamefont {G.-C.}\ \bibnamefont {Guo}},\ }\bibfield  {title} {\bibinfo
  {title} {Coherent control of an ultrabright single spin in hexagonal boron
  nitride at room temperature},\ }\href
  {https://doi.org/10.1038/s41467-023-38672-6} {\bibfield  {journal} {\bibinfo
  {journal} {Nature Communications}\ }\textbf {\bibinfo {volume} {14}},\
  \bibinfo {pages} {2893} (\bibinfo {year} {2023})}\BibitemShut {NoStop}%
\bibitem [{\citenamefont {Yang}\ \emph {et~al.}(2023)\citenamefont {Yang},
  \citenamefont {Zhu}, \citenamefont {Li}, \citenamefont {Zeng}, \citenamefont
  {Guo}, \citenamefont {Yu}, \citenamefont {Meng}, \citenamefont {Wang},
  \citenamefont {Xie}, \citenamefont {Zhou}, \citenamefont {Li}, \citenamefont
  {Xu}, \citenamefont {Gao}, \citenamefont {Liu}, \citenamefont {Wang},
  \citenamefont {Tang}, \citenamefont {Li},\ and\ \citenamefont
  {Guo}}]{Yang2023}%
  \BibitemOpen
  \bibfield  {author} {\bibinfo {author} {\bibfnamefont {Y.-Z.}\ \bibnamefont
  {Yang}}, \bibinfo {author} {\bibfnamefont {T.-X.}\ \bibnamefont {Zhu}},
  \bibinfo {author} {\bibfnamefont {Z.-P.}\ \bibnamefont {Li}}, \bibinfo
  {author} {\bibfnamefont {X.-D.}\ \bibnamefont {Zeng}}, \bibinfo {author}
  {\bibfnamefont {N.-J.}\ \bibnamefont {Guo}}, \bibinfo {author} {\bibfnamefont
  {S.}~\bibnamefont {Yu}}, \bibinfo {author} {\bibfnamefont {Y.}~\bibnamefont
  {Meng}}, \bibinfo {author} {\bibfnamefont {Z.-A.}\ \bibnamefont {Wang}},
  \bibinfo {author} {\bibfnamefont {L.-K.}\ \bibnamefont {Xie}}, \bibinfo
  {author} {\bibfnamefont {Z.-Q.}\ \bibnamefont {Zhou}}, \bibinfo {author}
  {\bibfnamefont {Q.}~\bibnamefont {Li}}, \bibinfo {author} {\bibfnamefont
  {J.-S.}\ \bibnamefont {Xu}}, \bibinfo {author} {\bibfnamefont {X.-Y.}\
  \bibnamefont {Gao}}, \bibinfo {author} {\bibfnamefont {W.}~\bibnamefont
  {Liu}}, \bibinfo {author} {\bibfnamefont {Y.-T.}\ \bibnamefont {Wang}},
  \bibinfo {author} {\bibfnamefont {J.-S.}\ \bibnamefont {Tang}}, \bibinfo
  {author} {\bibfnamefont {C.-F.}\ \bibnamefont {Li}},\ and\ \bibinfo {author}
  {\bibfnamefont {G.-C.}\ \bibnamefont {Guo}},\ }\bibfield  {title} {\bibinfo
  {title} {Laser direct writing of visible spin defects in hexagonal boron
  nitride for applications in spin-based technologies},\ }\href
  {https://doi.org/10.1021/acsanm.3c01047} {\bibfield  {journal} {\bibinfo
  {journal} {ACS Applied Nano Materials}\ }\textbf {\bibinfo {volume} {6}},\
  \bibinfo {pages} {6407} (\bibinfo {year} {2023})}\BibitemShut {NoStop}%
\bibitem [{\citenamefont {Auburger}\ and\ \citenamefont
  {Gali}(2021)}]{Auburger2021}%
  \BibitemOpen
  \bibfield  {author} {\bibinfo {author} {\bibfnamefont {P.}~\bibnamefont
  {Auburger}}\ and\ \bibinfo {author} {\bibfnamefont {A.}~\bibnamefont
  {Gali}},\ }\bibfield  {title} {\bibinfo {title} {Towards ab initio
  identification of paramagnetic substitutional carbon defects in hexagonal
  boron nitride acting as quantum bits},\ }\href
  {https://doi.org/10.1103/PhysRevB.104.075410} {\bibfield  {journal} {\bibinfo
   {journal} {Physical Review B}\ }\textbf {\bibinfo {volume} {104}},\ \bibinfo
  {pages} {075410} (\bibinfo {year} {2021})}\BibitemShut {NoStop}%
\bibitem [{\citenamefont {Jara}\ \emph {et~al.}(2021)\citenamefont {Jara},
  \citenamefont {Rauch}, \citenamefont {Botti}, \citenamefont {Marques},
  \citenamefont {Norambuena}, \citenamefont {Coto}, \citenamefont
  {{Castellanos-{\'A}guila}}, \citenamefont {Maze},\ and\ \citenamefont
  {Munoz}}]{Jara2021}%
  \BibitemOpen
  \bibfield  {author} {\bibinfo {author} {\bibfnamefont {C.}~\bibnamefont
  {Jara}}, \bibinfo {author} {\bibfnamefont {T.}~\bibnamefont {Rauch}},
  \bibinfo {author} {\bibfnamefont {S.}~\bibnamefont {Botti}}, \bibinfo
  {author} {\bibfnamefont {M.~A.~L.}\ \bibnamefont {Marques}}, \bibinfo
  {author} {\bibfnamefont {A.}~\bibnamefont {Norambuena}}, \bibinfo {author}
  {\bibfnamefont {R.}~\bibnamefont {Coto}}, \bibinfo {author} {\bibfnamefont
  {J.~E.}\ \bibnamefont {{Castellanos-{\'A}guila}}}, \bibinfo {author}
  {\bibfnamefont {J.~R.}\ \bibnamefont {Maze}},\ and\ \bibinfo {author}
  {\bibfnamefont {F.}~\bibnamefont {Munoz}},\ }\bibfield  {title} {\bibinfo
  {title} {First-principles identification of single photon emitters based on
  carbon clusters in hexagonal boron nitride},\ }\href
  {https://doi.org/10.1021/acs.jpca.0c07339} {\bibfield  {journal} {\bibinfo
  {journal} {The Journal of Physical Chemistry A}\ }\textbf {\bibinfo {volume}
  {125}},\ \bibinfo {pages} {1325} (\bibinfo {year} {2021})}\BibitemShut
  {NoStop}%
\bibitem [{\citenamefont {Golami}\ \emph {et~al.}(2022)\citenamefont {Golami},
  \citenamefont {Sharman}, \citenamefont {Ghobadi}, \citenamefont {Wein},
  \citenamefont {{Zadeh-Haghighi}}, \citenamefont {{Gomes da Rocha}},
  \citenamefont {Salahub},\ and\ \citenamefont {Simon}}]{Golami2022}%
  \BibitemOpen
  \bibfield  {author} {\bibinfo {author} {\bibfnamefont {O.}~\bibnamefont
  {Golami}}, \bibinfo {author} {\bibfnamefont {K.}~\bibnamefont {Sharman}},
  \bibinfo {author} {\bibfnamefont {R.}~\bibnamefont {Ghobadi}}, \bibinfo
  {author} {\bibfnamefont {S.~C.}\ \bibnamefont {Wein}}, \bibinfo {author}
  {\bibfnamefont {H.}~\bibnamefont {{Zadeh-Haghighi}}}, \bibinfo {author}
  {\bibfnamefont {C.}~\bibnamefont {{Gomes da Rocha}}}, \bibinfo {author}
  {\bibfnamefont {D.~R.}\ \bibnamefont {Salahub}},\ and\ \bibinfo {author}
  {\bibfnamefont {C.}~\bibnamefont {Simon}},\ }\bibfield  {title} {\bibinfo
  {title} {{{Ab initio}} and group theoretical study of properties of a carbon
  trimer defect in hexagonal boron nitride},\ }\href
  {https://doi.org/10.1103/PhysRevB.105.184101} {\bibfield  {journal} {\bibinfo
   {journal} {Physical Review B}\ }\textbf {\bibinfo {volume} {105}},\ \bibinfo
  {pages} {184101} (\bibinfo {year} {2022})}\BibitemShut {NoStop}%
\bibitem [{\citenamefont {Li}\ \emph {et~al.}(2022)\citenamefont {Li},
  \citenamefont {Smart},\ and\ \citenamefont {Ping}}]{Li2022}%
  \BibitemOpen
  \bibfield  {author} {\bibinfo {author} {\bibfnamefont {K.}~\bibnamefont
  {Li}}, \bibinfo {author} {\bibfnamefont {T.~J.}\ \bibnamefont {Smart}},\ and\
  \bibinfo {author} {\bibfnamefont {Y.}~\bibnamefont {Ping}},\ }\bibfield
  {title} {\bibinfo {title} {Carbon trimer as a 2 {{eV}} single-photon emitter
  candidate in hexagonal boron nitride: A first-principles study},\ }\href
  {https://doi.org/10.1103/PhysRevMaterials.6.L042201} {\bibfield  {journal}
  {\bibinfo  {journal} {Physical Review Materials}\ }\textbf {\bibinfo {volume}
  {6}},\ \bibinfo {pages} {L042201} (\bibinfo {year} {2022})}\BibitemShut
  {NoStop}%
\bibitem [{\citenamefont {McCamey}\ \emph {et~al.}(2010)\citenamefont
  {McCamey}, \citenamefont {van Schooten}, \citenamefont {Baker}, \citenamefont
  {Lee}, \citenamefont {Paik}, \citenamefont {Lupton},\ and\ \citenamefont
  {Boehme}}]{McCamey2010}%
  \BibitemOpen
  \bibfield  {author} {\bibinfo {author} {\bibfnamefont {D.~R.}\ \bibnamefont
  {McCamey}}, \bibinfo {author} {\bibfnamefont {K.~J.}\ \bibnamefont {van
  Schooten}}, \bibinfo {author} {\bibfnamefont {W.~J.}\ \bibnamefont {Baker}},
  \bibinfo {author} {\bibfnamefont {S.-Y.}\ \bibnamefont {Lee}}, \bibinfo
  {author} {\bibfnamefont {S.-Y.}\ \bibnamefont {Paik}}, \bibinfo {author}
  {\bibfnamefont {J.~M.}\ \bibnamefont {Lupton}},\ and\ \bibinfo {author}
  {\bibfnamefont {C.}~\bibnamefont {Boehme}},\ }\bibfield  {title} {\bibinfo
  {title} {Hyperfine-field-mediated spin beating in electrostatically bound
  charge carrier pairs},\ }\href
  {https://doi.org/10.1103/PhysRevLett.104.017601} {\bibfield  {journal}
  {\bibinfo  {journal} {Phys. Rev. Lett.}\ }\textbf {\bibinfo {volume} {104}},\
  \bibinfo {pages} {017601} (\bibinfo {year} {2010})}\BibitemShut {NoStop}%
\bibitem [{\citenamefont {Davies}(1988)}]{Davies1988}%
  \BibitemOpen
  \bibfield  {author} {\bibinfo {author} {\bibfnamefont {J.}~\bibnamefont
  {Davies}},\ }\bibfield  {title} {\bibinfo {title} {Optically-detected
  magnetic resonance studies of ii–vi compounds},\ }\href
  {https://doi.org/https://doi.org/10.1016/0022-0248(90)90782-G} {\bibfield
  {journal} {\bibinfo  {journal} {Journal of Crystal Growth}\ }\textbf
  {\bibinfo {volume} {86}},\ \bibinfo {pages} {599} (\bibinfo {year}
  {1988})}\BibitemShut {NoStop}%
\bibitem [{\citenamefont {Boehme}\ and\ \citenamefont
  {Lips}(2003)}]{Boehme2003}%
  \BibitemOpen
  \bibfield  {author} {\bibinfo {author} {\bibfnamefont {C.}~\bibnamefont
  {Boehme}}\ and\ \bibinfo {author} {\bibfnamefont {K.}~\bibnamefont {Lips}},\
  }\bibfield  {title} {\bibinfo {title} {Theory of time-domain measurement of
  spin-dependent recombination with pulsed electrically detected magnetic
  resonance},\ }\href {https://doi.org/10.1103/PhysRevB.68.245105} {\bibfield
  {journal} {\bibinfo  {journal} {Phys. Rev. B}\ }\textbf {\bibinfo {volume}
  {68}},\ \bibinfo {pages} {245105} (\bibinfo {year} {2003})}\BibitemShut
  {NoStop}%
\bibitem [{\citenamefont {McCamey}\ \emph {et~al.}(2008)\citenamefont
  {McCamey}, \citenamefont {Seipel}, \citenamefont {Paik}, \citenamefont
  {Walter}, \citenamefont {Borys}, \citenamefont {Lupton},\ and\ \citenamefont
  {Boehme}}]{McCamey2008}%
  \BibitemOpen
  \bibfield  {author} {\bibinfo {author} {\bibfnamefont {D.}~\bibnamefont
  {McCamey}}, \bibinfo {author} {\bibfnamefont {H.}~\bibnamefont {Seipel}},
  \bibinfo {author} {\bibfnamefont {S.-Y.}\ \bibnamefont {Paik}}, \bibinfo
  {author} {\bibfnamefont {M.}~\bibnamefont {Walter}}, \bibinfo {author}
  {\bibfnamefont {N.}~\bibnamefont {Borys}}, \bibinfo {author} {\bibfnamefont
  {J.}~\bibnamefont {Lupton}},\ and\ \bibinfo {author} {\bibfnamefont
  {C.}~\bibnamefont {Boehme}},\ }\bibfield  {title} {\bibinfo {title} {Spin
  rabi flopping in the photocurrent of a polymer light-emitting diode},\
  }\href@noop {} {\bibfield  {journal} {\bibinfo  {journal} {Nature Materials}\
  }\textbf {\bibinfo {volume} {7}},\ \bibinfo {pages} {723} (\bibinfo {year}
  {2008})}\BibitemShut {NoStop}%
\bibitem [{\citenamefont {Hall}\ \emph {et~al.}(2016)\citenamefont {Hall},
  \citenamefont {Kehayias}, \citenamefont {Simpson}, \citenamefont {Jarmola},
  \citenamefont {Stacey}, \citenamefont {Budker},\ and\ \citenamefont
  {Hollenberg}}]{Hall2016}%
  \BibitemOpen
  \bibfield  {author} {\bibinfo {author} {\bibfnamefont {{\relax
  LT}.}~\bibnamefont {Hall}}, \bibinfo {author} {\bibfnamefont
  {P.}~\bibnamefont {Kehayias}}, \bibinfo {author} {\bibfnamefont {{\relax
  DA}.}~\bibnamefont {Simpson}}, \bibinfo {author} {\bibfnamefont
  {A.}~\bibnamefont {Jarmola}}, \bibinfo {author} {\bibfnamefont
  {A.}~\bibnamefont {Stacey}}, \bibinfo {author} {\bibfnamefont
  {D.}~\bibnamefont {Budker}},\ and\ \bibinfo {author} {\bibfnamefont {{\relax
  LCL}.}~\bibnamefont {Hollenberg}},\ }\bibfield  {title} {\bibinfo {title}
  {Detection of nanoscale electron spin resonance spectra demonstrated using
  nitrogen-vacancy centre probes in diamond},\ }\href
  {https://doi.org/10.1038/ncomms10211} {\bibfield  {journal} {\bibinfo
  {journal} {Nature Communications}\ }\textbf {\bibinfo {volume} {7}},\
  \bibinfo {pages} {10211} (\bibinfo {year} {2016})}\BibitemShut {NoStop}%
\bibitem [{\citenamefont {Broadway}\ \emph {et~al.}(2018)\citenamefont
  {Broadway}, \citenamefont {Tetienne}, \citenamefont {Stacey}, \citenamefont
  {Wood}, \citenamefont {Simpson}, \citenamefont {Hall},\ and\ \citenamefont
  {Hollenberg}}]{Broadway2018}%
  \BibitemOpen
  \bibfield  {author} {\bibinfo {author} {\bibfnamefont {D.~A.}\ \bibnamefont
  {Broadway}}, \bibinfo {author} {\bibfnamefont {J.-P.}\ \bibnamefont
  {Tetienne}}, \bibinfo {author} {\bibfnamefont {A.}~\bibnamefont {Stacey}},
  \bibinfo {author} {\bibfnamefont {J.~D.}\ \bibnamefont {Wood}}, \bibinfo
  {author} {\bibfnamefont {D.~A.}\ \bibnamefont {Simpson}}, \bibinfo {author}
  {\bibfnamefont {L.~T.}\ \bibnamefont {Hall}},\ and\ \bibinfo {author}
  {\bibfnamefont {L.~C.}\ \bibnamefont {Hollenberg}},\ }\bibfield  {title}
  {\bibinfo {title} {Quantum probe hyperpolarisation of molecular nuclear
  spins},\ }\href {https://doi.org/10.1038/s41467-018-03578-1} {\bibfield
  {journal} {\bibinfo  {journal} {Nature Communications}\ }\textbf {\bibinfo
  {volume} {9}},\ \bibinfo {pages} {1246} (\bibinfo {year} {2018})}\BibitemShut
  {NoStop}%
\bibitem [{\citenamefont {Haykal}\ \emph {et~al.}(2022)\citenamefont {Haykal},
  \citenamefont {Tanos}, \citenamefont {Minotto}, \citenamefont {Durand},
  \citenamefont {Fabre}, \citenamefont {Li}, \citenamefont {Edgar},
  \citenamefont {Iv{\'a}dy}, \citenamefont {Gali}, \citenamefont {Michel},
  \citenamefont {Dr{\'e}au}, \citenamefont {Gil}, \citenamefont {Cassabois},\
  and\ \citenamefont {Jacques}}]{Haykal2022DecoherenceNitride}%
  \BibitemOpen
  \bibfield  {author} {\bibinfo {author} {\bibfnamefont {A.}~\bibnamefont
  {Haykal}}, \bibinfo {author} {\bibfnamefont {R.}~\bibnamefont {Tanos}},
  \bibinfo {author} {\bibfnamefont {N.}~\bibnamefont {Minotto}}, \bibinfo
  {author} {\bibfnamefont {A.}~\bibnamefont {Durand}}, \bibinfo {author}
  {\bibfnamefont {F.}~\bibnamefont {Fabre}}, \bibinfo {author} {\bibfnamefont
  {J.}~\bibnamefont {Li}}, \bibinfo {author} {\bibfnamefont {J.~H.}\
  \bibnamefont {Edgar}}, \bibinfo {author} {\bibfnamefont {V.}~\bibnamefont
  {Iv{\'a}dy}}, \bibinfo {author} {\bibfnamefont {A.}~\bibnamefont {Gali}},
  \bibinfo {author} {\bibfnamefont {T.}~\bibnamefont {Michel}}, \bibinfo
  {author} {\bibfnamefont {A.}~\bibnamefont {Dr{\'e}au}}, \bibinfo {author}
  {\bibfnamefont {B.}~\bibnamefont {Gil}}, \bibinfo {author} {\bibfnamefont
  {G.}~\bibnamefont {Cassabois}},\ and\ \bibinfo {author} {\bibfnamefont
  {V.}~\bibnamefont {Jacques}},\ }\bibfield  {title} {\bibinfo {title}
  {Decoherence of {{V}}{{{\textsubscript{B}}}}{\textsuperscript{-}} spin
  defects in monoisotopic hexagonal boron nitride},\ }\href
  {https://doi.org/10.1038/s41467-022-31743-0} {\bibfield  {journal} {\bibinfo
  {journal} {Nature Communications}\ }\textbf {\bibinfo {volume} {13}},\
  \bibinfo {pages} {1} (\bibinfo {year} {2022})}\BibitemShut {NoStop}%
\bibitem [{\citenamefont {Baber}\ \emph {et~al.}(2022)\citenamefont {Baber},
  \citenamefont {Malein}, \citenamefont {Khatri}, \citenamefont {Keatley},
  \citenamefont {Guo}, \citenamefont {Withers}, \citenamefont {Ramsay},\ and\
  \citenamefont {Luxmoore}}]{Baber2022}%
  \BibitemOpen
  \bibfield  {author} {\bibinfo {author} {\bibfnamefont {S.}~\bibnamefont
  {Baber}}, \bibinfo {author} {\bibfnamefont {R.~N.~E.}\ \bibnamefont
  {Malein}}, \bibinfo {author} {\bibfnamefont {P.}~\bibnamefont {Khatri}},
  \bibinfo {author} {\bibfnamefont {P.~S.}\ \bibnamefont {Keatley}}, \bibinfo
  {author} {\bibfnamefont {S.}~\bibnamefont {Guo}}, \bibinfo {author}
  {\bibfnamefont {F.}~\bibnamefont {Withers}}, \bibinfo {author} {\bibfnamefont
  {A.~J.}\ \bibnamefont {Ramsay}},\ and\ \bibinfo {author} {\bibfnamefont
  {I.~J.}\ \bibnamefont {Luxmoore}},\ }\bibfield  {title} {\bibinfo {title}
  {Excited state spectroscopy of boron vacancy defects in hexagonal boron
  nitride using time-resolved optically detected magnetic resonance},\ }\href
  {https://doi.org/10.1021/acs.nanolett.1c04366} {\bibfield  {journal}
  {\bibinfo  {journal} {Nano Letters}\ }\textbf {\bibinfo {volume} {22}},\
  \bibinfo {pages} {461} (\bibinfo {year} {2022})}\BibitemShut {NoStop}%
\bibitem [{\citenamefont {Fuchs}\ \emph {et~al.}(2011)\citenamefont {Fuchs},
  \citenamefont {Burkard}, \citenamefont {Klimov},\ and\ \citenamefont
  {Awschalom}}]{Fuchs2011}%
  \BibitemOpen
  \bibfield  {author} {\bibinfo {author} {\bibfnamefont {G.}~\bibnamefont
  {Fuchs}}, \bibinfo {author} {\bibfnamefont {G.}~\bibnamefont {Burkard}},
  \bibinfo {author} {\bibfnamefont {P.}~\bibnamefont {Klimov}},\ and\ \bibinfo
  {author} {\bibfnamefont {D.}~\bibnamefont {Awschalom}},\ }\bibfield  {title}
  {\bibinfo {title} {A quantum memory intrinsic to single nitrogen--vacancy
  centres in diamond},\ }\href@noop {} {\bibfield  {journal} {\bibinfo
  {journal} {Nature Physics}\ }\textbf {\bibinfo {volume} {7}},\ \bibinfo
  {pages} {789} (\bibinfo {year} {2011})}\BibitemShut {NoStop}%
\bibitem [{\citenamefont {Tetienne}\ \emph {et~al.}(2012)\citenamefont
  {Tetienne}, \citenamefont {Rondin}, \citenamefont {Spinicelli}, \citenamefont
  {Chipaux}, \citenamefont {Debuisschert}, \citenamefont {Roch},\ and\
  \citenamefont {Jacques}}]{Tetienne2012}%
  \BibitemOpen
  \bibfield  {author} {\bibinfo {author} {\bibfnamefont {J.-P.}\ \bibnamefont
  {Tetienne}}, \bibinfo {author} {\bibfnamefont {L.}~\bibnamefont {Rondin}},
  \bibinfo {author} {\bibfnamefont {P.}~\bibnamefont {Spinicelli}}, \bibinfo
  {author} {\bibfnamefont {M.}~\bibnamefont {Chipaux}}, \bibinfo {author}
  {\bibfnamefont {T.}~\bibnamefont {Debuisschert}}, \bibinfo {author}
  {\bibfnamefont {J.-F.}\ \bibnamefont {Roch}},\ and\ \bibinfo {author}
  {\bibfnamefont {V.}~\bibnamefont {Jacques}},\ }\bibfield  {title} {\bibinfo
  {title} {Magnetic-field-dependent photodynamics of single {{NV}} defects in
  diamond: An application to qualitative all-optical magnetic imaging},\ }\href
  {https://doi.org/10.1088/1367-2630/14/10/103033} {\bibfield  {journal}
  {\bibinfo  {journal} {New Journal of Physics}\ }\textbf {\bibinfo {volume}
  {14}},\ \bibinfo {pages} {103033} (\bibinfo {year} {2012})}\BibitemShut
  {NoStop}%
\bibitem [{\citenamefont {Chugh}\ \emph {et~al.}(2018)\citenamefont {Chugh},
  \citenamefont {Wong-Leung}, \citenamefont {Li}, \citenamefont {Lysevych},
  \citenamefont {Tan},\ and\ \citenamefont {Jagadish}}]{Chugh2018}%
  \BibitemOpen
  \bibfield  {author} {\bibinfo {author} {\bibfnamefont {D.}~\bibnamefont
  {Chugh}}, \bibinfo {author} {\bibfnamefont {J.}~\bibnamefont {Wong-Leung}},
  \bibinfo {author} {\bibfnamefont {L.}~\bibnamefont {Li}}, \bibinfo {author}
  {\bibfnamefont {M.}~\bibnamefont {Lysevych}}, \bibinfo {author}
  {\bibfnamefont {H.~H.}\ \bibnamefont {Tan}},\ and\ \bibinfo {author}
  {\bibfnamefont {C.}~\bibnamefont {Jagadish}},\ }\bibfield  {title} {\bibinfo
  {title} {Flow modulation epitaxy of hexagonal boron nitride},\ }\href
  {https://doi.org/10.1088/2053-1583/aad5aa} {\bibfield  {journal} {\bibinfo
  {journal} {2D Materials}\ }\textbf {\bibinfo {volume} {5}},\ \bibinfo {pages}
  {045018} (\bibinfo {year} {2018})}\BibitemShut {NoStop}%
\bibitem [{\citenamefont {Zhang}\ \emph {et~al.}(2022)\citenamefont {Zhang},
  \citenamefont {Guo}, \citenamefont {Wu}, \citenamefont {Wen}, \citenamefont
  {Yang}, \citenamefont {Jin}, \citenamefont {Zhang},\ and\ \citenamefont
  {Chang}}]{Zhang2022}%
  \BibitemOpen
  \bibfield  {author} {\bibinfo {author} {\bibfnamefont {G.}~\bibnamefont
  {Zhang}}, \bibinfo {author} {\bibfnamefont {F.}~\bibnamefont {Guo}}, \bibinfo
  {author} {\bibfnamefont {H.}~\bibnamefont {Wu}}, \bibinfo {author}
  {\bibfnamefont {X.}~\bibnamefont {Wen}}, \bibinfo {author} {\bibfnamefont
  {L.}~\bibnamefont {Yang}}, \bibinfo {author} {\bibfnamefont {W.}~\bibnamefont
  {Jin}}, \bibinfo {author} {\bibfnamefont {W.}~\bibnamefont {Zhang}},\ and\
  \bibinfo {author} {\bibfnamefont {H.}~\bibnamefont {Chang}},\ }\bibfield
  {title} {\bibinfo {title} {Above-room-temperature strong intrinsic
  ferromagnetism in 2d van der waals fe3gate2 with large perpendicular magnetic
  anisotropy},\ }\href@noop {} {\bibfield  {journal} {\bibinfo  {journal}
  {Nature Communications}\ }\textbf {\bibinfo {volume} {13}},\ \bibinfo {pages}
  {5067} (\bibinfo {year} {2022})}\BibitemShut {NoStop}%
\bibitem [{\citenamefont {Bedoya-Pinto}\ \emph {et~al.}(2021)\citenamefont
  {Bedoya-Pinto}, \citenamefont {Ji}, \citenamefont {Pandeya}, \citenamefont
  {Gargiani}, \citenamefont {Valvidares}, \citenamefont {Sessi}, \citenamefont
  {Taylor}, \citenamefont {Radu}, \citenamefont {Chang},\ and\ \citenamefont
  {Parkin}}]{Bedoya2021}%
  \BibitemOpen
  \bibfield  {author} {\bibinfo {author} {\bibfnamefont {A.}~\bibnamefont
  {Bedoya-Pinto}}, \bibinfo {author} {\bibfnamefont {J.-R.}\ \bibnamefont
  {Ji}}, \bibinfo {author} {\bibfnamefont {A.~K.}\ \bibnamefont {Pandeya}},
  \bibinfo {author} {\bibfnamefont {P.}~\bibnamefont {Gargiani}}, \bibinfo
  {author} {\bibfnamefont {M.}~\bibnamefont {Valvidares}}, \bibinfo {author}
  {\bibfnamefont {P.}~\bibnamefont {Sessi}}, \bibinfo {author} {\bibfnamefont
  {J.~M.}\ \bibnamefont {Taylor}}, \bibinfo {author} {\bibfnamefont
  {F.}~\bibnamefont {Radu}}, \bibinfo {author} {\bibfnamefont {K.}~\bibnamefont
  {Chang}},\ and\ \bibinfo {author} {\bibfnamefont {S.~S.~P.}\ \bibnamefont
  {Parkin}},\ }\bibfield  {title} {\bibinfo {title} {Intrinsic 2d-xy
  ferromagnetism in a van der waals monolayer},\ }\href
  {https://doi.org/10.1126/science.abd5146} {\bibfield  {journal} {\bibinfo
  {journal} {Science}\ }\textbf {\bibinfo {volume} {374}},\ \bibinfo {pages}
  {616} (\bibinfo {year} {2021})}\BibitemShut {NoStop}%
\bibitem [{\citenamefont {Huang}\ \emph {et~al.}(2023)\citenamefont {Huang},
  \citenamefont {Sun}, \citenamefont {Yan}, \citenamefont {Xie}, \citenamefont
  {Agarwal}, \citenamefont {Ye}, \citenamefont {Sung}, \citenamefont {Lu},
  \citenamefont {Zhou}, \citenamefont {Yan} \emph {et~al.}}]{Huang2023}%
  \BibitemOpen
  \bibfield  {author} {\bibinfo {author} {\bibfnamefont {M.}~\bibnamefont
  {Huang}}, \bibinfo {author} {\bibfnamefont {Z.}~\bibnamefont {Sun}}, \bibinfo
  {author} {\bibfnamefont {G.}~\bibnamefont {Yan}}, \bibinfo {author}
  {\bibfnamefont {H.}~\bibnamefont {Xie}}, \bibinfo {author} {\bibfnamefont
  {N.}~\bibnamefont {Agarwal}}, \bibinfo {author} {\bibfnamefont
  {G.}~\bibnamefont {Ye}}, \bibinfo {author} {\bibfnamefont {S.~H.}\
  \bibnamefont {Sung}}, \bibinfo {author} {\bibfnamefont {H.}~\bibnamefont
  {Lu}}, \bibinfo {author} {\bibfnamefont {J.}~\bibnamefont {Zhou}}, \bibinfo
  {author} {\bibfnamefont {S.}~\bibnamefont {Yan}}, \emph {et~al.},\ }\bibfield
   {title} {\bibinfo {title} {Revealing intrinsic domains and fluctuations of
  moir{\'e} magnetism by a wide-field quantum microscope},\ }\href@noop {}
  {\bibfield  {journal} {\bibinfo  {journal} {Nature Communications}\ }\textbf
  {\bibinfo {volume} {14}},\ \bibinfo {pages} {5259} (\bibinfo {year}
  {2023})}\BibitemShut {NoStop}%
\bibitem [{\citenamefont {Rizzato}\ \emph {et~al.}(2023)\citenamefont
  {Rizzato}, \citenamefont {Schalk}, \citenamefont {Mohr}, \citenamefont
  {Hermann}, \citenamefont {Leibold}, \citenamefont {Bruckmaier}, \citenamefont
  {Salvitti}, \citenamefont {Qian}, \citenamefont {Ji}, \citenamefont
  {Astakhov} \emph {et~al.}}]{Rizzato2023}%
  \BibitemOpen
  \bibfield  {author} {\bibinfo {author} {\bibfnamefont {R.}~\bibnamefont
  {Rizzato}}, \bibinfo {author} {\bibfnamefont {M.}~\bibnamefont {Schalk}},
  \bibinfo {author} {\bibfnamefont {S.}~\bibnamefont {Mohr}}, \bibinfo {author}
  {\bibfnamefont {J.~C.}\ \bibnamefont {Hermann}}, \bibinfo {author}
  {\bibfnamefont {J.~P.}\ \bibnamefont {Leibold}}, \bibinfo {author}
  {\bibfnamefont {F.}~\bibnamefont {Bruckmaier}}, \bibinfo {author}
  {\bibfnamefont {G.}~\bibnamefont {Salvitti}}, \bibinfo {author}
  {\bibfnamefont {C.}~\bibnamefont {Qian}}, \bibinfo {author} {\bibfnamefont
  {P.}~\bibnamefont {Ji}}, \bibinfo {author} {\bibfnamefont {G.~V.}\
  \bibnamefont {Astakhov}}, \emph {et~al.},\ }\bibfield  {title} {\bibinfo
  {title} {Extending the coherence of spin defects in hbn enables advanced
  qubit control and quantum sensing},\ }\href@noop {} {\bibfield  {journal}
  {\bibinfo  {journal} {Nature Communications}\ }\textbf {\bibinfo {volume}
  {14}},\ \bibinfo {pages} {5089} (\bibinfo {year} {2023})}\BibitemShut
  {NoStop}%
\bibitem [{\citenamefont {Ramsay}\ \emph {et~al.}(2023)\citenamefont {Ramsay},
  \citenamefont {Hekmati}, \citenamefont {Patrickson}, \citenamefont {Baber},
  \citenamefont {Arvidsson-Shukur}, \citenamefont {Bennett},\ and\
  \citenamefont {Luxmoore}}]{Ramsay2023}%
  \BibitemOpen
  \bibfield  {author} {\bibinfo {author} {\bibfnamefont {A.~J.}\ \bibnamefont
  {Ramsay}}, \bibinfo {author} {\bibfnamefont {R.}~\bibnamefont {Hekmati}},
  \bibinfo {author} {\bibfnamefont {C.~J.}\ \bibnamefont {Patrickson}},
  \bibinfo {author} {\bibfnamefont {S.}~\bibnamefont {Baber}}, \bibinfo
  {author} {\bibfnamefont {D.~R.}\ \bibnamefont {Arvidsson-Shukur}}, \bibinfo
  {author} {\bibfnamefont {A.~J.}\ \bibnamefont {Bennett}},\ and\ \bibinfo
  {author} {\bibfnamefont {I.~J.}\ \bibnamefont {Luxmoore}},\ }\bibfield
  {title} {\bibinfo {title} {Coherence protection of spin qubits in hexagonal
  boron nitride},\ }\href@noop {} {\bibfield  {journal} {\bibinfo  {journal}
  {Nature Communications}\ }\textbf {\bibinfo {volume} {14}},\ \bibinfo {pages}
  {461} (\bibinfo {year} {2023})}\BibitemShut {NoStop}%
\bibitem [{\citenamefont {Wolfowicz}\ \emph {et~al.}(2021)\citenamefont
  {Wolfowicz}, \citenamefont {Heremans}, \citenamefont {Anderson},
  \citenamefont {Kanai}, \citenamefont {Seo}, \citenamefont {Gali},
  \citenamefont {Galli},\ and\ \citenamefont
  {Awschalom}}]{Wolfowicz2021QuantumDefects}%
  \BibitemOpen
  \bibfield  {author} {\bibinfo {author} {\bibfnamefont {G.}~\bibnamefont
  {Wolfowicz}}, \bibinfo {author} {\bibfnamefont {F.~J.}\ \bibnamefont
  {Heremans}}, \bibinfo {author} {\bibfnamefont {C.~P.}\ \bibnamefont
  {Anderson}}, \bibinfo {author} {\bibfnamefont {S.}~\bibnamefont {Kanai}},
  \bibinfo {author} {\bibfnamefont {H.}~\bibnamefont {Seo}}, \bibinfo {author}
  {\bibfnamefont {A.}~\bibnamefont {Gali}}, \bibinfo {author} {\bibfnamefont
  {G.}~\bibnamefont {Galli}},\ and\ \bibinfo {author} {\bibfnamefont {D.~D.}\
  \bibnamefont {Awschalom}},\ }\bibfield  {title} {\bibinfo {title} {Quantum
  guidelines for solid-state spin defects},\ }\href
  {https://doi.org/10.1038/s41578-021-00306-y} {\bibfield  {journal} {\bibinfo
  {journal} {Nature Reviews Materials}\ }\textbf {\bibinfo {volume} {6}},\
  \bibinfo {pages} {906} (\bibinfo {year} {2021})}\BibitemShut {NoStop}%
\bibitem [{\citenamefont {Lillie}\ \emph {et~al.}(2020)\citenamefont {Lillie},
  \citenamefont {Broadway}, \citenamefont {Dontschuk}, \citenamefont
  {Scholten}, \citenamefont {Johnson}, \citenamefont {Wolf}, \citenamefont
  {Rachel}, \citenamefont {Hollenberg},\ and\ \citenamefont
  {Tetienne}}]{Lillie2020}%
  \BibitemOpen
  \bibfield  {author} {\bibinfo {author} {\bibfnamefont {S.~E.}\ \bibnamefont
  {Lillie}}, \bibinfo {author} {\bibfnamefont {D.~A.}\ \bibnamefont
  {Broadway}}, \bibinfo {author} {\bibfnamefont {N.}~\bibnamefont {Dontschuk}},
  \bibinfo {author} {\bibfnamefont {S.~C.}\ \bibnamefont {Scholten}}, \bibinfo
  {author} {\bibfnamefont {B.~C.}\ \bibnamefont {Johnson}}, \bibinfo {author}
  {\bibfnamefont {S.}~\bibnamefont {Wolf}}, \bibinfo {author} {\bibfnamefont
  {S.}~\bibnamefont {Rachel}}, \bibinfo {author} {\bibfnamefont {L.~C.~L.}\
  \bibnamefont {Hollenberg}},\ and\ \bibinfo {author} {\bibfnamefont {J.-P.}\
  \bibnamefont {Tetienne}},\ }\bibfield  {title} {\bibinfo {title} {Laser
  modulation of superconductivity in a cryogenic wide-field nitrogen-vacancy
  microscope},\ }\href {https://doi.org/10.1021/acs.nanolett.9b05071}
  {\bibfield  {journal} {\bibinfo  {journal} {Nano Letters}\ }\textbf {\bibinfo
  {volume} {20}},\ \bibinfo {pages} {1855} (\bibinfo {year}
  {2020})}\BibitemShut {NoStop}%
\bibitem [{\citenamefont {Chen}\ \emph {et~al.}(2021)\citenamefont {Chen},
  \citenamefont {Li}, \citenamefont {White}, \citenamefont {Nonahal},
  \citenamefont {Xu}, \citenamefont {Watanabe}, \citenamefont {Taniguchi},
  \citenamefont {Toth}, \citenamefont {Tran},\ and\ \citenamefont
  {Aharonovich}}]{Chen2021ACSINT}%
  \BibitemOpen
  \bibfield  {author} {\bibinfo {author} {\bibfnamefont {Y.}~\bibnamefont
  {Chen}}, \bibinfo {author} {\bibfnamefont {C.}~\bibnamefont {Li}}, \bibinfo
  {author} {\bibfnamefont {S.}~\bibnamefont {White}}, \bibinfo {author}
  {\bibfnamefont {M.}~\bibnamefont {Nonahal}}, \bibinfo {author} {\bibfnamefont
  {Z.-Q.}\ \bibnamefont {Xu}}, \bibinfo {author} {\bibfnamefont
  {K.}~\bibnamefont {Watanabe}}, \bibinfo {author} {\bibfnamefont
  {T.}~\bibnamefont {Taniguchi}}, \bibinfo {author} {\bibfnamefont
  {M.}~\bibnamefont {Toth}}, \bibinfo {author} {\bibfnamefont {T.~T.}\
  \bibnamefont {Tran}},\ and\ \bibinfo {author} {\bibfnamefont
  {I.}~\bibnamefont {Aharonovich}},\ }\bibfield  {title} {\bibinfo {title}
  {Generation of high-density quantum emitters in high-quality, exfoliated
  hexagonal boron nitride},\ }\href {https://doi.org/10.1021/acsami.1c14863}
  {\bibfield  {journal} {\bibinfo  {journal} {ACS Applied Materials \&
  Interfaces}\ }\textbf {\bibinfo {volume} {13}},\ \bibinfo {pages} {47283}
  (\bibinfo {year} {2021})}\BibitemShut {NoStop}%
\bibitem [{\citenamefont {Liu}\ \emph {et~al.}(2022{\natexlab{b}})\citenamefont
  {Liu}, \citenamefont {Mendelson}, \citenamefont {Abidi}, \citenamefont {Li},
  \citenamefont {Liu}, \citenamefont {Cai}, \citenamefont {Zhang},
  \citenamefont {You}, \citenamefont {Tamtaji}, \citenamefont {Wong},
  \citenamefont {Ding}, \citenamefont {Chen}, \citenamefont {Aharonovich},\
  and\ \citenamefont {Luo}}]{Liu2022ACS}%
  \BibitemOpen
  \bibfield  {author} {\bibinfo {author} {\bibfnamefont {H.}~\bibnamefont
  {Liu}}, \bibinfo {author} {\bibfnamefont {N.}~\bibnamefont {Mendelson}},
  \bibinfo {author} {\bibfnamefont {I.~H.}\ \bibnamefont {Abidi}}, \bibinfo
  {author} {\bibfnamefont {S.}~\bibnamefont {Li}}, \bibinfo {author}
  {\bibfnamefont {Z.}~\bibnamefont {Liu}}, \bibinfo {author} {\bibfnamefont
  {Y.}~\bibnamefont {Cai}}, \bibinfo {author} {\bibfnamefont {K.}~\bibnamefont
  {Zhang}}, \bibinfo {author} {\bibfnamefont {J.}~\bibnamefont {You}}, \bibinfo
  {author} {\bibfnamefont {M.}~\bibnamefont {Tamtaji}}, \bibinfo {author}
  {\bibfnamefont {H.}~\bibnamefont {Wong}}, \bibinfo {author} {\bibfnamefont
  {Y.}~\bibnamefont {Ding}}, \bibinfo {author} {\bibfnamefont {G.}~\bibnamefont
  {Chen}}, \bibinfo {author} {\bibfnamefont {I.}~\bibnamefont {Aharonovich}},\
  and\ \bibinfo {author} {\bibfnamefont {Z.}~\bibnamefont {Luo}},\ }\bibfield
  {title} {\bibinfo {title} {Rational control on quantum emitter formation in
  carbon-doped monolayer hexagonal boron nitride},\ }\href
  {https://doi.org/10.1021/acsami.1c21781} {\bibfield  {journal} {\bibinfo
  {journal} {ACS Applied Materials \& Interfaces}\ }\textbf {\bibinfo {volume}
  {14}},\ \bibinfo {pages} {3189} (\bibinfo {year}
  {2022}{\natexlab{b}})}\BibitemShut {NoStop}%
\bibitem [{\citenamefont {Mathur}\ \emph {et~al.}(2022)\citenamefont {Mathur},
  \citenamefont {Mukherjee}, \citenamefont {Gao}, \citenamefont {Luo},
  \citenamefont {McCullian}, \citenamefont {Li}, \citenamefont {Vamivakas},\
  and\ \citenamefont {Fuchs}}]{Mathur2022}%
  \BibitemOpen
  \bibfield  {author} {\bibinfo {author} {\bibfnamefont {N.}~\bibnamefont
  {Mathur}}, \bibinfo {author} {\bibfnamefont {A.}~\bibnamefont {Mukherjee}},
  \bibinfo {author} {\bibfnamefont {X.}~\bibnamefont {Gao}}, \bibinfo {author}
  {\bibfnamefont {J.}~\bibnamefont {Luo}}, \bibinfo {author} {\bibfnamefont
  {B.~A.}\ \bibnamefont {McCullian}}, \bibinfo {author} {\bibfnamefont
  {T.}~\bibnamefont {Li}}, \bibinfo {author} {\bibfnamefont {A.~N.}\
  \bibnamefont {Vamivakas}},\ and\ \bibinfo {author} {\bibfnamefont {G.~D.}\
  \bibnamefont {Fuchs}},\ }\bibfield  {title} {\bibinfo {title} {Excited-state
  spin-resonance spectroscopy of {{VB-}} defect centers in hexagonal boron
  nitride},\ }\href {https://doi.org/10.1038/s41467-022-30772-z} {\bibfield
  {journal} {\bibinfo  {journal} {Nature Communications}\ }\textbf {\bibinfo
  {volume} {13}},\ \bibinfo {pages} {3233} (\bibinfo {year}
  {2022})}\BibitemShut {NoStop}%
\bibitem [{\citenamefont {Mu}\ \emph {et~al.}(2022)\citenamefont {Mu},
  \citenamefont {Cai}, \citenamefont {Chen}, \citenamefont {Kenny},
  \citenamefont {Jiang}, \citenamefont {Ru}, \citenamefont {Lyu}, \citenamefont
  {Koh}, \citenamefont {Liu}, \citenamefont {Aharonovich},\ and\ \citenamefont
  {Gao}}]{Mu2022}%
  \BibitemOpen
  \bibfield  {author} {\bibinfo {author} {\bibfnamefont {Z.}~\bibnamefont
  {Mu}}, \bibinfo {author} {\bibfnamefont {H.}~\bibnamefont {Cai}}, \bibinfo
  {author} {\bibfnamefont {D.}~\bibnamefont {Chen}}, \bibinfo {author}
  {\bibfnamefont {J.}~\bibnamefont {Kenny}}, \bibinfo {author} {\bibfnamefont
  {Z.}~\bibnamefont {Jiang}}, \bibinfo {author} {\bibfnamefont
  {S.}~\bibnamefont {Ru}}, \bibinfo {author} {\bibfnamefont {X.}~\bibnamefont
  {Lyu}}, \bibinfo {author} {\bibfnamefont {T.~S.}\ \bibnamefont {Koh}},
  \bibinfo {author} {\bibfnamefont {X.}~\bibnamefont {Liu}}, \bibinfo {author}
  {\bibfnamefont {I.}~\bibnamefont {Aharonovich}},\ and\ \bibinfo {author}
  {\bibfnamefont {W.}~\bibnamefont {Gao}},\ }\bibfield  {title} {\bibinfo
  {title} {Excited-state optically detected magnetic resonance of spin defects
  in hexagonal boron nitride},\ }\href
  {https://doi.org/10.1103/PhysRevLett.128.216402} {\bibfield  {journal}
  {\bibinfo  {journal} {Physical Review Letters}\ }\textbf {\bibinfo {volume}
  {128}},\ \bibinfo {pages} {216402} (\bibinfo {year} {2022})}\BibitemShut
  {NoStop}%
\bibitem [{\citenamefont {Yu}\ \emph {et~al.}(2022)\citenamefont {Yu},
  \citenamefont {Sun}, \citenamefont {Wang}, \citenamefont {Zhang},
  \citenamefont {Ye}, \citenamefont {Zhou}, \citenamefont {Liu}, \citenamefont
  {Wang}, \citenamefont {Shi}, \citenamefont {Wang},\ and\ \citenamefont
  {Du}}]{Yu2022}%
  \BibitemOpen
  \bibfield  {author} {\bibinfo {author} {\bibfnamefont {P.}~\bibnamefont
  {Yu}}, \bibinfo {author} {\bibfnamefont {H.}~\bibnamefont {Sun}}, \bibinfo
  {author} {\bibfnamefont {M.}~\bibnamefont {Wang}}, \bibinfo {author}
  {\bibfnamefont {T.}~\bibnamefont {Zhang}}, \bibinfo {author} {\bibfnamefont
  {X.}~\bibnamefont {Ye}}, \bibinfo {author} {\bibfnamefont {J.}~\bibnamefont
  {Zhou}}, \bibinfo {author} {\bibfnamefont {H.}~\bibnamefont {Liu}}, \bibinfo
  {author} {\bibfnamefont {C.-J.}\ \bibnamefont {Wang}}, \bibinfo {author}
  {\bibfnamefont {F.}~\bibnamefont {Shi}}, \bibinfo {author} {\bibfnamefont
  {Y.}~\bibnamefont {Wang}},\ and\ \bibinfo {author} {\bibfnamefont
  {J.}~\bibnamefont {Du}},\ }\bibfield  {title} {\bibinfo {title}
  {Excited-state spectroscopy of spin defects in hexagonal boron nitride},\
  }\href {https://doi.org/10.1021/acs.nanolett.1c04841} {\bibfield  {journal}
  {\bibinfo  {journal} {Nano Letters}\ }\textbf {\bibinfo {volume} {22}},\
  \bibinfo {pages} {3545} (\bibinfo {year} {2022})}\BibitemShut {NoStop}%
\bibitem [{\citenamefont {Murzakhanov}\ \emph {et~al.}(2021)\citenamefont
  {Murzakhanov}, \citenamefont {Yavkin}, \citenamefont {Mamin}, \citenamefont
  {Orlinskii}, \citenamefont {Mumdzhi}, \citenamefont {Gracheva}, \citenamefont
  {Gabbasov}, \citenamefont {Smirnov}, \citenamefont {Davydov},\ and\
  \citenamefont {Soltamov}}]{Murzakhanov2021}%
  \BibitemOpen
  \bibfield  {author} {\bibinfo {author} {\bibfnamefont {F.~F.}\ \bibnamefont
  {Murzakhanov}}, \bibinfo {author} {\bibfnamefont {B.~V.}\ \bibnamefont
  {Yavkin}}, \bibinfo {author} {\bibfnamefont {G.~V.}\ \bibnamefont {Mamin}},
  \bibinfo {author} {\bibfnamefont {S.~B.}\ \bibnamefont {Orlinskii}}, \bibinfo
  {author} {\bibfnamefont {I.~E.}\ \bibnamefont {Mumdzhi}}, \bibinfo {author}
  {\bibfnamefont {I.~N.}\ \bibnamefont {Gracheva}}, \bibinfo {author}
  {\bibfnamefont {B.~F.}\ \bibnamefont {Gabbasov}}, \bibinfo {author}
  {\bibfnamefont {A.~N.}\ \bibnamefont {Smirnov}}, \bibinfo {author}
  {\bibfnamefont {V.~Y.}\ \bibnamefont {Davydov}},\ and\ \bibinfo {author}
  {\bibfnamefont {V.~A.}\ \bibnamefont {Soltamov}},\ }\bibfield  {title}
  {\bibinfo {title} {Creation of negatively charged boron vacancies in
  hexagonal boron nitride crystal by electron irradiation and mechanism of
  inhomogeneous broadening of boron vacancy-related spin resonance lines},\
  }\href {https://doi.org/10.3390/nano11061373} {\bibfield  {journal} {\bibinfo
   {journal} {Nanomaterials (Basel, Switzerland)}\ }\textbf {\bibinfo {volume}
  {11}},\ \bibinfo {pages} {1373} (\bibinfo {year} {2021})}\BibitemShut
  {NoStop}%
\bibitem [{\citenamefont {Bonamente}(2017)}]{Bonamenta2017}%
  \BibitemOpen
  \bibfield  {author} {\bibinfo {author} {\bibfnamefont {M.}~\bibnamefont
  {Bonamente}},\ }\bibfield  {title} {\bibinfo {title} {Goodness of fit and
  parameter uncertainty},\ }in\ \href
  {https://doi.org/10.1007/978-1-4939-6572-4_10} {\emph {\bibinfo {booktitle}
  {Statistics and {{Analysis}} of {{Scientific Data}}}}},\ \bibinfo {series and
  number} {Graduate {{Texts}} in {{Physics}}},\ \bibinfo {editor} {edited by\
  \bibinfo {editor} {\bibfnamefont {M.}~\bibnamefont {Bonamente}}}\ (\bibinfo
  {publisher} {{Springer}},\ \bibinfo {address} {{New York, NY}},\ \bibinfo
  {year} {2017})\ pp.\ \bibinfo {pages} {177--193}\BibitemShut {NoStop}%
\bibitem [{\citenamefont {Lee}\ \emph {et~al.}(2011)\citenamefont {Lee},
  \citenamefont {Paik}, \citenamefont {McCamey}, \citenamefont {Yu},
  \citenamefont {Burn}, \citenamefont {Lupton},\ and\ \citenamefont
  {Boehme}}]{Lee2010}%
  \BibitemOpen
  \bibfield  {author} {\bibinfo {author} {\bibfnamefont {S.-Y.}\ \bibnamefont
  {Lee}}, \bibinfo {author} {\bibfnamefont {S.-Y.}\ \bibnamefont {Paik}},
  \bibinfo {author} {\bibfnamefont {D.~R.}\ \bibnamefont {McCamey}}, \bibinfo
  {author} {\bibfnamefont {J.}~\bibnamefont {Yu}}, \bibinfo {author}
  {\bibfnamefont {P.~L.}\ \bibnamefont {Burn}}, \bibinfo {author}
  {\bibfnamefont {J.~M.}\ \bibnamefont {Lupton}},\ and\ \bibinfo {author}
  {\bibfnamefont {C.}~\bibnamefont {Boehme}},\ }\bibfield  {title} {\bibinfo
  {title} {Tuning hyperfine fields in conjugated polymers for coherent organic
  spintronics},\ }\href {https://doi.org/10.1021/ja108352d} {\bibfield
  {journal} {\bibinfo  {journal} {Journal of the American Chemical Society}\
  }\textbf {\bibinfo {volume} {133}},\ \bibinfo {pages} {2019} (\bibinfo {year}
  {2011})}\BibitemShut {NoStop}%
\bibitem [{\citenamefont {Hall}\ \emph {et~al.}(2014)\citenamefont {Hall},
  \citenamefont {Cole},\ and\ \citenamefont {Hollenberg}}]{Hall2014}%
  \BibitemOpen
  \bibfield  {author} {\bibinfo {author} {\bibfnamefont {L.~T.}\ \bibnamefont
  {Hall}}, \bibinfo {author} {\bibfnamefont {J.~H.}\ \bibnamefont {Cole}},\
  and\ \bibinfo {author} {\bibfnamefont {L.~C.~L.}\ \bibnamefont
  {Hollenberg}},\ }\bibfield  {title} {\bibinfo {title} {Analytic solutions to
  the central-spin problem for nitrogen-vacancy centers in diamond},\ }\href
  {https://doi.org/10.1103/PhysRevB.90.075201} {\bibfield  {journal} {\bibinfo
  {journal} {Physical Review B}\ }\textbf {\bibinfo {volume} {90}},\ \bibinfo
  {pages} {075201} (\bibinfo {year} {2014})}\BibitemShut {NoStop}%
\bibitem [{\citenamefont {Wood}\ \emph {et~al.}(2016)\citenamefont {Wood},
  \citenamefont {Broadway}, \citenamefont {Hall}, \citenamefont {Stacey},
  \citenamefont {Simpson}, \citenamefont {Tetienne},\ and\ \citenamefont
  {Hollenberg}}]{Wood2016}%
  \BibitemOpen
  \bibfield  {author} {\bibinfo {author} {\bibfnamefont {J.~D.~A.}\
  \bibnamefont {Wood}}, \bibinfo {author} {\bibfnamefont {D.~A.}\ \bibnamefont
  {Broadway}}, \bibinfo {author} {\bibfnamefont {L.~T.}\ \bibnamefont {Hall}},
  \bibinfo {author} {\bibfnamefont {A.}~\bibnamefont {Stacey}}, \bibinfo
  {author} {\bibfnamefont {D.~A.}\ \bibnamefont {Simpson}}, \bibinfo {author}
  {\bibfnamefont {J.-P.}\ \bibnamefont {Tetienne}},\ and\ \bibinfo {author}
  {\bibfnamefont {L.~C.~L.}\ \bibnamefont {Hollenberg}},\ }\bibfield  {title}
  {\bibinfo {title} {Wide-band nanoscale magnetic resonance spectroscopy using
  quantum relaxation of a single spin in diamond},\ }\href
  {https://doi.org/10.1103/PhysRevB.94.155402} {\bibfield  {journal} {\bibinfo
  {journal} {Physical Review B}\ }\textbf {\bibinfo {volume} {94}},\ \bibinfo
  {pages} {155402} (\bibinfo {year} {2016})}\BibitemShut {NoStop}%
\bibitem [{\citenamefont {Scholten}\ \emph {et~al.}(2021)\citenamefont
  {Scholten}, \citenamefont {Healey}, \citenamefont {Robertson}, \citenamefont
  {Abrahams}, \citenamefont {Broadway},\ and\ \citenamefont
  {Tetienne}}]{Scholten2021}%
  \BibitemOpen
  \bibfield  {author} {\bibinfo {author} {\bibfnamefont {S.~C.}\ \bibnamefont
  {Scholten}}, \bibinfo {author} {\bibfnamefont {A.~J.}\ \bibnamefont
  {Healey}}, \bibinfo {author} {\bibfnamefont {I.~O.}\ \bibnamefont
  {Robertson}}, \bibinfo {author} {\bibfnamefont {G.~J.}\ \bibnamefont
  {Abrahams}}, \bibinfo {author} {\bibfnamefont {D.~A.}\ \bibnamefont
  {Broadway}},\ and\ \bibinfo {author} {\bibfnamefont {J.-P.}\ \bibnamefont
  {Tetienne}},\ }\bibfield  {title} {\bibinfo {title} {Widefield quantum
  microscopy with nitrogen-vacancy centers in diamond: Strengths, limitations,
  and prospects},\ }\href {https://doi.org/10.1063/5.0066733} {\bibfield
  {journal} {\bibinfo  {journal} {Journal of Applied Physics}\ }\textbf
  {\bibinfo {volume} {130}},\ \bibinfo {pages} {150902} (\bibinfo {year}
  {2021})}\BibitemShut {NoStop}%
\bibitem [{\citenamefont {Broadway}\ \emph
  {et~al.}(2020{\natexlab{a}})\citenamefont {Broadway}, \citenamefont
  {Scholten}, \citenamefont {Tan}, \citenamefont {Dontschuk}, \citenamefont
  {Lillie}, \citenamefont {Johnson}, \citenamefont {Zheng}, \citenamefont
  {Wang}, \citenamefont {Oganov}, \citenamefont {Tian}, \citenamefont {Li},
  \citenamefont {Lei}, \citenamefont {Wang}, \citenamefont {Hollenberg},\ and\
  \citenamefont {Tetienne}}]{BroadwayAM2020}%
  \BibitemOpen
  \bibfield  {author} {\bibinfo {author} {\bibfnamefont {D.~A.}\ \bibnamefont
  {Broadway}}, \bibinfo {author} {\bibfnamefont {S.~C.}\ \bibnamefont
  {Scholten}}, \bibinfo {author} {\bibfnamefont {C.}~\bibnamefont {Tan}},
  \bibinfo {author} {\bibfnamefont {N.}~\bibnamefont {Dontschuk}}, \bibinfo
  {author} {\bibfnamefont {S.~E.}\ \bibnamefont {Lillie}}, \bibinfo {author}
  {\bibfnamefont {B.~C.}\ \bibnamefont {Johnson}}, \bibinfo {author}
  {\bibfnamefont {G.}~\bibnamefont {Zheng}}, \bibinfo {author} {\bibfnamefont
  {Z.}~\bibnamefont {Wang}}, \bibinfo {author} {\bibfnamefont {A.~R.}\
  \bibnamefont {Oganov}}, \bibinfo {author} {\bibfnamefont {S.}~\bibnamefont
  {Tian}}, \bibinfo {author} {\bibfnamefont {C.}~\bibnamefont {Li}}, \bibinfo
  {author} {\bibfnamefont {H.}~\bibnamefont {Lei}}, \bibinfo {author}
  {\bibfnamefont {L.}~\bibnamefont {Wang}}, \bibinfo {author} {\bibfnamefont
  {L.~C.~L.}\ \bibnamefont {Hollenberg}},\ and\ \bibinfo {author}
  {\bibfnamefont {J.-P.}\ \bibnamefont {Tetienne}},\ }\bibfield  {title}
  {\bibinfo {title} {Imaging domain reversal in an ultrathin van der waals
  ferromagnet},\ }\href {https://doi.org/10.1002/adma.202003314} {\bibfield
  {journal} {\bibinfo  {journal} {Advanced Materials}\ }\textbf {\bibinfo
  {volume} {32}},\ \bibinfo {pages} {2003314} (\bibinfo {year}
  {2020}{\natexlab{a}})}\BibitemShut {NoStop}%
\bibitem [{\citenamefont {Broadway}\ \emph
  {et~al.}(2020{\natexlab{b}})\citenamefont {Broadway}, \citenamefont {Lillie},
  \citenamefont {Scholten}, \citenamefont {Rohner}, \citenamefont {Dontschuk},
  \citenamefont {Maletinsky}, \citenamefont {Tetienne},\ and\ \citenamefont
  {Hollenberg}}]{Broadway2020PA}%
  \BibitemOpen
  \bibfield  {author} {\bibinfo {author} {\bibfnamefont {D.}~\bibnamefont
  {Broadway}}, \bibinfo {author} {\bibfnamefont {S.}~\bibnamefont {Lillie}},
  \bibinfo {author} {\bibfnamefont {S.}~\bibnamefont {Scholten}}, \bibinfo
  {author} {\bibfnamefont {D.}~\bibnamefont {Rohner}}, \bibinfo {author}
  {\bibfnamefont {N.}~\bibnamefont {Dontschuk}}, \bibinfo {author}
  {\bibfnamefont {P.}~\bibnamefont {Maletinsky}}, \bibinfo {author}
  {\bibfnamefont {J.-P.}\ \bibnamefont {Tetienne}},\ and\ \bibinfo {author}
  {\bibfnamefont {L.}~\bibnamefont {Hollenberg}},\ }\bibfield  {title}
  {\bibinfo {title} {Improved current density and magnetization reconstruction
  through vector magnetic field measurements},\ }\href
  {https://doi.org/10.1103/PhysRevApplied.14.024076} {\bibfield  {journal}
  {\bibinfo  {journal} {Phys. Rev. Appl.}\ }\textbf {\bibinfo {volume} {14}},\
  \bibinfo {pages} {024076} (\bibinfo {year} {2020}{\natexlab{b}})}\BibitemShut
  {NoStop}%
\bibitem [{\citenamefont {Rondin}\ \emph {et~al.}(2014)\citenamefont {Rondin},
  \citenamefont {Tetienne}, \citenamefont {Hingant}, \citenamefont {Roch},
  \citenamefont {Maletinsky},\ and\ \citenamefont {Jacques}}]{Rondin2014}%
  \BibitemOpen
  \bibfield  {author} {\bibinfo {author} {\bibfnamefont {L.}~\bibnamefont
  {Rondin}}, \bibinfo {author} {\bibfnamefont {J.-P.}\ \bibnamefont
  {Tetienne}}, \bibinfo {author} {\bibfnamefont {T.}~\bibnamefont {Hingant}},
  \bibinfo {author} {\bibfnamefont {J.-F.}\ \bibnamefont {Roch}}, \bibinfo
  {author} {\bibfnamefont {P.}~\bibnamefont {Maletinsky}},\ and\ \bibinfo
  {author} {\bibfnamefont {V.}~\bibnamefont {Jacques}},\ }\bibfield  {title}
  {\bibinfo {title} {Magnetometry with nitrogen-vacancy defects in diamond},\
  }\href {https://doi.org/10.1088/0034-4885/77/5/056503} {\bibfield  {journal}
  {\bibinfo  {journal} {Reports on Progress in Physics}\ }\textbf {\bibinfo
  {volume} {77}},\ \bibinfo {pages} {056503} (\bibinfo {year}
  {2014})}\BibitemShut {NoStop}%
\bibitem [{\citenamefont {Barry}\ \emph {et~al.}(2020)\citenamefont {Barry},
  \citenamefont {Schloss}, \citenamefont {Bauch}, \citenamefont {Turner},
  \citenamefont {Hart}, \citenamefont {Pham},\ and\ \citenamefont
  {Walsworth}}]{Barry2020}%
  \BibitemOpen
  \bibfield  {author} {\bibinfo {author} {\bibfnamefont {J.~F.}\ \bibnamefont
  {Barry}}, \bibinfo {author} {\bibfnamefont {J.~M.}\ \bibnamefont {Schloss}},
  \bibinfo {author} {\bibfnamefont {E.}~\bibnamefont {Bauch}}, \bibinfo
  {author} {\bibfnamefont {M.~J.}\ \bibnamefont {Turner}}, \bibinfo {author}
  {\bibfnamefont {C.~A.}\ \bibnamefont {Hart}}, \bibinfo {author}
  {\bibfnamefont {L.~M.}\ \bibnamefont {Pham}},\ and\ \bibinfo {author}
  {\bibfnamefont {R.~L.}\ \bibnamefont {Walsworth}},\ }\bibfield  {title}
  {\bibinfo {title} {Sensitivity optimization for nv-diamond magnetometry},\
  }\href {https://doi.org/10.1103/RevModPhys.92.015004} {\bibfield  {journal}
  {\bibinfo  {journal} {Rev. Mod. Phys.}\ }\textbf {\bibinfo {volume} {92}},\
  \bibinfo {pages} {015004} (\bibinfo {year} {2020})}\BibitemShut {NoStop}%
\bibitem [{\citenamefont {Healey}\ \emph {et~al.}(2020)\citenamefont {Healey},
  \citenamefont {Stacey}, \citenamefont {Johnson}, \citenamefont {Broadway},
  \citenamefont {Teraji}, \citenamefont {Simpson}, \citenamefont {Tetienne},\
  and\ \citenamefont {Hollenberg}}]{Healey2020}%
  \BibitemOpen
  \bibfield  {author} {\bibinfo {author} {\bibfnamefont {A.~J.}\ \bibnamefont
  {Healey}}, \bibinfo {author} {\bibfnamefont {A.}~\bibnamefont {Stacey}},
  \bibinfo {author} {\bibfnamefont {B.~C.}\ \bibnamefont {Johnson}}, \bibinfo
  {author} {\bibfnamefont {D.~A.}\ \bibnamefont {Broadway}}, \bibinfo {author}
  {\bibfnamefont {T.}~\bibnamefont {Teraji}}, \bibinfo {author} {\bibfnamefont
  {D.~A.}\ \bibnamefont {Simpson}}, \bibinfo {author} {\bibfnamefont {J.-P.}\
  \bibnamefont {Tetienne}},\ and\ \bibinfo {author} {\bibfnamefont {L.~C.~L.}\
  \bibnamefont {Hollenberg}},\ }\bibfield  {title} {\bibinfo {title}
  {Comparison of different methods of nitrogen-vacancy layer formation in
  diamond for wide-field quantum microscopy},\ }\href
  {https://doi.org/10.1103/PhysRevMaterials.4.104605} {\bibfield  {journal}
  {\bibinfo  {journal} {Phys. Rev. Mater.}\ }\textbf {\bibinfo {volume} {4}},\
  \bibinfo {pages} {104605} (\bibinfo {year} {2020})}\BibitemShut {NoStop}%
\bibitem [{\citenamefont {Healey}\ \emph {et~al.}(2022)\citenamefont {Healey},
  \citenamefont {Rahman}, \citenamefont {Scholten}, \citenamefont {Robertson},
  \citenamefont {Abrahams}, \citenamefont {Dontschuk}, \citenamefont {Liu},
  \citenamefont {Hollenberg}, \citenamefont {Lu},\ and\ \citenamefont
  {Tetienne}}]{HealeyACS2022}%
  \BibitemOpen
  \bibfield  {author} {\bibinfo {author} {\bibfnamefont {A.~J.}\ \bibnamefont
  {Healey}}, \bibinfo {author} {\bibfnamefont {S.}~\bibnamefont {Rahman}},
  \bibinfo {author} {\bibfnamefont {S.~C.}\ \bibnamefont {Scholten}}, \bibinfo
  {author} {\bibfnamefont {I.~O.}\ \bibnamefont {Robertson}}, \bibinfo {author}
  {\bibfnamefont {G.~J.}\ \bibnamefont {Abrahams}}, \bibinfo {author}
  {\bibfnamefont {N.}~\bibnamefont {Dontschuk}}, \bibinfo {author}
  {\bibfnamefont {B.}~\bibnamefont {Liu}}, \bibinfo {author} {\bibfnamefont
  {L.~C.~L.}\ \bibnamefont {Hollenberg}}, \bibinfo {author} {\bibfnamefont
  {Y.}~\bibnamefont {Lu}},\ and\ \bibinfo {author} {\bibfnamefont {J.-P.}\
  \bibnamefont {Tetienne}},\ }\bibfield  {title} {\bibinfo {title} {Varied
  magnetic phases in a van der waals easy-plane antiferromagnet revealed by
  nitrogen-vacancy center microscopy},\ }\href
  {https://doi.org/10.1021/acsnano.2c04132} {\bibfield  {journal} {\bibinfo
  {journal} {ACS Nano}\ }\textbf {\bibinfo {volume} {16}},\ \bibinfo {pages}
  {12580} (\bibinfo {year} {2022})}\BibitemShut {NoStop}%
\end{thebibliography}%

\clearpage
\onecolumngrid

\begin{center}

\textbf{\large Supplementary Information for the manuscript ``Multi-species optically addressable spin defects in a van der Waals material''}

\end{center}
\makeatletter
\renewcommand{\theequation}{S\arabic{equation}}
\renewcommand{\thefigure}{S\arabic{figure}}

\section{Experimental setup}\label{sec:setup}

The experiments reported in this work were carried out on a custom-built wide-field fluorescence microscope. The typical experimental setup and conditions are indicated below.

\begin{figure*}[h!]
\centering
\includegraphics[width=0.4\textwidth]{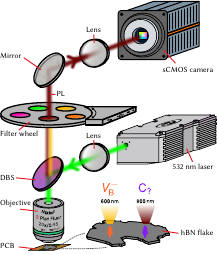}
\caption{\add{\textbf{Simplified optical setup.}
When measuring $\VB$ or $\C$ defects, only the PL emission filter is changed to maximise the signal-to-background ratio for each spin species.}}
\label{FigSI_setup}
\end{figure*}

Optical excitation from a continuous-wave (CW) $\lambda = 532$\,nm laser (Laser Quantum Opus 2 W) was gated using an acousto-optic modulator (Gooch \& Housego R35085-5) and focused using a widefield lens ($f=400$~mm) to the back aperture of the objective lens (Nikon S Plan Fluor ELWD 20x, NA = 0.45). 
The photoluminescence (PL) from the sample is separated from the excitation light with a dichroic beam splitter (DBS) and a 550\,nm longpass filter. 
The PL is either sent to a spectrometer (Ocean Insight Maya2000-Pro) or imaged with a scientific CMOS camera (Andor Zyla 5.5-W USB3) using a $f=300$\,mm tube lens. 
When imaging, additional shortpass and longpass optical filters were inserted to only collect a given wavelength range. \add{A simplified schematic of the setup is shown in Fig.~\ref{FigSI_setup} to emphasize that only the PL emission filters are changed when measuring the $\VB$ or $\C$ defects.}  
The laser spot diameter ($1/e^2$) at the sample was about 50\,$\upmu$m and the total CW laser power up to 500\,mW, which gives a maximum intensity of about 0.5\,mW$/\upmu$m$^2$ in the centre of the spot. 
For Fig.~4, the laser power at the sample was reduced to 25\,mW to minimise laser-induced heating. 

Radiofrequency (RF) driving was provided by a signal generator (Windfreak SynthNV PRO) gated using an IQ modulator (Texas Instruments TRF37T05EVM) and amplified (Mini-Circuits HPA-50W-63+). 
A pulse pattern generator (SpinCore PulseBlasterESR-PRO 500\,MHz) was used to gate the excitation laser and RF, as well as for triggering the camera. 
The output of the amplifier was connected to a printed circuit board (PCB) equipped with a coplanar waveguide and terminated by a 50\,$\Omega$ termination. 
The hBN sample was placed above the coplanar waveguide, either in direct contact (Fig.~1-3) or on a quartz coverslip placed on the PCB (Fig.~4). 

The external magnetic field was applied using a permanent magnet, and the measurements were performed at room temperature in ambient atmosphere. 
The only exception is Fig.~1(e), for which the sample was placed in a closed-cycle cryostat with a base temperature of $T\approx5$\,K~\cite{Lillie2020} which allowed us to apply a calibrated magnetic field using the enclosed superconducting vector magnet.

\section{hBN powder samples} \label{sec:powder_charac}

In this section, we detail the preparation of the hBN powder sample used in Fig.~1 of the main text, and present additional characterisation of this powder as well as of another hBN powder sample used later in the SI to demonstrate dual-spin-species imaging (Sec.~\ref{sec:powder_imaging}).

\subsection{Sample preparation} \label{sec:powder_prep}

The hBN powders used in this work were sourced from Graphene Supermarket (BN Ultrafine Powder), with a specified purity of 99.0\%. 
Two batches of powder nominally identical but purchased at different times were used: `Powder 1' was purchased in 2022 whereas `Powder 2' was purchased in 2017. 
The as-received powders were electron irradiated with a beam energy of 2\,MeV and a variable dose between $1\times10^{18}$\,cm$^{-2}$ and $5\times10^{18}$~cm$^{-2}$. 
No annealing or further processing was performed.
The experiments reported in Fig.~1 used Powder~1 with a dose of $2\times10^{18}$\,cm$^{-2}$, whereas the experiments reported in Fig.~\ref{FigSI_dual_FeGaTe} used Powder~2 with a dose of $1\times10^{18}$\,cm$^{-2}$. 

\begin{figure*}[htb!]
\centering
\includegraphics[width=0.8\textwidth]{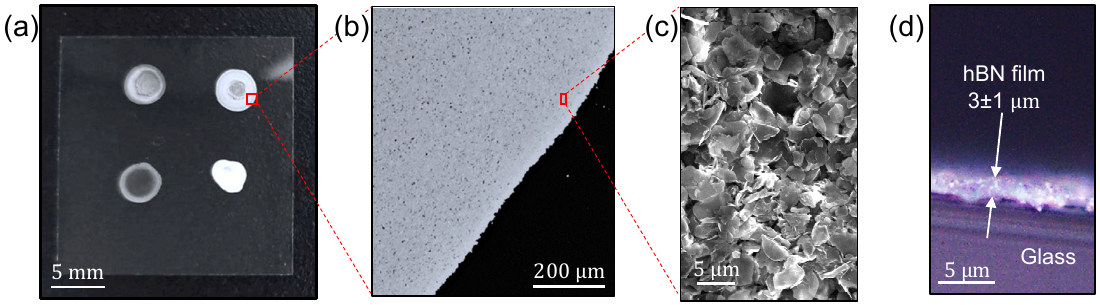}
\caption{\textbf{hBN powder film preparation.}
(a)~Photograph of hBN powder films formed on a glass coverslip. 
(b)~Magnified view of (a). 
(c)~Scanning electron microscopy image of a film. 
(d)~Optical micrograph of a cross-section of a film, from which the film thickness can be estimated.}
\label{FigSI_powder_films}
\end{figure*}

To form thin films of the hBN powders, the powder was suspended in isopropyl alcohol (IPA) at a concentration of 20\,mg/mL and sonicated for 30\,min using a horn-sonicator.
The sediment from the suspension was drawn using a pipette then drop-cast on a glass coverslip or directly onto the PCB, generally forming a relatively continuous film as shown in Fig.~\ref{FigSI_powder_films}(a,b). 
Although individual flakes have a thickness of $\lesssim10$\,nm and a lateral size of $\sim100$\,nm~\cite{Robertson2023}, a scanning electron microscopy image [Fig.~\ref{FigSI_powder_films}(c)] reveals aggregates with a size of $\sim1\,\upmu$m. 
To estimate the thickness of the films, we cleaved the coverslip across the film and inspected the cross-section using an optical microscope [Fig.~\ref{FigSI_powder_films}(d)], indicating a typical film thickness of a few micrometers. 

\subsection{Optical characterisation} 

\begin{figure*}[t!]
\centering
\includegraphics[width=0.7\textwidth]{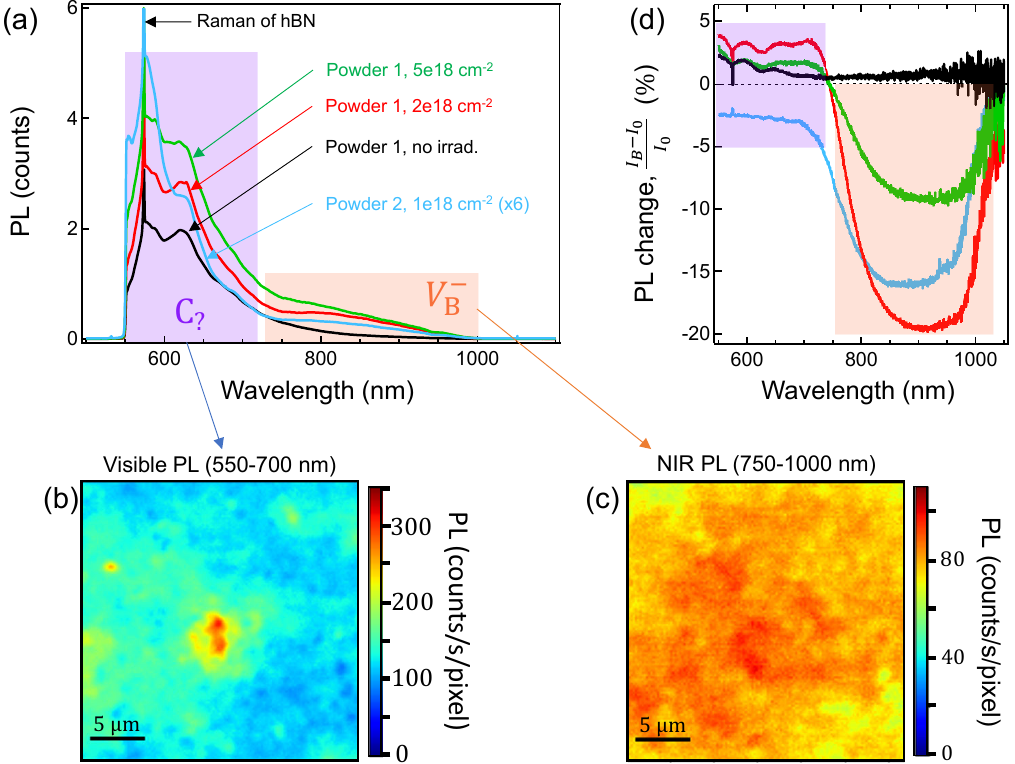}
\caption{\textbf{hBN powder optical characterisation.}
(a)~PL spectra under $\lambda=532$\,nm laser excitation of different powder samples. 
The values indicated in the legend correspond to the electron irradiation dose. 
A 550\,nm longpass filter is used to block the excitation light.
(b,c)~Widefield PL images of a film of Powder 1 (irradiation dose $5\times10^{18}$~cm$^{-2}$) under laser illumination ($\lambda=532$\,nm), filtering for (b) the $\C$ defects (PL emission band $\lambda=550-700$\,nm), and (c) the $\VB$ defects ($\lambda=750-1000$\,nm). 
The two images correspond to the same region.
(d)~Relative PL change $(I_B-I_0)/I_0$ due to the application of a magnetic field $B_0\approx120$\,mT as a function of emission wavelength, for the same samples as in (a), with the same colour coding. $I_B$ ($I_0$) is the PL signal with (without) the magnetic field.}
\label{FigSI_powder_optical}
\end{figure*}

The PL spectra of the different powders under green laser excitation ($\lambda=532$\,nm) are plotted in Fig.~\ref{FigSI_powder_optical}(a). 
Considering first Powder 1 samples that received different doses of electron irradiation, we observe the appearance of a broad near-infrared (NIR) emission peak centered around 820\,nm, with an amplitude correlating with the irradiation dose, indicating the successful creation of $\VB$ defects. 
In addition to the characteristic $\VB$ emission, all samples emit PL in the visible region ($\lambda=550-700$\,nm) including in the non-irradiated (as-received) powder, with a tail extending in the NIR (up to 900 nm). 
Similar PL emission was previously observed in various hBN samples~\cite{Mendelson2021,Chen2021ACSINT,Liu2022ACS,Yang2023} and was attributed to carbon-related defects, often referred to in the literature as the 2\,eV emitters. 
In the main text, we call the visible emitters present in our samples $\C$ defects to remind that they are likely carbon-related but that their exact structure remains unknown. 

Interestingly, it can be seen in Fig.~\ref{FigSI_powder_optical}(a) that electron irradiation increases the amplitude of the visible PL signal, by approximately two fold for the highest dose ($5\times10^{18}$~cm$^{-2}$) compared to the non-irradiated case. 
Compared to the Powder 1 samples, Powder 2 (irradiated to $1\times10^{18}$\,cm$^{-2}$) emits significantly less visible PL (there is a multiplying factor of $6$ in Fig.~\ref{FigSI_powder_optical}(a) for Powder 2) and has a slightly different spectral distribution, indicating that the concentration and specific optical properties of the $\C$ defects is variable from batch to batch. 
To assess the spatial homogeneity of the films, we recorded widefield PL images of an irradiated Powder 1 sample [Fig.~\ref{FigSI_powder_optical}(b,c)]. 
While the $\VB$ emission is relatively uniform [Fig.~\ref{FigSI_powder_optical}(c)], the visible emission appears more patchy [Fig.~\ref{FigSI_powder_optical}(b)], indicating that the density of $\C$ defects varies within a given sample.
Nevertheless, the visible emission is continuous for the micrometer-thick films studied here, i.e.\ there is no dark region with no visible PL. 
This feature is important to allow for magnetic imaging using powder films as demonstrated in Fig.~\ref{FigSI_dual_FeGaTe}.  

As a first test to probe whether the defects responsible for the observed PL emission are spin-active, we applied an external magnetic field ($B\approx120$\,mT) and observed its effect on the PL spectrum [Fig.~\ref{FigSI_powder_optical}(d)]. 
In the NIR region (750-1000\,nm), the PL drops by 5-20\% in all samples except the non-irradiated powder which experiences less than 1\% change. 
This field-induced PL quenching is due to spin mixing of the $\VB$ defects~\cite{Baber2022,Mathur2022,Mu2022,Yu2022}, which in a powder have a randomly oriented quantization axis with respect to the field direction. 
In the visible region (550-700\,nm), all samples respond to the applied magnetic field to varying degrees. 
Surprisingly, the Powder 1 samples exhibit a PL increase (by 1-4\%) while the Powder 2 sample sees a decreased PL (by 2-3\%). 
The field-induced PL change extends over the entire 550-700\,nm range. 

\subsection{Spin characterisation} 

\begin{figure*}[t!]
\centering
\includegraphics[width=1\textwidth]{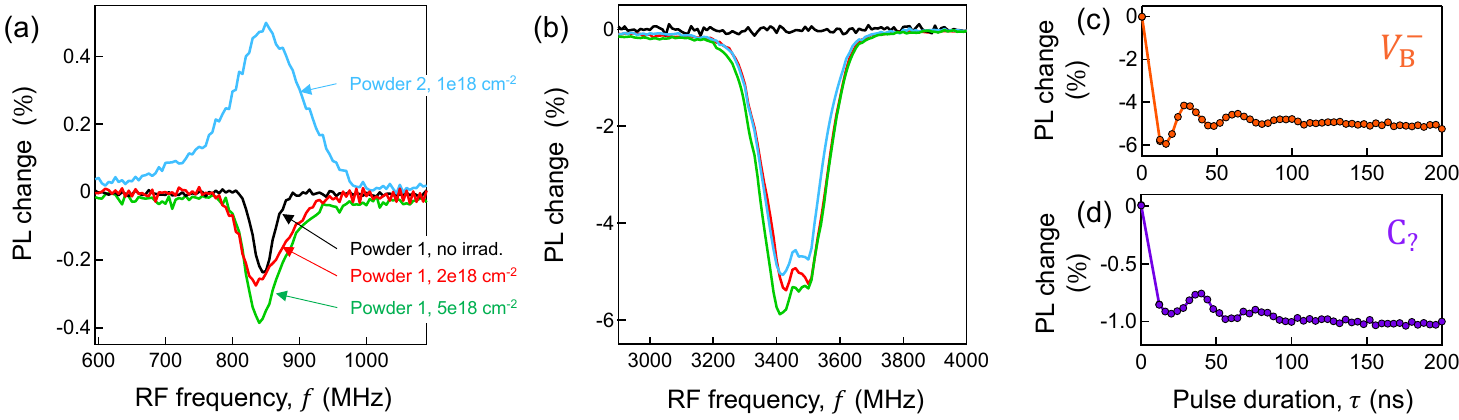}
\caption{\textbf{hBN powder spin characterisation.}
(a)~ODMR spectra of the $\C$ defects (PL emission band 550-700\,nm) at $B_0\approx30$\,mT for all powder samples.
(b)~ODMR spectra of the $\VB$ defects (PL emission band 750-1000\,nm) at $B_0=0$ for all samples, colour coded as in (a).
In (a,b), the ODMR spectra were acquired in CW mode, and the RF driving strength was kept roughly identical across samples corresponding to a Rabi frequency of about 20 MHz.
(c,d)~Rabi oscillations of (c) the $\VB$ defects and (d) the $\C$ defects in Powder 1 (irradiation dose $5\times10^{18}$~cm$^{-2}$).}
\label{FigSI_powder_ODMR}
\end{figure*}

\begin{figure*}[t!]
\centering
\includegraphics[width=1.0\textwidth]{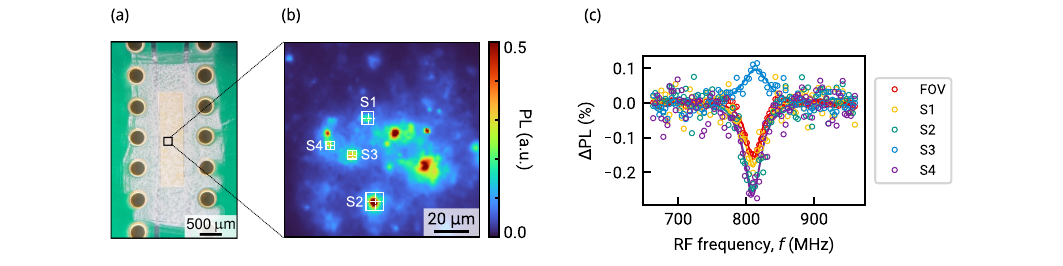}
\caption{\textbf{Spatial inhomogeneity of $\C$ spin contrast sign in hBN powders.}
(a)~Photograph of a sparse film of hBN powder. 
(b)~PL image filtering for the  emission band 550-700\,nm. 
(c)~ODMR spectra from selected regions indicated in (b).}
\label{FigSI_ODMR_sign}
\end{figure*}

ODMR spectra of the same set of powder samples are shown in Fig.~\ref{FigSI_powder_ODMR}(a,b) for the $\C$ and $\VB$ defects, respectively. 
For the $\C$ case where only the visible PL (550-700\,nm) is collected [Fig.~\ref{FigSI_powder_ODMR}(a)], the ODMR spectra show a single resonance at frequency $f_r=g_C\mu_BB_0/h$ in all cases ($g_C\approx2$), but for Powder 1 samples the contrast is negative whereas it is positive for the Powder 2 sample. 
The sign of the ODMR contrast anti-correlates with the sign of the PL change due to applying a magnetic field [Fig.~\ref{FigSI_powder_optical}(d)].
That is, where applying a static magnetic field increases (decreases) the PL, then driving the spin transition under this applied static field decreases (increases) the PL. 
This observation points to a mechanism whereby the spin dependence of the optical transitions is enabled or enhanced by the static field (consistent with the spin pair theory discussed later) rather than reduced due to spin mixing, as is the case for the $\VB$ defect or the nitrogen-vacancy centre in diamond~\cite{Tetienne2012}. 
We also observed opposite signs from different spatial locations within the same powder sample [Fig.~\ref{FigSI_ODMR_sign}], indicating that the variability is inherent to the material itself. The variability in sign of the ODMR contrast within and between samples is currently not understood. 

Figure~\ref{FigSI_powder_ODMR}(a) further reveals that the ODMR contrast of the $\C$ defects increases slightly with increased irradiation dose (Powder 1 samples), from about 0.2\% for the non-irradiated powder to 0.4\% for the highest irradiation dose in these measurements (with up to 1\% observed under stronger RF driving). 
Meanwhile, the linewidth (FWHM) of the resonance is also increased, from about 30 MHz to 60 MHz. 
Powder 2 has a positive contrast of about 0.5\% here (up to 1\% under stronger RF driving) and a significantly larger linewidth of about 200 MHz (in CW ODMR), under identical laser and RF driving conditions.  
Pulsed ODMR measurements generally led to a significant line narrowing down to about 30\,MHz, relatively consistent across all powder samples.

The $\VB$ ODMR spectra at $B_0=0$ [Fig.~\ref{FigSI_powder_ODMR}(b)] show very little difference between the various powder samples, with a contrast of $\approx5\%$, except for the non-irradiated sample which exhibits no contrast, confirming the absence of $\VB$ defects in this case. 

We were able to observe Rabi oscillations in all samples for both spin species where present. 
Example Rabi oscillations from Powder 1 are shown in Fig.~\ref{FigSI_powder_ODMR}(c,d) for $\VB$ and $\C$, respectively.

\section{hBN single-crystal samples} \label{sec:crystal_charac}

In this section, we detail the preparation and additional characterisation of the hBN single-crystal samples used in Fig. 2 and 3 of the main text.

\subsection{Sample preparation}

The starting whole hBN crystals were sourced from HQ Graphene, and had a thickness of $\sim100\,\upmu$m and a lateral size of $\sim1\,$mm. 
The as-received crystals were electron irradiated with a beam energy of 2\,MeV and a dose between $2\times10^{18}$\,cm$^{-2}$ and $1\times10^{19}$\,cm$^{-2}$.
No annealing or further processing was performed. 
In Fig.~3, a whole bulk crystal was placed on the PCB and measured. 
In Fig.~2, a thinner flake ($\sim1\,\mu$m) was exfoliated from a bulk crystal using scotch-tape and transferred to the PCB. 
Exfoliation of a thin flake allowed us to achieve a stronger and more uniform RF driving compared to using the whole crystal.

\subsection{Bulk crystal characterisation} \label{sec:bulk}

\begin{figure*}[b!]
\centering
\includegraphics[width=0.8\textwidth]{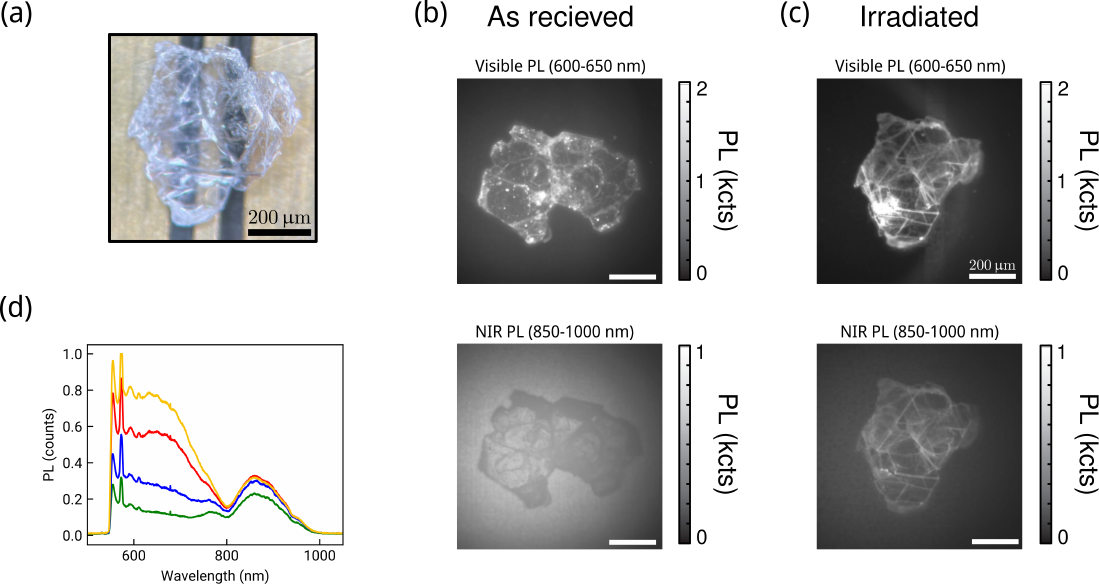}
\caption{
\textbf{Optical characterisation of whole hBN crystals.}
(a)~Optical photograph of an irradiated ($2\times10^{18}$\,cm$^{-2}$ at 2\,MeV) bulk crystal.
(b)~PL images of an as-received (non irradiated) crystal under green LED illumination, taken with a 600-650\,nm bandpass filter (top) and 850\,nm longpass filter (bottom). 
There is sparse visible emission, while the NIR emission is negligible.
(c)~PL images of the irradiated crystal shown in (a). Visible emission ($\C$) correlates with wrinkles [visible in (a)] in the crystal, whereas the NIR emission ($\VB$) is more uniform albeit reduced in magnitude.
The wrinkles are still visible in the NIR image due to the long tail of the $\C$ emission profile.
(d)~PL spectra taken at various locations for the irradiated crystal, under focused laser excitation ($\lambda=532\,$nm).
The NIR emission is relatively uniform, whereas the visible emission varies in magnitude by a factor 5 between the wrinkles (maximum emission) and the regions free of wrinkles (minimum emission).
}
\label{FigSI_bulk}
\end{figure*}

A photograph of a typical bulk crystal studied is shown in Fig.~\ref{FigSI_bulk}(a). Domain boundaries are visible, which are believed to be wrinkles that formed during the crystal growth process.

PL images of an as-received crystal [Fig.~\ref{FigSI_bulk}(b)] show no signature of $\VB$ emission but there is some PL in the visible range from isolated spots (possibly single emitters) as well as from extended features. 
ODMR can be detected from this PL, indicating the presence of a low density of $\C$ defects.

After irradiation, the visible PL is overall more intense and homogeneous [top image in Fig.~\ref{FigSI_bulk}(c)], suggesting that the electron irradiation plays a role in activating a high density of $\C$ defects. 
There is also enhanced emission from the wrinkles, which appear up to $\sim5$ brighter than the regions in between. 
This could be due to optical effects or indicate a higher density of active optical emitters in the wrinkles. 
Nevertheless, we were able to detect ODMR from the $\C$ defects in all regions of the crystal independent of brightness. 
In addition, the electron irradiation results in NIR PL from $\VB$ defects [bottom image in Fig.~\ref{FigSI_bulk}(c)]. 
There is a correlation with the wrinkles again, but this is mainly due to the NIR tail of the $\C$ emission, whereas the PL from the $\VB$ defects is relatively uniform throughout the crystal. 
This is confirmed by examining PL spectra from various localised regions in the crystal including the bright wrinkles as well as darker regions, a few examples are shown in Fig.~\ref{FigSI_bulk}(d).  

For the experiments presented in Fig.~3 of the main text, we used a crystal irradiated with the highest dose of $1\times10^{19}$\,cm$^{-2}$, in order to create a dense ensemble of $\VB$ defects. 
In Ref.~\cite{Murzakhanov2021}, Murzakhanov et al.\ characterised a bulk hBN crystal from the same supplier after electron irradiation with a dose of $6\times10^{18}$\,cm$^{-2}$ at 2\,MeV. 
Through careful analysis of EPR data, they determined a concentration of $\VB$ spins in their sample of $6\times10^{17}$\,cm$^{-3}$. 
Given our sample is nominally identical with only a slightly higher irradiation dose (by a factor of 1.67), we can assume that to a good approximation the $\VB$ density in our sample is increased by the same factor, which gives a density of $1\times10^{18}$\,cm$^{-3}$. 
This value was used for the analysis of the cross-relaxation experiment, see Sec.~\ref{sec:CR}. 
For the measurements presented in Fig.~3 of the main text, we used a region of the crystal close to a wrinkle in order to get bright visible emission, which is assumed to correlate with a high density of $\C$ defect. 
The latter density is a priori unknown.

\subsection{Exfoliated flake characterisation}

\begin{figure*}[b!]
\centering
\includegraphics[width=1.0\textwidth]{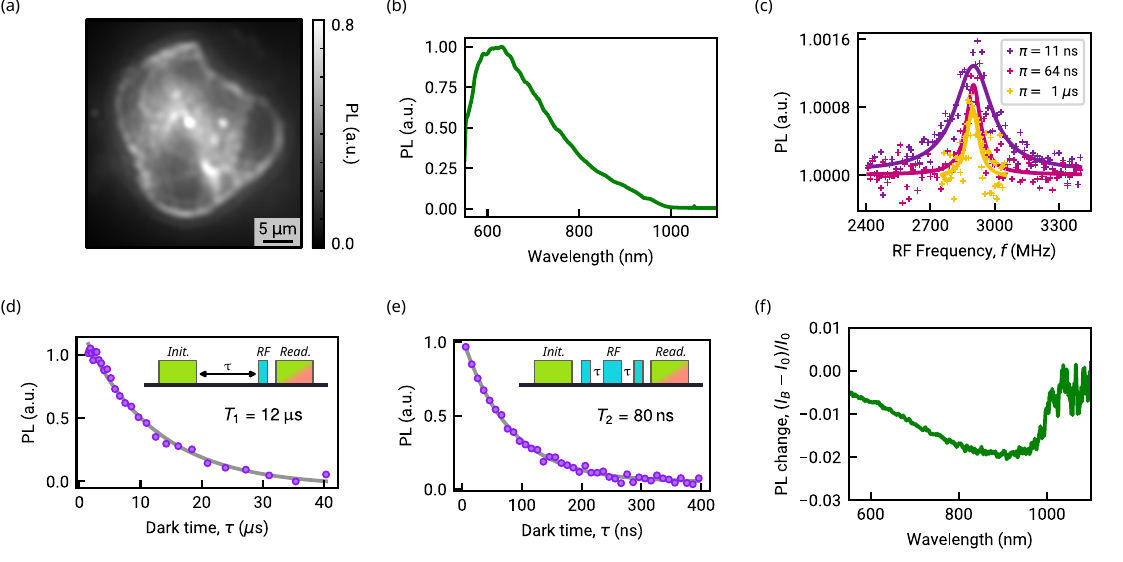}
\caption{\textbf{Spin characteriation of $\C$ defect in exfoliated flakes.}
(a)~Widefield PL image (emission filter $\lambda=550-700$\,nm) of a hBN flake under laser illumination ($\lambda=532$\,nm).
(b)~PL spectrum of the entire flake under $\lambda=532$\,nm laser excitation.
(c)~Pulsed ODMR spectra of the $\C$ defects at $B_0\approx100$\,mT for different RF $\pi$-pulse durations. 
(d)~Spin relaxation curve of the $\C$ defects. The pulse sequence is shown as an inset.
(e) Spin decoherence curve of the $\C$ defects. The pulse sequence (Hahn echo) is shown as an inset.
(f)~Relative PL change $(I_B-I_0)/I_0$ due to the application of a magnetic field $B_0\approx120$\,mT (applied perpendicular to the $c$-axis of the hBN flake) as a function of emission wavelength.}
\label{FigSI_flake}
\end{figure*}

For the experiments presented in Fig.~2 of the main text, we used a thin flake exfoliated from a bulk crystal similar to that shown in Fig.~\ref{FigSI_bulk}(a,c), irradiated at a dose of $2\times10^{18}$\,cm$^{-2}$. 
The PL image (550-700\,nm) of a typical flake is shown in Fig.~\ref{FigSI_flake}(a), exhibiting relatively uniform visible emission throughout the flake, with some brighter spots and edges but no pronounced wrinkle feature as was seen in Fig.~\ref{FigSI_bulk}(c). 
The PL spectrum averaged over the entire flake [Fig.~\ref{FigSI_flake}(b)] is dominated by the visible emission with only a subtle shoulder in the NIR due to the $\VB$ defects, indicating a relatively large density of $\C$ defects in this flake. 

Pulsed ODMR spectra of the $\C$ defects at $B_0=100$\,mT [Fig.~\ref{FigSI_flake}(c)] reveal a positive contrast of up to 0.15\% (with a short RF $\pi$-pulse) and a linewidth down to 70\,MHz (with a long RF $\pi$-pulse to avoid power broadening). 
The spin lifetime of the $\C$ defects is found to be $T_1\approx12\,\upmu$s [Fig.~\ref{FigSI_flake}(d)] and the spin coherence time in a Hahn echo sequence is $T_2\approx80$\,ns [Fig.~\ref{FigSI_flake}(e)]. 

We also measured the PL change due to applying a magnetic field $B_0=120$\,mT perpendicular to the $c$-axis [Fig.~\ref{FigSI_flake}(f)], showing a small PL reduction both in the NIR (as expected for $\VB$) and in the visible. 
Thus, for this flake we find the same anti-correlation between ODMR contrast and PL change due to a static field as we observed in the powder samples.

\section{hBN MOVPE-grown samples} \label{sec:MOVPE_charac}

In this section, we give further details on the hBN film used in Fig. 4 of the main text. 

The film was grown by metal-organic vapour-phase epitaxy (MOVPE) on a 2-inch sapphire wafer, using precursors triethylboron (TEB) and ammonia. 
The growth details can be found in Ref.~\cite{Chugh2018}. 
The film used in our work was grown with a TEB flow of 30\,$\mu$mol/min, and has a thickness of about 40\,nm determined by atomic force microscopy. 
In Ref.~\cite{Mendelson2021}, it was shown that this TEB flow leads to a dense and uniform ensemble of carbon-related optical emitters, including the ODMR-active $\C$ defects studied in this paper. 

\begin{figure*}[h!]
\centering
\includegraphics[width=0.7\textwidth]{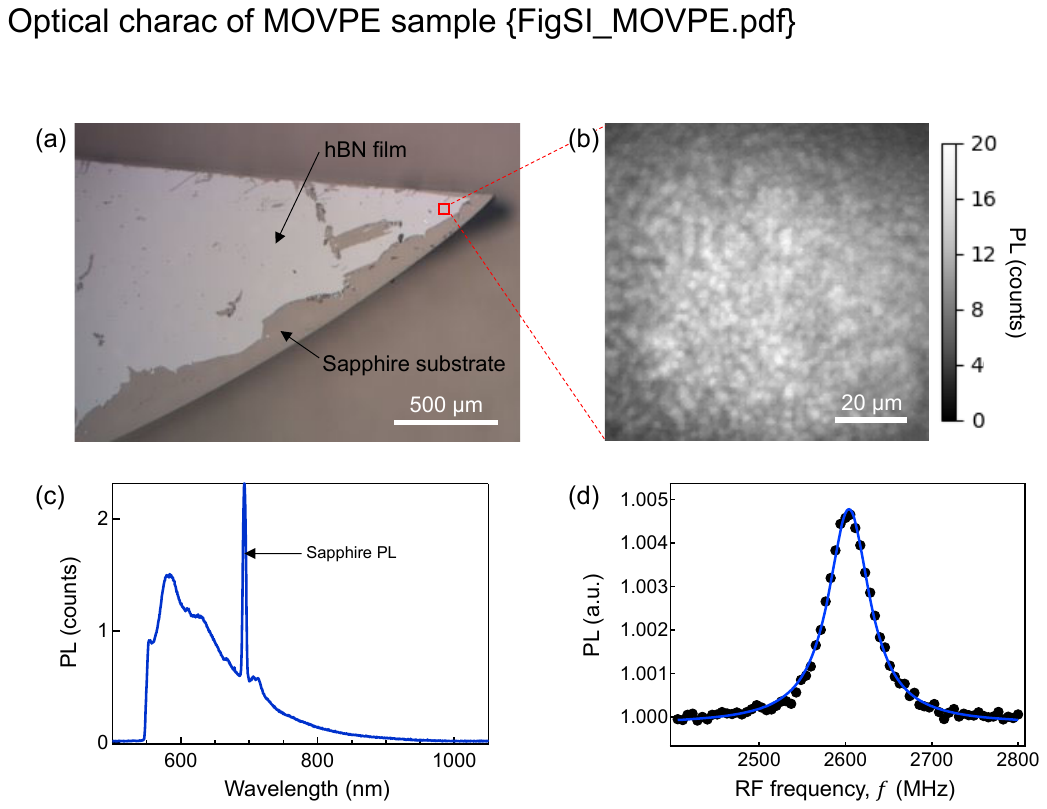}
\caption{\textbf{Characteriation of the MOVPE-grown hBN film.}
(a)~Optical micrograph of the hBN film on the sapphire substrate. 
(b)~Widefield PL image (emission filter $\lambda=550-700$\,nm) of the film under laser illumination ($\lambda=532$\,nm).
(c)~PL spectrum of the film under $\lambda=532$\,nm laser excitation. 
The characteristic peak due to the sapphire substrate (near 700\,nm) is indicated.
(d)~CW-ODMR spectrum of the $\C$ defects at $B_0\approx90$\,mT. 
The solid line is a Lorentzian fit.}
\label{FigSI_MOVPE}
\end{figure*}

An optical micrograph of the hBN film on the sapphire substrate, as obtained after growth and cleaving into a smaller sample, is shown in Fig.~\ref{FigSI_MOVPE}(a). 
The film covers the entire substrate except for small missing regions as a result of sample handling, and near the edge of the original wafer. 
The PL image [Fig.~\ref{FigSI_MOVPE}(b)] shows a relatively uniform signal, as reported previously~\cite{Mendelson2021}. 
The PL spectrum [Fig.~\ref{FigSI_MOVPE}(c)] resembles that of the hBN powder [see Fig.~\ref{FigSI_powder_optical}(a)], with a broad peak around 600\,nm and a tail extending into the NIR region.
Notice the sharp peak around 700 nm, which comes from the sapphire substrate. 
A CW-ODMR spectrum of the $\C$ defects is shown in Fig.~\ref{FigSI_MOVPE}(d), exhibiting a positive contrast of 0.5\%, and a FWHM of 60\,MHz.     

\section{Determining the spin multiplicity of the $\C$ defect} \label{sec:multiplicity}

In this section, we give the theoretical background for Fig. 2(e) of the main text from which the $S=\sfrac{1}{2}$ character of the $\C$ defect was determined, we analyse the possibility of experimental errors (which are ruled out), and finally we present evidence suggesting that the $\C$ defect involves a weakly coupled pair of $S=\sfrac{1}{2}$ particles. 

\subsection{Simple Rabi theory} \label{sec:Rabi}

We consider the generic static Hamiltonian applicable to a spin system of any multiplicity $S$,
\begin{equation}
    \mathcal{H}_0 = hD S_z^2 + g\mu_B \vect{S} \cdot \vect{B}_0\,,
\end{equation}
where $\vect{B}_0$ is the bias static magnetic field and $\vect{S}=(S_x,S_y,S_z)$ are the unitless spin operators. 
During a Rabi experiment, we apply an RF field (frequency $\omega$) in resonance with a spin transition (frequency $\omega_r$) given by $\mathcal{H}_0$ to coherently cycle population between two initial and final states $\ket{i}$ and $\ket{f}$ (eigenstates of $\mathcal{H}_0$). 
This RF driving adds a term in the Hamiltonian of the form
\begin{equation}
    \mathcal{H}_1(t) = g\mu_B \vect{S} \cdot \vect{B}_1 \cos(\omega t)\,,
\end{equation}
where $\vect{B}_1$ is the RF magnetic field vector.
At resonance ($\omega=\omega_r$) and under the rotating wave approximation ($g\mu_B B_1\ll \hbar\omega_r$), Rabi nutations between $\ket{i}$ and $\ket{f}$ occur at a Rabi flopping frequency $\Omega$ given by
\begin{equation} \label{eq:Rabi}
    h \Omega = g\mu_B\bra{f}  \vect{S} \cdot \vect{B}_1 \ket{i}\,.
\end{equation}
We assume that the bias field is aligned with the defect's intrinsic quantization axis ($z$) and that the RF field is orthogonal (pointing in the $x$ direction), 
\begin{align}
    \vect{B}_0 &= B_0 \hat{\vect{z}}\\
    \vect{B}_1 &= B_1 \hat{\vect{x}}~.
\end{align}
Considering the `highest' spin transition from $m_S=S-1$ to $m_S=S$, we obtain
\begin{align}
    h \Omega &= g\mu_B \bra{S} \vect{S}\cdot \vect{B}_1 \ket{S-1}\\
    &= g\mu_B B_1 \sqrt{\frac{S}{2}}\\
    &= h \Omega_{\sfrac{1}{2}} \sqrt{2S}\,,
\end{align}
where we introduced $\Omega_{\sfrac{1}{2}}$, the Rabi frequency of a $S=\sfrac{1}{2}$ spin. 
We see that when driving the $\ket{0}\rightarrow\ket{+1}$ transition of a $S=1$ spin the Rabi frequency is a factor $\sqrt{2}$ larger than the $S=\sfrac{1}{2}$ case, and when driving the $\ket{+1/2}\rightarrow\ket{+3/2}$ transition of a $S=3/2$ spin it is a factor $\sqrt{3}$ larger.

The ratio determined above is relevant when comparing the $\VB$ defect (a $S=1$ spin where we drive a single spin transition at a time, e.g.\ $\ket{0}\rightarrow\ket{+1}$) to the $S=\sfrac{1}{2}$ scenario of the $\C$ defect. This is the grey dashed line displayed in Fig.~2(e) of the main text. 
However, we are also interested in the a priori plausible scenario where the $\C$ defect has $S=1$ or $S=3/2$. 
In these cases, the formula derived above does not apply because we know from the ODMR spectrum that the zero-field splitting term in $\mathcal{H}_0$ is negligible compared to the natural linewidth and the RF broadening (i.e.\ $g\mu_BB_1\gg hD$), in which case we would be driving multiple spin transitions at once.

For instance, consider the $S=1$ scenario. 
If the spin is initialised in $\ket{0}$, and we then drive the $\ket{0}\rightarrow\ket{\pm1}$ transitions simultaneously, then it can be shown that the population in the $\ket{0}$ state oscillates at a Rabi frequency $\Omega=2\Omega_{\sfrac{1}{2}}$. 
For a $S=3/2$ system initialised in a mixture of $\ket{+1/2}$ and $\ket{-1/2}$ where we drive the $\ket{\pm1/2}\rightarrow\ket{\pm3/2}$ transitions simultaneously, the Rabi frequency is also $\Omega=2\Omega_{\sfrac{1}{2}}$. 
These two cases correspond to the blue solid line in Fig. 2(e) of the min text. 

Finally, we need to consider the case of multiple weakly coupled spins. 
For instance, consider a pair of $S=\sfrac{1}{2}$ particles with negligible exchange or dipolar coupling between them. 
In this case, the two spins are generally driven one at a time such that the Rabi frequency of the pair is $\Omega=\Omega_{\sfrac{1}{2}}$, just like an isolated $S=\sfrac{1}{2}$ particle. 
These scenarios correspond to the red solid line in Fig. 2(e) of the main text. We will return to the spin pair case in Sec.~\ref{sec:pair} and show that it can be distinguished from an isolated $S=\sfrac{1}{2}$ by further analysis of the Rabi measurements. 

\subsection{Rabi measurements}

\begin{figure*}[t!]
\centering
\includegraphics[width=1.0\textwidth]{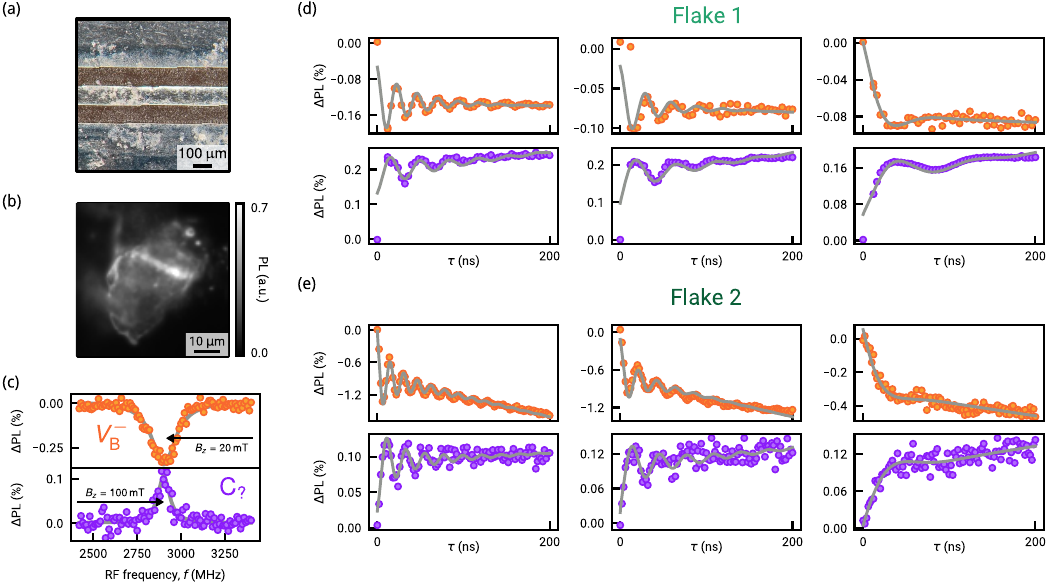}
\caption{\textbf{Rabi measurement of spin multiplicity.} 
(a)~Optical photograph of hBN flakes exfoliated onto a coplanar waveguide.
(b)~Widefield PL image (emission filter $\lambda = 550-700$\,nm)  of Flake 1. 
Flake 2 is shown in Fig.~\ref{FigSI_flake}(a).
(c)~ODMR measurement of Flake 2.
Ensembles of $\VB$ (top) and $\C$ (bottom) defects, at different fields to match the driving frequency and thus power.
(d,e)~Rabi measurements for Flakes 1 (d) and 2 (e), with fits (grey; see text). 
Each column is a pair of measurements on the $\VB$ (top) and $\C$ (bottom) defects at the same RF power, all driving at 2.9\,GHz.
The righter-most (lowest RF power) measurement series show highly damped flopping and so cannot be fit accurately; these points are not included in Fig.~2(e), but give a Rabi frequency ratio close to 1.
}
\label{FigSI_spin_multiplicity}
\end{figure*}

We undertake this Rabi measurement comparison with hBN flakes exfoliated directly onto a coplanar waveguide [Fig.~\ref{FigSI_spin_multiplicity}(a,b)] to maximise the driving RF power ($B_1$).
The out-of-plane magnetic field strength $B_0$ (provided by a permanent magnet) is tuned manually for the measurement on each defect type, to obtain a resonant frequency of 2.9\,GHz in both cases [Fig.~\ref{FigSI_spin_multiplicity}(c)].
For the $\C$ defect ($\ket{-1/2} \to \ket{+1/2}$) this corresponds to a field of 100\,mT, and 20\,mT for the $\VB$ defect ($\ket{0} \to \ket{-1}$), avoiding anti-crossings and the cross-relaxation resonance.
Rabi measurements are then undertaken at the respective fields at the same driving frequency [Fig.~\ref{FigSI_spin_multiplicity}(d,e)].
Each column pair in (d,e) is a Rabi measurement at a given RF driving power (decreasing from left to right), for flakes 1 (d) and 2 (e). 
The flopping of the $\C$ ensemble in each measurement can be immediately identified as slower than the $\VB$ ensemble, indicative of a lower spin multiplicity.
We quantify this effect by fitting the Rabi measurement to an exponentially decayed cosine function with a free decay time, amplitude, frequency, phase and vertical offset.
The free phase accounts for an average over detunings (see next section).
The fit Rabi frequencies are displayed in main text Fig.~2(e), with the fit error estimated from a Monte-Carlo bootstrap method~\cite{Bonamenta2017}, added to a systematic error in field misalignment (see below).
We will now address some alternative explanations for this difference, before ruling them out. 

\subsection{Effect of detuning}

\begin{figure*}[t!]
\centering
\includegraphics[width=0.8\textwidth]{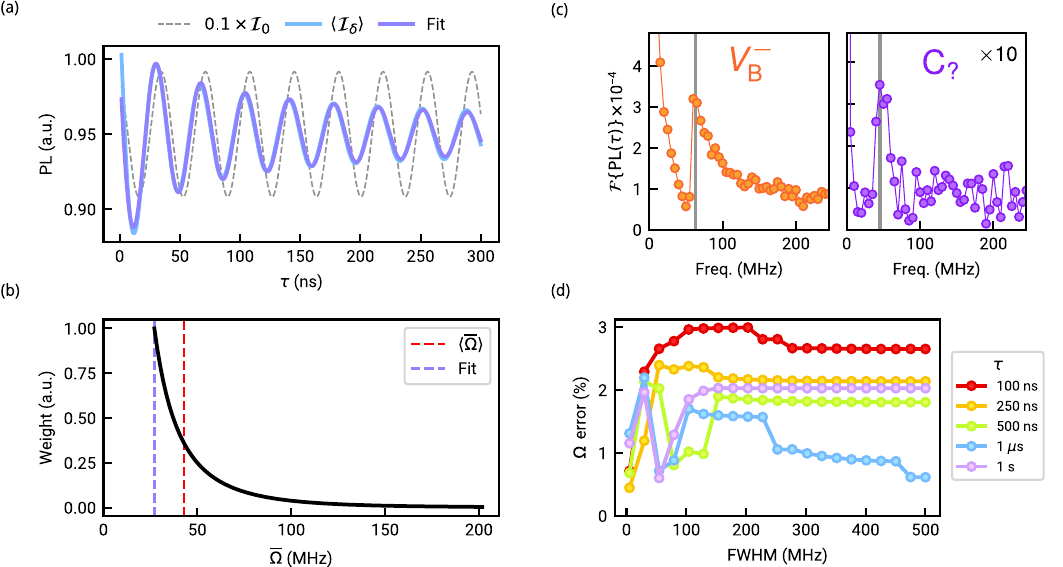}
\caption{\textbf{Effect of linewidth/damping on Rabi measurement.} 
(a)~Modelled Rabi curves, without considering detuning (grey dashed; rescaled by factor 0.1), averaging over detunings up to 200\,MHz linewidth (FWHM; blue line), and a fit (purple; see text).
(b)~Lorentzian weight (vertical axis) on generalised Rabi components (horizontal axis) for the simulation in (a).
The expected generalised Rabi frequency (red dashed) and fit frequency from (a) (purple dashed) are annotated.
(c)~Fourier transform of Rabi measurements in Fig.~\ref{FigSI_spin_multiplicity}(e) left column, with grey lines representing the Rabi frequencies estimated from the fits.
(d)~Systematic Rabi estimation error (vertical axis; see text) for various ODMR linewidths (horizontal axis) and added decay times ($\tau$).
The estimation error is below the fit and field angle errors ($\approx 10\%$).
}
\label{FigSI_Rabi_sim}
\end{figure*}

The simple theory described in Sec.~\ref{sec:Rabi} assumed an idealised spin system with well defined transitions and no dephasing effects. 
In reality, the transitions have a finite linewidth (related to the spin dephasing time $T_2^*$) and may also encompass multiple hyperfine lines. 
The $\VB$ defect's $\ket{0} \to \ket{-1}$ electron spin transition has a particularly large linewidth, with 7 hyperfine lines 47\,MHz apart~\cite{GottschollNM2020}. 
A finite linewidth implies that the RF driving is detuned from resonance most of the time, which may result in a faster apparent flopping rate.

To model this effect, we use the generalised equation for the Rabi curve in the presence of a detuning $\delta$ (in rad/s),
\begin{equation}
    {\mathcal I}_\delta(t) = {\mathcal I}_0(0) \frac{\Omega^2}{\overline{\Omega}^2} \cos(2 \pi \overline{\Omega} t)~,
\end{equation}
where the generalised Rabi frequency reads
\begin{equation}
    \overline{\Omega} = \sqrt{\Omega^2 + \delta^2}~.
\end{equation}
Note that any $\delta$ increases $\overline{\Omega}$ with respect to the bare $\Omega$, as well as reducing the oscillation amplitude.
We numerically model this generalised Rabi curve by averaging over a 200\,MHz Lorentzian linewidth to approximately match the $\VB$ lineshape.
The resultant generalised curve [Fig.~\ref{FigSI_Rabi_sim}(a) blue] has a reduced amplitude compared with the no-detuning curve [Fig.~\ref{FigSI_Rabi_sim}(a) grey dashed], as well as a phase offset and a decay envelope, due to the interference of different $\overline{\Omega}$.
Despite this averaging, a function consisting of a cosine function with a free phase, frequency, amplitude, and with an exponential decay envelope, fits the data well [Fig.~\ref{FigSI_Rabi_sim}(a) purple].
The similar flopping frequency (within 0.01\%) between the simple and detuning-averaged curves is at first surprising, as the expected generalised Rabi frequency based on the calculated $\overline{\Omega}$ distribution is a factor of 1.6 higher [Fig.~\ref{FigSI_Rabi_sim}(b)].
The bias of the fit to larger amplitude components, however, ensures the zero-detuning component is fit.
The Lorentzian-distributed generalised Rabi frequency components can also be seen in Fourier-transformed experimental curves [Fig.~\ref{FigSI_Rabi_sim}(c)], with a tail to the higher frequency components after an initial peak that lines up with the frequency estimated by the fit.
Note that the tail for the $\C$ defect is compressed, as it has a narrower ($0.1\times$) intrinsic linewidth than the $\VB$.

The general Rabi curve shows a decay envelope from the averaging over detunings, where the model includes no additional decay envelope due to other dephasing effects.
In Fig.~\ref{FigSI_Rabi_sim}(d) we explore the frequency estimation error as a function of characteristic decay times $\tau$ and ODMR lineshape widths (FWHM).
Here the estimation error is defined as the relative error between the fit frequency and the input bare Rabi frequency $\Omega$.
The estimation error is below 3\% for all simulations, below the $\approx10\%$ error estimated from the fit statistics and systematic field misalignments (see below).

\subsection{Effect of field misalignments}

In Sec.~\ref{sec:Rabi} we assumed the static field $\vect{B}_0$ to be exactly aligned with the intrinsic quantization axis of the $\VB$ defect (the $c$-axis of the hBN crystal, which was assumed parallel to $z$), and the RF field $\vect{B}_1$ to be exactly perpendicular to the $z$-axis, as is nominally the case experimentally. 
Here we model the effect of misalignments away from these nominal conditions.

\begin{figure*}[t!]
\centering
\includegraphics[width=0.95\textwidth]{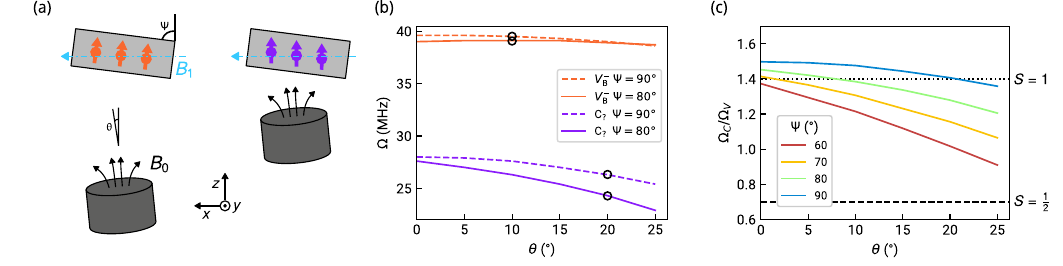}
\caption{\textbf{Modelling Rabi frequency versus field misalignment.} 
(a)~Cartoon depicting the measurement conditions for Fig.~\ref{FigSI_spin_multiplicity}; the RF field (blue) is in-plane, with the bias magnet moved closer for the $\C$ measurement.
The bias field misalignment ($\theta$) and flake tilt angle ($\Psi$) are defined as the polar angle from $z$ along $x$.
(b)~Modelled Rabi frequency as a function of $\theta$ for different values of $\Psi$.
The Rabi frequency drops as the driving RF field becomes less orthogonal to the spin axis, with the $\VB$ defect more protected by its large ZFS. 
The bias field strength $B_0$ is 20\,mT and 100\,mT for the $\VB$ (orange) and $\C$ (purple) defects, respectively, and the RF field strength is $B_1=1$\,mT.
Here the $\C$ defect is assumed to have $S=\sfrac{1}{2}$.
For the main text Fig.~2(e) the errorbars in $\Omega$ are calculated assuming worst-case $\Psi=10\degree$, $\theta=10\degree$ for the lower field $\VB$ measurement, and $\theta=20\degree$ for the higher field $\C$ measurement (circled). 
(c)~Rabi frequency ratio ($\Omega_C/\Omega_V$) as a function of $\vect{B}_0$ misalignment for the $\C$ defect measurement ($\theta$), assuming fixed $\vect{B}_0$ misalignment for the $\VB$ measurement ($10\degree$), for various flake tilt angles $\Psi$. 
For this simulation the $\C$ defect is assumed to have spin $S=1$ with $D=0$.
Each flake measured would need to be $30\degree$ misaligned from $z$ ($\Psi=60\degree$) and the bias field $\theta=30\degree$ misaligned to reproduce the ratio expected for a spin $S=\sfrac{1}{2}$, far beyond the systematic alignment error estimated ($10\degree$).
}
\label{FigSI_misaligned_field}
\end{figure*}

First, we consider the effect of a $\vect{B}_0$ misalignment. 
The two Rabi experiments were undertaken with the same RF conditions (power, geometry), however the bias magnet was moved closer to the sample in $z$ for the $\C$ measurement [Fig.~\ref{FigSI_misaligned_field}(a)].
Figure~\ref{FigSI_misaligned_field}(b) (dashed lines) plots the calculated Rabi frequency for the two defects as a function of the bias field's polar angle ($\theta$).
The Rabi frequency scales with the projection of the RF driving field $\vect{B}_1$ onto the defect's quantization axis according to Eq.~\ref{eq:Rabi}.
Therefore, the Rabi frequency drops as the driving RF field becomes less orthogonal to the spin axis.
As the zero-field splitting (ZFS) of the $\VB$ defect protects the spin sublevels from mixing, it maintains a higher orthogonality with the RF driving field and thus a higher Rabi frequency.

Next, we consider the effect of a $\vect{B}_1$ (RF) misalignment, modelled as a tilt of the flake's $c$-axis relative to $x$ (angle $\Psi$) in Fig.~\ref{FigSI_misaligned_field}(a).
We use a coplanar RF waveguide designed to provide an in-plane $\vect{B}_1$ ($\Psi=90\degree$) above the central stripline, but the flake may form a small angle with the surface.
From $\VB$ ODMR measurements on various flakes we estimate the crystal tilt/RF misalignment angle $\Psi$ to be less than 10\degree.
Figure~\ref{FigSI_misaligned_field}(b) displays the bare Rabi frequency under this worst-case tilted-flake assumption, as a function of polar bias angle (solid lines).
We assume an initial bias field misalignment for the $\VB$ measurement to be 10\degree, and then 20\degree\ misalignment for the $\C$ measurement as the magnet is brought closer (from 10\,cm to 1\,cm).
We estimate the relative error in Rabi frequency due to this field misalignment uncertainty for Fig.~2(e) in the main text from the percentage error between the solid and dashed lines in Fig.~\ref{FigSI_misaligned_field}(b) at these bias field polar angles (annotated circles). 
This corresponds to a 10\% maximum error for the $\C$ case, well below what would be required to be compatible with the $S=1$ or $S=3/2$ scenarios.  

To see how much misalignment would be required to make the $\C$ defect look like a spin $S=\sfrac{1}{2}$ if it was in fact a spin $S=1$ for example, in Fig.~\ref{FigSI_misaligned_field}(c) we model the experiment in Fig.~\ref{FigSI_spin_multiplicity} assuming the $\C$ defect has $S=1$ and $D=0$, and assume the worst-case field misalignment of 10\degree\ for the $\VB$ measurement, before varying the angle of misalignment of the $\C$ defect, for various RF misalignments $\Psi$.
With no misalignment, the Rabi frequency ratio $\Omega_C/\Omega_V$ is $\sqrt{2}\approx1.4$, but this ratio drops as misalignment is increased. 
To recover the $1/\sqrt{2}\approx0.7$ ratio measured experimentally and expected for a spin $S=\sfrac{1}{2}$, one would require to have a combination of unrealistically large misalignments, for instance 30\degree\ misalignment of the flake or RF field ($\Psi=60\degree$) and a $\theta=30\degree$ bias field misalignment for the $\C$ measurement.
These angles are beyond our estimates for the systematic error in alignments so rule out a misaligned bias field as the cause of the Rabi frequency difference.

\subsection{Evidence for weakly coupled spin pairs} \label{sec:pair}

As proposed in the main text, the spin-half behaviour observed in the Rabi oscillations could arise from a weakly coupled spin pair system, which may provide a more natural explanation for ODMR compared to an isolated $S=\sfrac{1}{2}$ system, by invoking different photon absorption/emission probabilities for the singlet and triplet configurations of the spin pair~\cite{Davies1988,Boehme2003,McCamey2008}. Here we present the results of an experimental test of the spin pair theory, which follows experiments performed on organic semiconductors~\cite{McCamey2010,Lee2010}. 

In these works, polaron pairs (electron and hole) form a weakly coupled spin pair exhibiting spin-dependent recombination, enabling electrically or optically detected magnetic resonance (EDMR or ODMR)~\cite{Boehme2003}. 
Weak coupling means that the exchange and dipolar interactions between the two spins are sufficiently small that only one resonance at $f_r\approx g\mu_BB_0/h$ is generally visible in the EDMR/ODMR spectrum. 
This implies that the hyperfine broadening of the magnetic resonances ($B_{\rm hyp}$) is larger than the difference between the two magnetic resonance frequencies of the system. 
However, it is possible to evidence the spin pair nature of the system by observing a beat in the Rabi oscillations under sufficiently strong driving ($B_1>B_{\rm hyp}$) ~\cite{McCamey2010,Lee2010}. 
In the following, we test whether we can observe this spin-beat in the case of the $\C$ defect in hBN.

The principle of the spin-beat experiment is as follows. 
For sufficiently weak driving ($B_1\ll B_{\rm hyp}$), the RF field is never simultaneously on resonance with both spin transitions, i.e.\ it drives only one spin at a time on average, and so the observed Rabi frequency is that of a single $S=\sfrac{1}{2}$ particle, $\Omega=\Omega_{\sfrac{1}{2}}$ (as defined in Sec.~\ref{sec:Rabi}). 
On the other hand, when $B_1\gg B_{\rm hyp}$ the two spins oscillate in synchrony. 
As a result, if the readout (the PL intensity in our case) is sensitive to the triplet vs singlet content of the state of the spin pair, or equivalently to whether the two spins are parallel or antiparallel, then  that readout will oscillate twice as fast, at $2\Omega_{\sfrac{1}{2}}$. 
When $B_1\sim B_{\rm hyp}$, the Rabi oscillations show both the fundamental frequency $\Omega_{\sfrac{1}{2}}$ and the second harmonic $2\Omega_{\sfrac{1}{2}}$, causing a spin beat~\cite{McCamey2010,Lee2010}. 
This spin-beat in the Rabi oscillations would not occur if the readout was only sensitive to the spins being up or down independent of each other, and so is a signature of the fact that the two spins are read out (and initialised) as a pair, i.e.\ that optical pumping creates correlations between the spins.

\begin{figure*}[t!]
\centering
\includegraphics[width=1\textwidth]{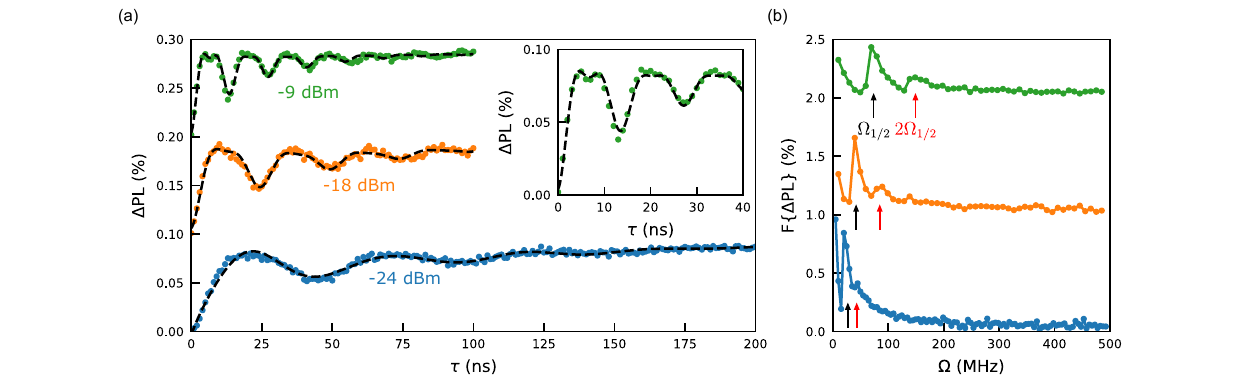}
\caption{\textbf{Observation of a spin-beat in the Rabi oscillations.} 
(a)~Rabi oscillations recorded at 3 different microwave powers. 
The curves are arbitrarily offset and all fit with a guide-to-the-eye, two frequency damped oscillation model. 
A zoomed in portion of the highest power measurement is shown in the inset. 
(b)~Fourier transforms of the Rabi measurements in (a). 
}
\label{FigSI_spin_pairs}
\end{figure*}

For the hBN flake measured here, the smallest ODMR linewidth is about 70\,MHz [corresponding to $B_{\rm hyp}$, see ODMR spectrum in Fig.~\ref{FigSI_flake}(c)] and so we need to have $\Omega_{\sfrac{1}{2}}\sim70$\,MHz in order to observe the spin-beat. 
Fig.~\ref{FigSI_spin_pairs}(a) shows Rabi oscillations recorded under increasing RF driving powers with the maximum power corresponding to the desired regime $\Omega_{\sfrac{1}{2}}\sim70$\,MHz. 
Across all powers the oscillations exhibit both an exponentially damped behaviour and an additional exponential decay, but as the power is increased there is also an onset of a second frequency component. 
We fit the oscillations with the phenomenological model, 
\begin{equation}
    f(t) = A e^{-k_1 t} \cos{(\Omega_1 t)} + B (e^{-k_1 t} + e^{-k_2 t}) \cos{(\Omega_2 t)} + C e^{-k_3 t} + D 
\end{equation}
where we have assumed the two oscillating terms have a shared damping term, with one of them carrying an additional dampening component. 
The additional exponential decay was chosen to have a decay rate independent of the damping components though in principle there may be some relation between them. 
Importantly the two frequencies, $\Omega_1$ and $\Omega_2$ are both free parameters. 
For the highest power (green data), we observe a beat frequency in the oscillation and two distinct frequencies are able to be extracted from the fit. 
We find $\Omega_1 = 146$\,MHz and $\Omega_2 = 73$\,MHz (i.e.\ satisfying $\Omega_1 = 2 \Omega_2$) with comparable amplitudes ($|A| \sim |B|$), consistent with the spin-beat. 
The spin-beat is also observed for the middle power (orange data) but with a smaller second harmonic component. 
At the lowest power (blue data), the spin-beat is no longer observable and the data is well fit with a single frequency.  

To better resolve the frequency components we take the fast Fourier transform of the Rabi curves [Fig.~\ref{FigSI_spin_pairs}(b)]. 
We also indicate the expected frequencies $\Omega_{\sfrac{1}{2}}$ and $2\Omega_{\sfrac{1}{2}}$, where $\Omega_{\sfrac{1}{2}}$ was calibrated using a $\VB$ Rabi measurement as in Fig.~\ref{FigSI_spin_multiplicity}. 
As anticipated, two peaks at $\Omega_{\sfrac{1}{2}}$ and $2\Omega_{\sfrac{1}{2}}$ are resolved, with the amplitude of the second harmonic increasing as $B_1$ becomes comparable to $B_{\rm hyp}$, consistent with the spin-beat.

Altogether, the observations presented in Fig.~\ref{FigSI_spin_pairs} support the suggestion that the $\C$ defect is a weakly coupled spin pair in its electronic ground state, with an optical initialisation/readout mechanism sensitive to the collective spin state of the pair (parallel vs antiparallel).

\section{Cross-relaxation experiment}

\subsection{Full dataset} \label{sec:dataCR}

\begin{figure*}[t!]
\centering
\includegraphics[width=1.0\textwidth]{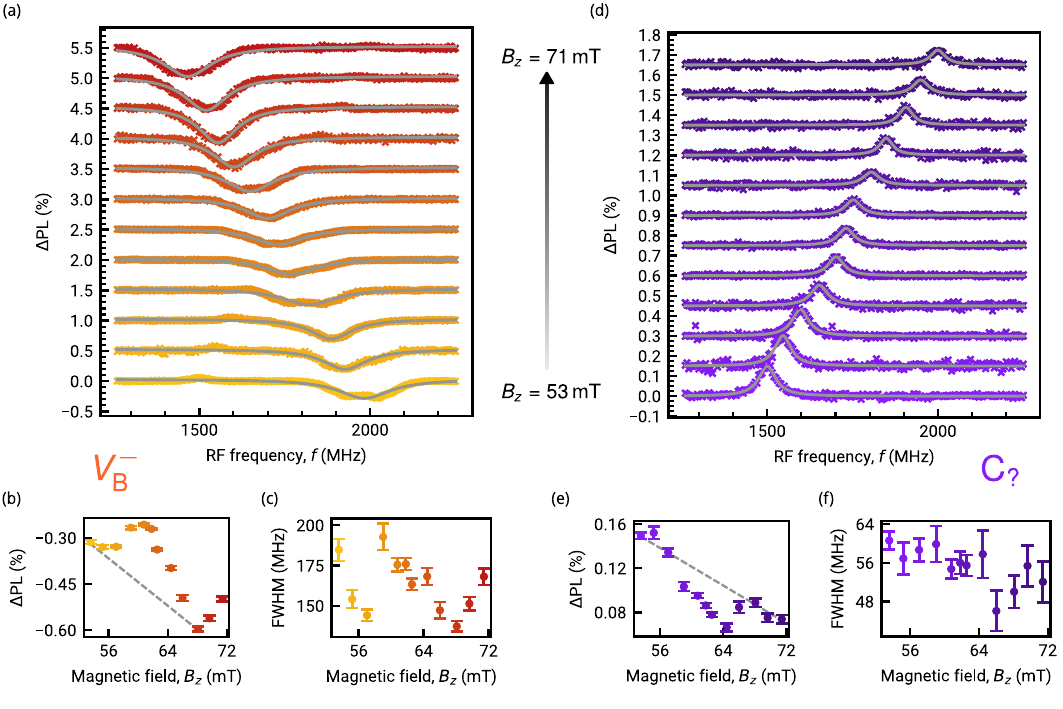}
\caption{
\textbf{Full cross-relaxation ODMR dataset.}
(a)~ODMR measurement of an ensemble of $\VB$ defects in a single crystal, for various magnetic field strengths, as shown in Fig.~3(c).
The spectrum (orange crosses) and Lorentzian fit (grey line) for each field value is vertically offset by 0.5\%.
A small positive contrast resonance is observed, due to the tail of the $\C$ defect's PL spectrum in the NIR where we measure the $\VB$.
(b)~Contrast inferred from the fits in (a), as a function of magnetic field (points).
The grey line depicts a linear slope due to RF power variations across the frequency range; the decrease in contrast magnitude in the last two points is ascribed to spin-mixing close to the ESLAC (76\,mT; \cite{Baber2022}).
(c)~Resonance linewidths (FWHM) extracted from the fits in (a), as a function of magnetic field (points).
(d-f)~Equivalent measurements to (a-c) taken on the $\C$ defect ensemble.
The spectra in (d) are vertically offset by 0.15\%.
}
\label{FigSI_CR_fits}
\end{figure*}

In the main text we displayed [Fig.~3(e)] a change in spin contrast at the CR condition.
Here we plot the full dataset and explain the data normalisation and background subtraction procedure.
An ODMR spectrum at each bias field strength for the pair of defects is shown in Fig.~\ref{FigSI_CR_fits}(a,d).
These spectra are fit with a simple Lorentzian lineshape, and the resonance contrast [Fig.~\ref{FigSI_CR_fits}(b,e)] and linewidth [Fig.~\ref{FigSI_CR_fits}(c,f)] plot versus field.
The contrast for both defects show a decrease in contrast magnitude with RF frequency, as a background to the CR resonance lineshape.
We identify this background trend as being caused by an RF power spectrum variation, which we approximate with a linear slope [grey lines in Fig.~\ref{FigSI_CR_fits}(b,e)].
For the $\C$ defects we estimate the linear slope with the first and the last point.
In the $\VB$ case we do not include the two highest magnetic field points, identifying their decrease in contrast magnitude with the excited-state level anticrossing (ESLAC) at the nearby field of 76\,mT~\cite{Baber2022}.
To compare to a theoretical CR lineshape (see next section) we calculated the spin contrast as the magnitude of the ODMR resonance contrast, normalised to unity off-resonance after the background subtraction.

\subsection{Modelling}\label{sec:CR}

In this section we model the expected spin contrast drop due to cross-relaxation between the two defect species measured in Fig.~3 of the main text.
We will initially consider the measurement of the $\C$ defect in a $\VB$ bath, before discussing the $\VB$ measurement.
Broadly, we will determine the additional relaxation rate experienced by the $\C$ spins as a function of detuning from the CR resonance condition, given the defect species' dephasing rates, as an ensemble average over the probe-target ($\C$-$\VB$) separation distance distribution.
Given this cross-relaxation rate, we will model the spin contrast of the pulsed ODMR measurement, accounting for the effect of spin relaxation during the laser pulse as well as during the dark time $\tau$.

\begin{figure*}[b!]
\centering
\includegraphics[width=1.0\textwidth]{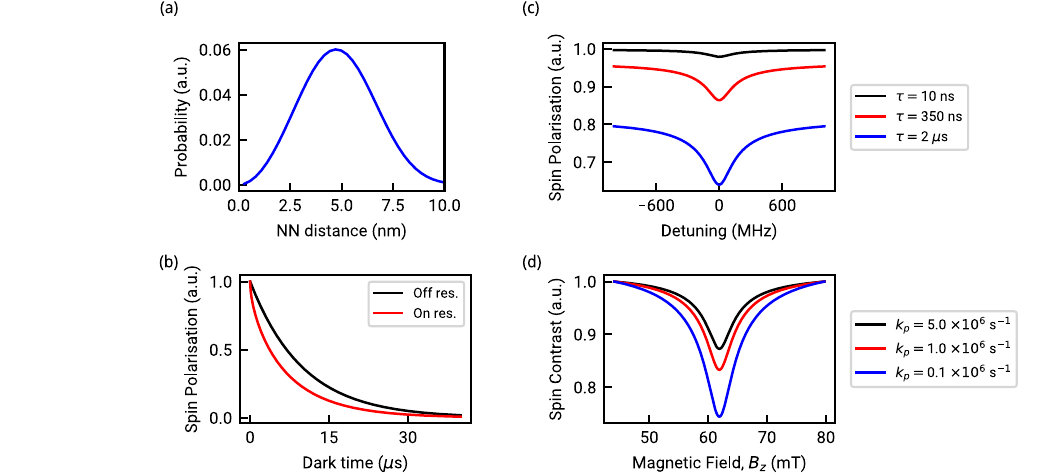}
\caption{
\textbf{Modelling cross-relaxation measurement on the $\C$ defect in a $\VB$ bath.}
(a)~Distribution of distances to the nearest $\VB$ neighbour of the $\C$ defects. 
(b)~Calculated ensemble-averaged spin relaxation curve of the $\C$ defects on and off the cross-relaxation resonance, assuming perfect initialisation of the $\C$-$\VB$ pair. 
(c)~Calculated ensemble-averaged cross-relaxation spectrum of $\C$ as a function of detuning, for a given dark time ($\tau$), assuming perfect initialisation. 
(d)~Calculated cross-relaxation spectrum as a function of magnetic field taking into account the finite spin initialisation by the laser pulse due to cross-relaxation, for different optical pumping rates $k_p$. Each spectrum was normalised to 1 away from the resonance.
}
\label{FigSI_CR}
\end{figure*}

The probability distribution for the nearest-neighbour distance in a spin ensemble of density $n$ is~\cite{Hall2014}
\begin{equation} \label{eq:proba}
    P(r) = 4 \pi n r^2 \exp \left( - \frac{4 \pi n r^3}{3} \right)~.
\end{equation}
If we take $n$ to be the density of $\VB$ spins ($n=[\VB]$), then for any given $\C$ spin Eq.~\ref{eq:proba} also describes the distance of the $\C$ spin (probe spin) to the nearest $\VB$ spin (target).
As discussed in Sec.~\ref{sec:bulk} we estimate the $\VB$ density in our sample to be $[\VB] = 10^{18}$\,cm$^{-3}$~\cite{Murzakhanov2021}.
The calculated nearest-neighbour distribution is shown in Fig.~\ref{FigSI_CR}(a), peaked below 5\,nm and with a significant population below 2\,nm separation. 

We will only consider the effect of the nearest $\VB$ (which is a good approximation for spin relaxation calculations~\cite{Hall2016}) but will average the results over the probability distribution Eq.~\ref{eq:proba} to model the measurement of an ensemble of $\C$ spins. 
Thus, we only need to model a two-spin system, namely a $\C$-$\VB$ pair separated by a distance~$r$. 
Following \textcite{Wood2016}, the additional relaxation rate due to cross-relaxation in the strong dephasing regime ($W_{\rm d}\ll\Sigma_2$~\cite{Hall2016}, justified later) is
\begin{equation}
    \Gamma_1^{\rm CR} = W_{\rm d}^2\frac{\Sigma_2}{\Sigma_2^2 + \Delta ^2}~,
    \label{eq:cr_gamma}
\end{equation}
where $W_{\rm d}$ is the dipole-dipole coupling strength, $\Sigma_2  = \Gamma_2\{\C\} + \Gamma_2\{\VB\}$ is the sum of the defect dephasing rates, and $ \Delta = \omega\{\C\} - \omega\{\VB\}$ is the detuning from the CR resonance condition. $W_{\rm d}$ is given by 
\begin{equation}
    W_{\rm d} = \left(\frac{\mu_0 h \tilde{\gamma}^2}{2 \sqrt{2}}\right) \left(\frac{3 \sin^2 \theta}{r^3}\right)~,
    \label{eq:cr_gamma2}
\end{equation}
where $\tilde{\gamma}=28$\,GHz/T is the reduced gyromagnetic ratio (taken to be identical for $\C$ and $\VB$ for simplicity), and $\theta$ is the angle between the quantization axis (magnetic field direction) and the $\C$-$\VB$ direction. 
The dephasing rate $\Gamma_2$ of each spin ensemble is related to the natural linewidth observed in a pulsed ODMR measurement~\cite{Hall2016}. 
We will take $\Gamma_2\{\C\}=2\pi\times50$\,MHz and $\Gamma_2\{\VB\}=2\pi\times160$\,MHz, as determined for the sample considered.

The total relaxation rate experienced by the $\C$ spin is then $\Gamma_1=\Gamma_1^{\rm bkg}+\Gamma_1^{\rm CR}$ where $\Gamma_1^{\rm bkg}$ is the background relaxation away for the CR resonance. We will take $1/\Gamma_1^{\rm bkg}=10\,\upmu$s which is a typical $T_1$ value for $\C$ in our samples, see Fig.~\ref{FigSI_flake}(d).
Assuming perfect initialisation of the $\C$-$\VB$ pair in the interacting state, we can plot an ensemble-averaged relaxation curve of the $\C$ spins [Fig.~\ref{FigSI_CR}(b)], on and off resonance. 
We see that the effect of CR is felt even at short dark times $\tau<1\,\upmu$s where the spin polarisation decays sharply. 
This sharp decay can be understood by considering the distance distribution, Fig.~\ref{FigSI_CR}(a). 
At the most probable distance of 4.7 nm, the added relaxation rate on resonance (averaging over an isotropic angular distribution) is $\Gamma_1^{\rm CR}\approx4\times10^4$\,s$^{-1}$, but this increases to $\Gamma_1^{\rm CR}\approx6\times10^6$\,s$^{-1}$ at 2 nm distance, which is much larger than $\Gamma_1^{\rm bkg}=10^5$\,s$^{-1}$ and causes some $\C$ spins to relax over a time scale of $\sim100$\,ns. Note that the dipole-dipole coupling strength is $W_{\rm d}/2\pi\sim1-10$\,MHz, much smaller than the dephasing rate of the coupled system ($\Sigma_2/2\pi\approx210$\,MHz), which justifies the strong dephasing regime assumption.  

If we now vary the detuning $\Delta$ and calculate the remaining spin polarisation after a fixed dark time $\tau$, we can construct a CR spectrum, shown in Fig.~\ref{FigSI_CR}(c) for different $\tau$ values. 
We see that the CR resonance is pronounced even for a relatively modest $\tau=350$\,ns, which is the value used experimentally, leading to a 10\% contrast. 
A longer $\tau$ (e.g.\ 2\,$\upmu$s) can double the contrast of the CR resonance, but this implies a longer measurement sequence which in practice does not lead to an improved signal-to-noise ratio.

Experimentally, to probe the spin polarisation we used a pulsed ODMR sequence comprised of a laser pulse of duration $t_p=1.2\,\upmu$s, a dark time $\tau=350$\,ns, and an RF $\pi$-pulse. The laser pulse serves not only to read out the spin polarisation but also to re-set it to some initial value. 
However, at the CR resonance the relaxation rate is sufficiently high that it reduces the level of polarisation reached at the end of the laser pulse. 
Consequently, the ODMR contrast at the CR resonance is reduced even further. 

To account for this effect, we modelled the spin dynamics during the entire pulsed ODMR sequence, following a similar treatment to \textcite{Robertson2023} (refer to equations S12-S19 of their paper). 
In brief, we consider a two-level model (the two spin states $\ket{\uparrow}$ and $\ket{\downarrow}$ of the $\C$ defect) where the laser adds a rate $k_p$ driving the transition $\ket{\downarrow}\rightarrow\ket{\uparrow}$, while the relaxation rate $\Gamma_1$ is always on. 
The RF $\pi$-pulse is assumed to instantly swap the two spin populations. 
The measured PL is related to the spin populations during the laser pulse, which allows us to derive an expression for the PL measured from the ODMR sequence with the $\pi$-pulse (signal $S$) as well as the PL measured with the same sequence without the $\pi$-pulse (reference $R$) which we use experimentally for normalisation.
The ODMR contrast is defined as $\mathcal{C}=1-R/S$.

Using this model, we can compute the ODMR contrast (${\mathcal{C}}$) versus detuning ($\Delta$) spectrum, and finally ${\mathcal{C}}$ versus $B_z$ as presented in Fig.~3(e) of the main text. 
Additional simulated spectra are shown in Fig.~\ref{FigSI_CR}(d) for different values of the pumping rate $k_p$. 
As expected, the finite pumping rate $k_p$ enhances the contrast of the CR resonance, e.g.\ to 17\% for $k_p\approx10^6$\,s$^{-1}$ and 25\% for $k_p\approx10^5$\,s$^{-1}$, matching the experimentally observed contrast. 
We independently estimated the experimental value of $k_p$ to be in the range $k_p\approx10^5-10^6$\,s$^{-1}$ based on a separate measurement of $\mathcal{C}$ as a function of $t_p$ following the analysis in \textcite{Robertson2023}. For the curve displayed in the main text, we assumed $k_p\approx10^5$\,s$^{-1}$.

To model the CR measurement with the $\VB$ ensemble as the probe, we apply the same model as described above except that at the CR resonance the $\VB$ spins are coupled with the entire bath of $S=1/2$ spins, which includes the $\C$ defects but also other paramagnetic defects that are optically inactive. 
Since this density is a priori unknown, we leave it as a free parameter. 
All the other parameters are taken as above, except for the pumping rate $k_p$ which was determined to be $k_p\approx5\times10^6$\,s$^{-1}$~\cite{Robertson2023}. We find a best fit to the data using $[S=1/2]=5\times10^{18}$\,cm$^{-3}$.

\section{Magnetic imaging with the MOVPE-grown hBN film} \label{sec:magnetic_imaging}

\subsection{Sample preparation}

\begin{figure*}[t!]
\centering
\includegraphics[width=0.8\textwidth]{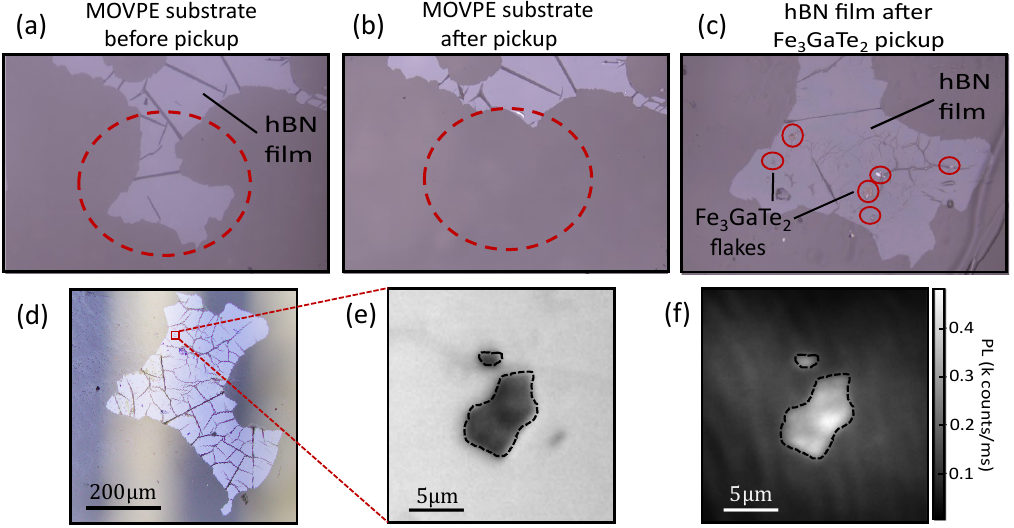}
\caption{{\bf Sample preparation for the magnetic imaging experiment.} 
(a,b)~Optical micrographs of the MOVPE-grown hBN film on the original sapphire substrate (a) before and (b) after picking up with the PC/PDMS stamp. 
The red dashed circle indicates the region of the film that was picked up. 
(c)~Optical micrograph of the stamp after picking up the Fe$_3$GaTe$_2$ flakes (circled in red). 
(d)~Optical micrograph of the hBN/Fe$_3$GaTe$_2$ sample after release on the quartz substrate and positioning above the coplanar waveguide (seen in the background). 
(e)~Reflection image of the region used for magnetic imaging, with the Fe$_3$GaTe$_2$ flake appearing dark. 
(f)~PL image of the same region, where the Fe$_3$GaTe$_2$ flake now appears bright as it reflects the PL from the $\C$ defects.}
\label{FigSI_MOVPE_transfer}
\end{figure*}

For the magnetic imaging experiment reported in Fig.~4 of the main text, we used the 40-nm-thick MOVPE-grown hBN film described in Sec.~\ref{sec:MOVPE_charac}. As a test magnetic sample, we used a Fe$_3$GaTe$_2$ crystal purchased from PrMat.  Fe$_3$GaTe$_2$ microflakes were first exfoliated in an Argon-filled glove box onto a Si/SiO$_2$ substrate using the scotch-tape technique. Then, the hBN film and Fe$_3$GaTe$_2$ flakes were picked up in sequence by a polycarbonate/polydimethylsiloxane (PC/PDMS) stamp [Fig.~\ref{FigSI_MOVPE_transfer}(a-c)] and released onto a quartz substrate. Finally, the PC film was dissolved with chloroform.

The quartz substrate was then positioned above a coplanar waveguide to facilitate ODMR measurements [Fig.~\ref{FigSI_MOVPE_transfer}(d)]. A Fe$_3$GaTe$_2$ flake was located from the reflection image [Fig.~\ref{FigSI_MOVPE_transfer}(e)], and then via the PL image [Fig.~\ref{FigSI_MOVPE_transfer}(f)] where the flake typically appears brighter than the background because it reflects additional PL from the $\C$ defects.

\subsection{Magnetic imaging with the $\C$ defects}

ODMR spectra were acquired using the widefield microscope setup described in Sec.~\ref{sec:setup}, which allowed us to form magnetic field images with a diffraction-limited spatial resolution~\cite{Scholten2021}. 
The objective lens used here has a numerical aperture of ${\rm NA}=0.45$, so considering a central PL emission wavelength of $\lambda=600$\,nm, the FWHM of the point-spread function (an Airy disk in the diffraction-limited case) is approximately $\lambda/2{\rm NA}\approx670$\,nm. 
The magnitude of the applied magnetic field was fixed to $B_0\approx35$\,mT, and the direction was varied from the out-of-plane direction ($\theta_B=0^\circ$) to the in-plane direction ($\theta_B=90^\circ$) by manually rotating a permanent magnet. 

The ODMR spectra were fit with a Lorentzian function to extract the resonance frequency $f_r$, which was converted to a magnetic field $B_{\rm tot}=hf_r/g_C\mu_B$. 
The total magnetic field is $\vec{B}_{\rm tot}=\vec{B}_0+\vec{B}_{\rm s}$ where $\vec{B}_{\rm s}$ is the sample's stray field. 
In the limit $|\vec{B}_{\rm s}|\ll|\vec{B}_0|$ (which is satisfied here), we can use the scalar approximation $B_{\rm tot}=B_0+\Delta B$ where $\Delta B$ is the projection of $\vec{B}_{\rm s}$ onto the direction of the external field $\vec{B}_0$.
To obtain a map of $\Delta B$, we take the raw map $B_{\rm tot}$ and subtract the background value away from the magnetic flake, which serves as an estimate of $B_0$. 
To remove imaging artefacts that are sometimes present away from the flakes (and therefore non-physical), we perform an additional background subtraction following the procedure described in Ref.~\cite{BroadwayAM2020}.  

\begin{figure*}[h!]
\centering
\includegraphics[width=0.8\textwidth]{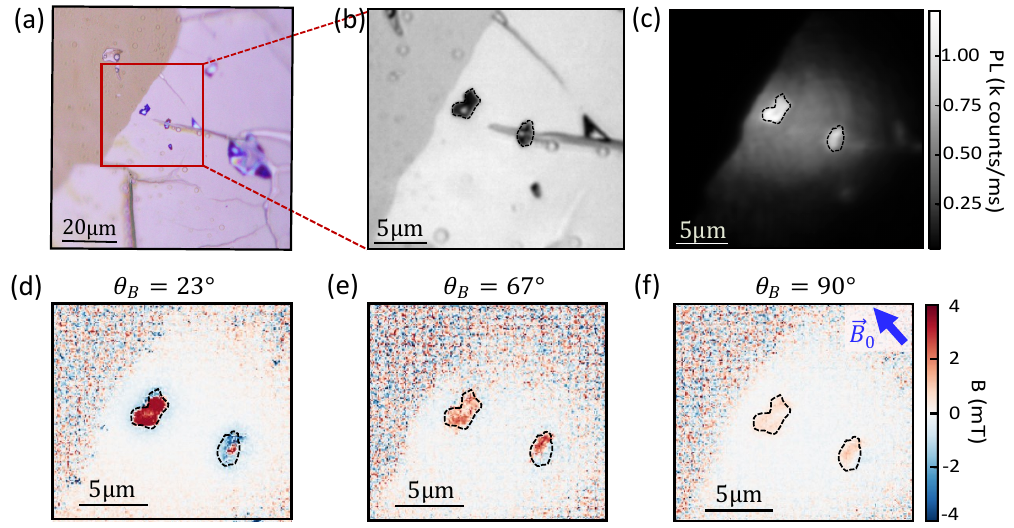}
\caption{{\bf Magnetic imaging of additional Fe$_3$GaTe$_2$ flakes.} 
(a)~True-color optical micrograph of Fe$_3$GaTe$_2$ flakes covered by the 40-nm hBN film. 
(b)~Reflection image of two Fe$_3$GaTe$_2$ flakes selected for magnetic imaging (indicated by dashed lines). 
(c)~Corresponding PL image, with the Fe$_3$GaTe$_2$ flakes appearing bright due to reflection of the PL from the $\C$ defects. 
(d-f)~Magnetic field maps recorded for various angles of the applied field $\vec{B}_0$: (d) $\theta_B=23^\circ$, (e) $\theta_B=67^\circ$, (f) $\theta_B=90^\circ$. 
The in-plane projection of $\vec{B}_0$ is indicated by the blue arrow in (f).}
\label{FigSI_MOVPE_more_flakes}
\end{figure*}

In Fig.~4 of the main text, 3 magnetic images of the same Fe$_3$GaTe$_2$ flake taken with 3 different angles $\theta_B$ were shown. 
The corresponding PL image is shown in Fig.~\ref{FigSI_MOVPE_transfer}(f). 
In Fig.~\ref{FigSI_MOVPE_more_flakes}, we show images obtained for two additional Fe$_3$GaTe$_2$ flakes, again for different angles $\theta_B$. 
A similar qualitative behaviour to the flake studied in Fig.~4 is observed. 
That is, as $\vec{B}_0$ is rotated away from the easy axis of magnetisation, the magnitude of the stray field is reduced overall, indicating a partial demagnetisation of the flakes, while the symmetry of the residual stray field pattern (positive and negative field on opposite edges) suggests that the remaining net magnetisation still points along the easy axis. 
These statements are supported by numerical simulations of the stray field, see next section. 

\subsection{Stray field simulations}

The measured stray field maps were simulated by computing the stray field from a uniformly magnetized flake averaged over the thickness of the hBN film, and then applying a convolution with the point-spread function of our optical microscope. 
We assumed a thickness of 50\,nm for the magnetic Fe$_3$GaTe$_2$ flake and 40\,nm for the hBN film. 
We also assumed a standoff of 10\,nm between the magnetic flake and the hBN film, to account for the roughness of the hBN/Fe$_3$GaTe$_2$ interface and because the PL from the $\C$ defects closest to the Fe$_3$GaTe$_2$ flake are likely to be quenched. 

For the point-spread function, for simplicity we assumed a Gaussian function with a 1-$\mu$m FWHM. 
This is more than the diffraction limit (670-nm FWHM) but led to a better match to the experiment in the line profile comparison [Fig.~4(f) of the main text]. 
Using a 670-nm FWHM convolution predicts a larger magnetic field near the edges of the flake than observed experimentally. 
However, this apparent discrepancy can be explained by the unrealistic assumption of a perfectly uniform magnetisation. 
In Sec.~\ref{sec:mag_recon}, we use the magnetic field data to reconstruct the magnetisation map, and find that the magnetisation in fact decreases towards the edges.

\begin{figure*}[t!]
\centering
\includegraphics[width=0.95\textwidth]{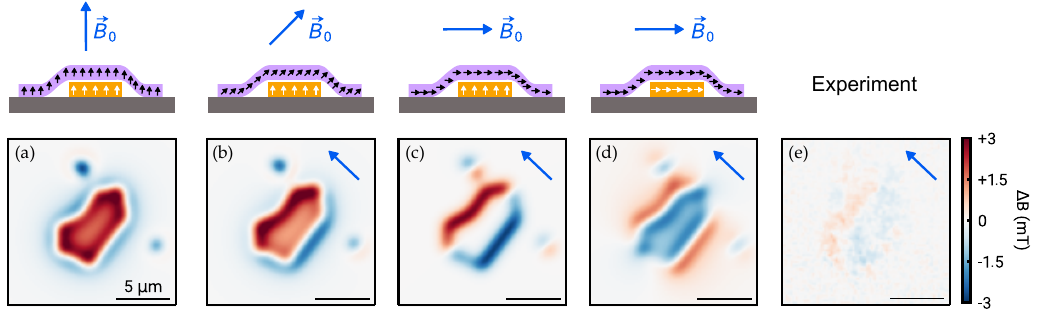}
\caption{{\bf Simulations of the magnetic stray field from a Fe$_3$GaTe$_2$ flake.} 
(a-c)~Stray field maps for a flake magnetized along the $z$ direction for various angles of the projection axis $\vec{B}_0$: (a) $\theta_B=0^\circ$, (b) $\theta_B=45^\circ$, (c) $\theta_B=90^\circ$. 
(d)~Stray field map for a flake magnetized in-plane with $\theta_B=90^\circ$. 
The in-plane projection of $\vec{B}_0$ is indicated by the blue arrow, which also sets the direction of the in-plane magnetisation in (d). 
(e)~Measured stray field map at $\theta_B=90^\circ$, reproduced from Fig.~4(h) of the main text.}
\label{FigSI_magnetic_sims}
\end{figure*}

Simulations of the flake imaged in Fig.~4 of the main text are shown in Fig.~\ref{FigSI_magnetic_sims}(a-c) for different angles $\theta_B$, assuming the flake is uniformly magnetized along the $z$ direction. 
Here a magnetisation value of $M_s=160$\,kA/m was used, which was found to match the data well for the $\theta_B=0^\circ$ case. 
This value is in reasonably good agreement with the reported value for bulk Fe$_3$GaTe$_2$ at room temperature ($M_s\approx300$\,kA/m~\cite{Zhang2022}). 

In the simulations, we see that the symmetry of the stray field changes as $\theta_B$ is increased, but the maximum magnitude remains roughly unchanged. This is in contrast with the experiment (Fig.~4 as well as Fig.~\ref{FigSI_MOVPE_more_flakes}) where the stray field becomes weaker overall as $\theta_B$ is increased. 
This can be explained by partial demagnetisation of the flake, which means that multiple domains of opposite magnetisation are formed \add{(up/down domains, which is energetically favourable in the absence of an out-of-plane magnetic field)}, with a characteristic domain size below our spatial resolution. 
This is consistent with Ref.~\cite{Zhang2022} where magnetic force microscopy images at room temperature in the absence of external field revealed sub-micron domains. \add{The zero-field condition in Ref.~\cite{Zhang2022} is essentially equivalent to our in-plane field condition in terms of magnetic energy, and the multi-domain structure they observed would indeed give a nearly vanishing net stray field at our spatial resolution.} 
The symmetry of the residual stray field pattern observed in our experiments under an in-plane field ($\theta_B=90^\circ$), with positive and negative fields on opposite edges, remains consistent with a residual net magnetisation pointing along the easy axis ($z$) rather than canting towards the direction of the applied field, see simulations of the extreme cases in Fig.~\ref{FigSI_magnetic_sims}(c,d) compared with the experiment [reproduced in Fig.~\ref{FigSI_magnetic_sims}(e)]. 
This out-of-plane net magnetisation is consistent with the strong perpendicular magnetic anisotropy of Fe$_3$GaTe$_2$~\cite{Zhang2022}, \add{and with the above interpretation of multiple up/down domains}. 

\subsection{Magnetisation reconstruction}
\label{sec:mag_recon}

From a given magnetic field map projected along a known direction (e.g. $B_z$), it is possible to reconstruct both other magnetic field components ($B_x$ and $B_y$) and, when certain assumptions are valid, the underlying source of the magnetic fields~\cite{Broadway2020PA}. 
Here, we employ the assumption that the magnetisation of the flake has a uniform direction ($z$-direction). 
This is a valid assumption as it is known that Fe$_3$GaTe$_2$ has a large perpendicular anisotropy, which orientates the magnetisation along the $z$-axis of the material~\cite{Zhang2022}. 

\begin{figure}[b!]
\centering
\includegraphics[width=1\textwidth]{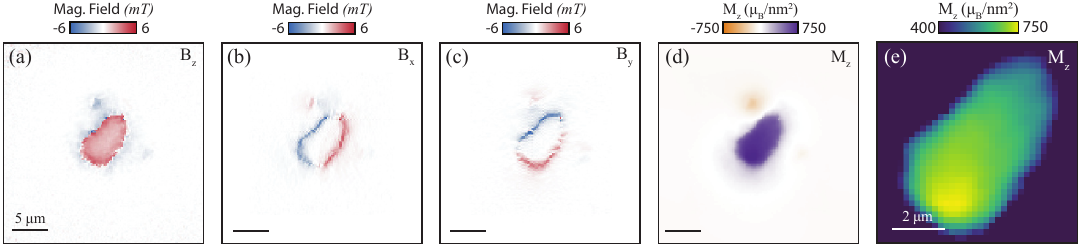}
\caption{{\bf Reconstructed magnetisation map of a Fe$_3$GaTe$_2$ flake.} 
(a)~Raw magnetic image taken in the $z$ direction, i.e.\ $B_z$ map. 
This is the same image as Fig.~4(d) of the main text. 
(b,c)~Reconstructed magnetic field components $B_x$ and $B_y$ respectively. 
(d)~Reconstructed areal magnetisation along the $z$ direction, i.e.\ $M_z$ map. 
(e)~Magnified and re-scaled version of (d) emphasising magnetisation variations within the main flake.}
\label{FigSI_mag_recon}
\end{figure}

In Fig.~\ref{FigSI_mag_recon}, we show the results from one such reconstruction process. We take the originally measured $B_z$ magnetic field image shown in Fig.~4(d) of the main text [reproduced in Fig.~\ref{FigSI_mag_recon}(a)], transform it into Fourier space and extract the $B_x$ and $B_y$ magnetic field components [Fig.~\ref{FigSI_mag_recon}(b,c)]. 
Likewise, we can convert the original magnetic field image $B_z$ into a magnetisation image $M_z$ [Fig.~\ref{FigSI_mag_recon}(d)]. 
The reconstruction method under these assumptions is detailed in Ref.~\cite{Broadway2020PA}. 

The resulting $M_z$ map [Fig.~\ref{FigSI_mag_recon}(d)] successfully reconstructs the magnetisation of the main flake (which is magnetised along the direction of the applied field, i.e.\ $M_z>0$), and also reconstructs a smaller flake above the main flake, which seems to be magnetised in the opposite direction. 
A re-scaled version allows us to better analyse the variations across the main flake [Fig.~\ref{FigSI_mag_recon}(e)]. 
We see that the magnetisation decreases from $\approx750\,\mu_B/{\rm nm}^2$ near the bottom of the flake, to $\approx500\,\mu_B/{\rm nm}^2$ towards the top and edges. 
Note that this is the areal magnetisation, i.e.\ the density of magnetic moments per unit area. Assuming the flake is 50 nm thick, the range above corresponds to a volume magnetisation of 90-140 kA/m.  

\subsection{Magnetic sensitivity}

To facilitate comparison with other magnetic sensors, we calculate the magnetic sensitivity using the usual figure of merit for a shot-noise-limited CW-ODMR measurement~\cite{Rondin2014,Barry2020},
\begin{equation}
    \eta_B \approx \frac{h}{g_C \mu_B}\frac{\Delta \nu}{\mathcal{C}\sqrt{\mathcal{R}}},
\end{equation}
where $\mathcal{R}$ is the photon count rate, $\mathcal{C}$ is the ODMR contrast, and $\Delta \nu$ is the ODMR linewidth. 
In our experimental setup, with the 40-nm MOVPE hBN film we have $\mathcal{R}\approx10^7$\,photons/s for a (500\,nm)$^2$ pixel (just under the diffraction limit), under the maximum laser power available. 
The ODMR linewidth is $\Delta \nu\approx60$\,MHz and the maximum contrast is $\mathcal{C}\approx0.5\%$. 
This gives a sensitivity of $\eta_B\approx130\,\mu$T$/\sqrt{\rm Hz}$ from a single pixel.

This sensitivity is an order of magnitude better than for $\VB$ defects in hBN flakes of comparable thickness~\cite{HealeyNP2023,HuangNC2022}, mainly thanks to a much higher photon count rate $\mathcal{R}$ for the $\C$ defects. 
On the other hand, state-of-the-art layers of nitrogen-vacancy centres in diamond of similar thickness exhibit a superior per-pixel sensitivity of $\eta_B\approx10\,\mu$T$/\sqrt{\rm Hz}$~\cite{Healey2020}, due to a better ODMR contrast and linewidth and a similar photon count rate to our MOVPE hBN film. 
Thus, magnetic imaging with $\C$ defects in MOVPE hBN films already appears reasonably competitive compared to the state of the art, with an order of magnitude inferior sensitivity (without any material or protocol optimisation for the hBN case) but a much easier fabrication and a unique ominidirectional magnetometry capability.     
 
We note that in the magnetic imaging experiments reported in Fig.~4 of the main text, the laser power and RF power were lowered in order to minimize heating of the magnetic sample, and so the photon count rate $\mathcal{R}$ and the contrast $\mathcal{C}$ were below the maximum values. 
Moreover, the RF frequency was scanned rather than fixed to a single optimised value. 
As a result, the actual magnetic sensitivity in these experiments was about an order of magnitude larger, $\eta_B\approx1$\,mT$/\sqrt{\rm Hz}$ per pixel, as estimated from the pixel-to-pixel noise observed in the images.

\add{The stray field from a monolayer ferromagnet is typically in the range 10-100\,$\mu$T at our spatial resolution~\cite{HealeyACS2022}. Thus, successful imaging of a monolayer would require acquisition times of the order of a day under the conditions of Fig.~4, which is feasible although longer than the few hours used to acquire the images reported here. With an experiment optimised to reach the projected sensitivity of $\eta_B\sim100\,\mu$T$/\sqrt{\rm Hz}$ per pixel, acquisition times for monolayer imaging would be reduced to a few hours.}

\section{Demonstration of dual-spin-species multi-modal imaging} \label{sec:dual_imaging}

In this section, we demonstrate the possibility of performing multi-modal imaging using both spin species ($\C$ defects and $\VB$ defects) within a single hBN system. 
While the MOVPE-grown hBN film used in the previous section did not contain $\VB$ defects (the creation of $\VB$ defects in MOVPE films will be studied in future work), we have shown in Fig.~1 and 2 of the main text that both spin species can be engineered to co-exist in hBN powders and exfoliated flakes, respectively. 
Below, we use each of these systems to demonstrate dual-spin imaging.

\subsection{With an hBN powder film} \label{sec:powder_imaging}

\begin{figure*}[t!]
\centering
    \includegraphics[width=1\textwidth]{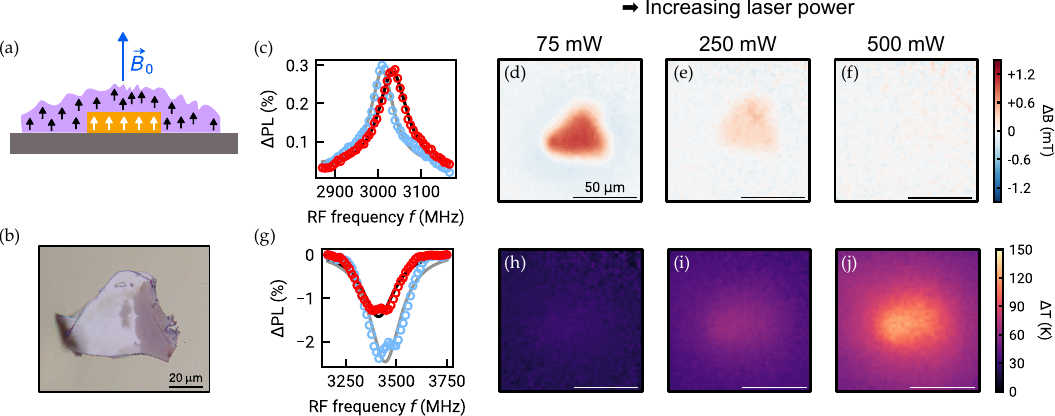}
\caption{\textbf{Dual-spin imaging with an hBN powder film.}
(a)~Schematic of the experiment. 
A hBN powder film (a few microns thick) is formed on top of a Fe$_3$GaTe$_2$ magnetic flake. 
(b)~Optical micrograph of the Fe$_3$GaTe$_2$ flake prior to depositing the hBN powder. 
(c)~ODMR spectra of the $\C$ defects. 
Here an external field $B_0\approx100$\,mT is applied along the $z$ direction. 
(d-f)~Stray field maps of the Fe$_3$GaTe$_2$ flake for 3 different laser powers, causing different levels of laser-induced heating. 
The ODMR spectrum on (red) and off (blue) of the flake are compared in (c).
(g)~ODMR spectra of the $\VB$ defects in zero magnetic field ($B_0=0$). 
(h-j)~Temperature maps ($\Delta T$ relative to room temperature) for the same 3 different laser powers as in (d-f).
The ODMR spectrum on (red) and off (blue) of the flake are again compared in (g).}
\label{FigSI_dual_FeGaTe}
\end{figure*}

Similar to the experiments described in Sec.~\ref{sec:magnetic_imaging}, we used a Fe$_3$GaTe$_2$ flake as a test magnetic sample, but replaced the 40-nm-thick MOVPE hBN film with a hBN powder film (a few micron thick) formed by drop casting as described in Sec.~\ref{sec:powder_prep} [Fig.~\ref{FigSI_dual_FeGaTe}(a)]. 
We first performed magnetic imaging of a Fe$_3$GaTe$_2$ flake [see optical micrograph in Fig.~\ref{FigSI_dual_FeGaTe}(b)] under an applied field $B_0\approx100$\,mT along the easy axis of magnetisation ($z$ axis). 
An ODMR spectrum of the $\C$ defects in the powder film is shown in Fig.~\ref{FigSI_dual_FeGaTe}(c), and resulting stray field maps are shown in Fig.~\ref{FigSI_dual_FeGaTe}(d-f). 

Three different magnetic images were acquired using different laser powers, causing different levels of laser-induced heating of the Fe$_3$GaTe$_2$ flake. 
At the lowest laser power [Fig.~\ref{FigSI_dual_FeGaTe}(d)], the flake is uniformly magnetized and there is no sign of laser-induced heating, as reducing the laser power further did not change the stray field magnitude. 
As the laser power is increased, the stray field magnitude reduces and vanishes at the highest power [Fig.~\ref{FigSI_dual_FeGaTe}(f)], indicating that the flake's temperature is near or above the Curie temperature ($T_c\approx350$\,K for bulk Fe$_3$GaTe$_2$~\cite{Zhang2022}).

Next, we use the $\VB$ defects to measure the local temperature in the same conditions. 
Here the same region of the sample is imaged under the same conditions as above, the only difference being that the external field is removed ($B_0=0$) and the NIR PL is now collected, to enable ODMR measurements of the $\VB$ defects in the powder. 
The $\VB$ ODMR spectrum is shown in Fig.~\ref{FigSI_dual_FeGaTe}(g), which is fit with a Lorentzian function to estimate the zero-field splitting $D$, then converted to a temperature change $\Delta T$ relative to the very low laser power case, following the same procedure as in Ref.~\cite{HealeyNP2023}. 
The resulting $\Delta T$ maps are shown in Fig.~\ref{FigSI_dual_FeGaTe}(h-j), recorded with the same three laser powers as in Fig.~\ref{FigSI_dual_FeGaTe}(d-f). 
While there is no measurable temperature increase at the lowest laser power, we observe a local increase in the centre of the image (coinciding with the centre of the laser spot) reaching a maximum of $\Delta T\approx100$\,K at the maximum laser power, which explains why the magnetism of the flake was fully quenched, see Fig.~\ref{FigSI_dual_FeGaTe}(f). 

We note that the measured temperature is that of the hBN powder rather than that of the Fe$_3$GaTe$_2$ flake, however, because the powder film is in physical contact with the sample one can expect a relatively good thermal coupling and hence a relatively faithful reading of the sample's temperature. 
Moreover, since the film itself is relatively porous [see Fig.~\ref{FigSI_powder_films}(c)], lateral thermal conduction within the film is likely to have a minimal impact on the heat transport in the overall system, making the hBN powder film a minimally invasive temperature sensor.

Importantly, there is no apparent cross-talk between the magnetic images obtained with the $\C$ defects [Fig.~\ref{FigSI_dual_FeGaTe}(d-f)] and the temperature maps obtained with the $\VB$ defects [Fig.~\ref{FigSI_dual_FeGaTe}(h-j)]. 
This implies that the temperature variations across the image do not measurably affect the resonance frequency $f_r$ of the $\C$ defects. 
In other words, magnetometry with the $\C$ defects seems to be robust against temperature changes in the conditions of Fig.~\ref{FigSI_dual_FeGaTe}. 
This is somewhat expected, since $f_r=g_C\mu_BB_0/h$ is purely a Zeeman splitting and $g_C\approx2$ contains no measurable orbital contribution, and so $f_r$ should not depend on temperature to first order.

\subsection{With an exfoliated hBN flake}

\begin{figure*}[t!]
\centering
\includegraphics[width=1\textwidth]{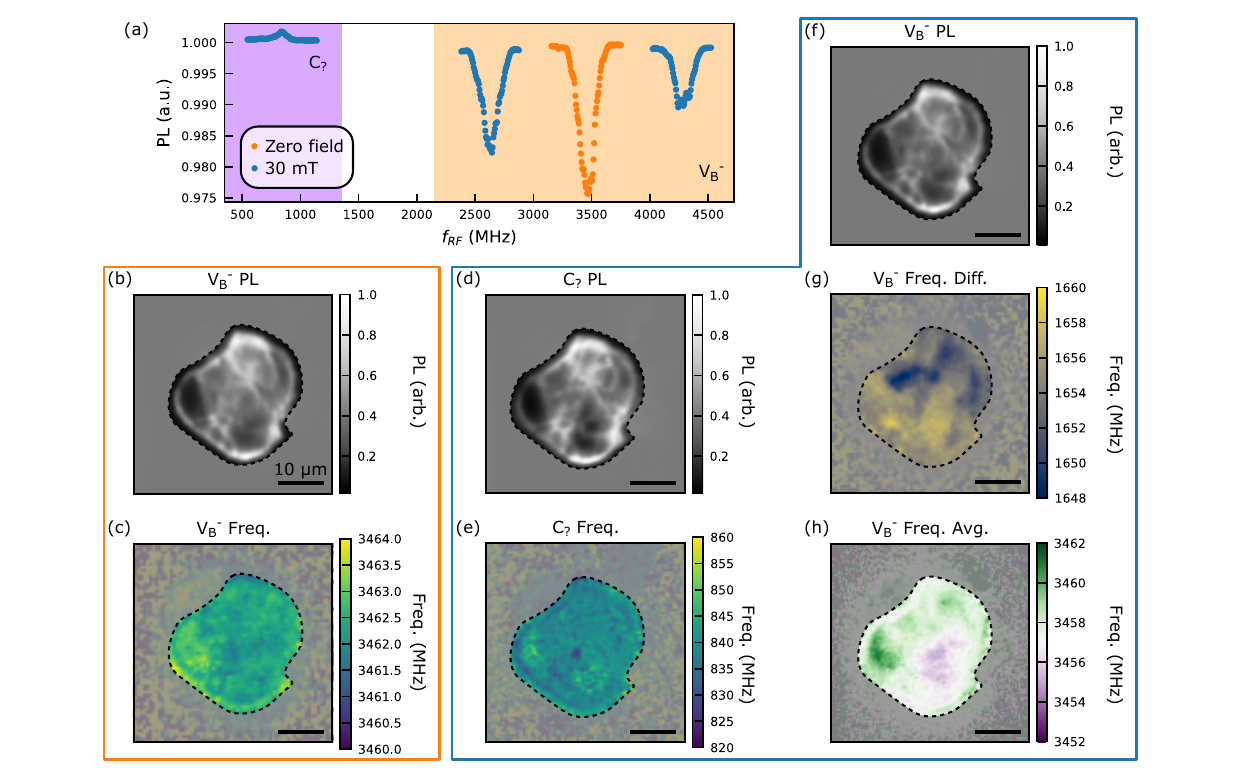}
\caption{\textbf{Dual-spin imaging with an exfoliated hBN flake.}
(a)~ODMR spectrum of a hBN flake under $B_0=0$ (orange) and under $B_0=30$\,mT applied along the $z$ direction (blue). 
The spectrum is constructed from both the $\C$ (purple region) and $\VB$ (orange region) defects. 
(b)~PL image for the $\VB$ from an exfoliated flake of hBN. 
(c)~ODMR frequency map of $\VB$ at $B_0=0$, corresponding to a map of the zero-field splitting $D$, which depends on the local temperature and strain. 
(d,e)~Same as (b,c) except images are of the $\C$ defect under $B_0=30$\,mT, and so (e) corresponds to a magnetic field map. 
(f)~PL image for the $\VB$ under $B_0=30$\,mT, nominally identical to (b). 
(g) Frequency difference map for the two $\VB$ ODMR peaks, $\Delta f = f_2-f_1$, corresponding to the magnetic field projected along the local $c$-axis of the flake. 
(h)~Average frequency map for the two $\VB$ ODMR peaks, $\bar{f} = (f_1+f_2)/2$, corresponding to the zero-field splitting $D$.}
\label{FigSI_multi_img}
\end{figure*}

Dual-spin imaging can also be performed using a thin flake exfoliated from an electron-irradiated bulk hBN crystal, prepared as described in Sec.~\ref{sec:crystal_charac}.
Here we image a hBN flake without any test magnetic sample present, with the aim of illustrating the pros and cons of a flake versus a powder film or MOVPE film for the purpose of imaging. 

Compared to the powder, here an external magnetic field can be applied along the $\VB$ symmetry axis ($c$-axis of the hBN crystal) to cleanly split the ODMR into two peaks. The applied field magnitude $B_0\approx30$\,mT is chosen for a simple case where the $\C$ ODMR peak is at a lower frequency than the lowest frequency peak of the $\VB$ spectrum thus giving the combined $\C$-$\VB$ spectrum three peaks at $840$\,MHz, $2625$\,MHz, and $4275$\,MHz respectively [blue spectrum in Fig.~\ref{FigSI_multi_img}(a)]. 

Before applying an external field, we first record an ODMR spectrum from the $\VB$ defects at $B_0=0$ [orange ODMR spectrum in Fig.~\ref{FigSI_multi_img}(a)].
Fitting the spectrum with a single Lorentzian function allows us to estimate the zero-field splitting $D$, which carries information about the local temperature as well as strain~\cite{Gottscholl2021NC}. 
There is also a small splitting of the central peak caused by strain and electric fields~\cite{Lyu2022,Gong2023}, ignored in this section. 
The $\VB$ PL image of the flake is shown in Fig.~\ref{FigSI_multi_img}(b) and the corresponding $D$ map in Fig.~\ref{FigSI_multi_img}(c). 
The PL image features some bright lines attributed to wrinkles, but is sufficiently uniform otherwise to allow ODMR mapping through the entire flake. 
Meanwhile, the $D$ map reveals some features that are likely due to strain from within the flake and/or interaction with the substrate. 
Thus, hBN flakes are generally well suited for strain mapping when temperature variations are negligible, in contrast to hBN powder films which are mechanically loosely coupled to the substrate and so are better suited for temperature mapping.

Applying a bias field $B_0$ enables magnetometry with the $\C$ defects, see PL image and corresponding $\C$ frequency map in Fig.~\ref{FigSI_multi_img}(d,e), respectively. 
Here the frequency map is indicative of the total magnetic field, which is nominally uniform across the flake. 
Meanwhile, the $\VB$ ODMR spectrum at the same field $B_0$ is fit to extract the two resonance frequencies $f_1$ and $f_2$. 
The frequency difference $\Delta f = f_2-f_1$ is related to the total magnetic field projection along the local $c$-axis ($B_c$) via $\Delta f=2g_V\mu_BB_c/h$, shown in Fig.~\ref{FigSI_multi_img}(g). 
This measurements is therefore sensitive to spatial variations in the $c$-axis direction due e.g.\ to ripples in the flake, which cause variations in the projected field $B_c$. 
This explains why there are extended features in the projected field map Fig.~\ref{FigSI_multi_img}(g) that are not seen in the total field map Fig.~\ref{FigSI_multi_img}(e). 
The mean frequency $\bar{f} = (f_1+f_2)/2$ is approximately equal to the zero-field splitting $D$, i.e.\ $\bar{f}\approx D$ (exact for a perfectly aligned bias field), shown in Fig.~\ref{FigSI_multi_img}(h). 
This image should be nominally identical to Fig.~\ref{FigSI_multi_img}(c) but additional features may arise due to the variations in the projected field $B_c$ mentioned above, which in turn cause small (second-order) shifts in $\bar{f}$ due to the Zeeman effect. 
Thus, these measurements illustrate that careful analysis is required to properly extract information from $\VB$ measurements in flakes, whereas magnetometry with the $\C$ defects is intrinsically robust to geometrical imperfections of the host flake.
Nevertheless, the results in Fig.~\ref{FigSI_multi_img} demonstrate the feasibility of dual-spin imaging using hBN exfoliated flakes.   


\end{document}